\documentclass[12pt]{iopart}

\pdfoutput=1

\usepackage{iopams}
\usepackage{cite}
\usepackage{siunitx}
\usepackage{nicefrac}
\usepackage{graphicx}
\usepackage{tabulary}
\usepackage{xcolor}

\let\vec\bi

\begin{document}
\title[X-ray lenses for nanometer focusing]{A ray-trace analysis of X-ray multilayer Laue lenses for nanometer focusing}

\author{H N Chapman$^{1,2,3}$ and S Bajt$^{3,4}$}
\address{$^ 1$ Center for Free-Electron Laser Science, DESY,
  Notkestrasse 85, 22607 Hamburg, Germany}
\address{$^ 2$ Department of Physics, Universität
  Hamburg, Luruper Chaussee 149, 22761 Hamburg, Germany}
\address{$^ 3$ Hamburg Centre for Ultrafast Imaging, Luruper Chaussee 149, 22761 Hamburg, Germany}
\address{$^ 4$ DESY, Notkestrasse 85, 22607 Hamburg, Germany}
\ead{henry.chapman@desy.de}

\begin{abstract}
Thick diffractive optical elements offer a promising way to achieve focusing or imaging at
a resolution approaching \SI{1}{\nano\meter} for X-ray wavelengths shorter than about
\SI{0.1}{\nano\meter}.  Efficient focusing requires that these are fabricated with
structures that vary in period and orientation so that rays obey Bragg's law over the
entire lens aperture and give rise to constructive interference at the focus.  Here the
analysis method of ray-tracing of thick diffractive optical elements is applied to such
lenses to optimise their designs and to investigate their operating and manufacturing
tolerances.  Expressions are provided of the fourth-order series expansions of the wavefront aberrations and transmissions of
both axi-symmetric lenses and pairs of crossed lenses
that each focuses in only one dimension like a cylindrical lens.
We find that aplanatic zone-plate designs, whereby aberrations are corrected over a large
field of view, can be achieved by axi-symmetric lenses but not the crossed lenses.  We investigate the
performance of \SI{1}{\nano\meter}-resolution lenses with focal lengths of about
\SI{1}{\milli\meter} and show their fields of view are mainly limited by the acceptance
angle of Bragg diffraction, and that aberrations can limit the performance of lenses with longer focal
lengths.  We apply the ray-tracing formalism for a tolerancing analysis of imperfect lenses
and examine some strategies for the correction of their aberrations.
\end{abstract}

%
\vspace{2pc}
%
%
%
%

\section{Introduction}
\label{sec:intro}
Thick diffractive optical elements for X-ray wavelengths, including so-called
sputter-sliced zone plates and multilayer Laue lenses (MLLs), are fabricated by depositing
alternating layers of materials onto a substrate to achieve the required diameter or
height of the lens, and then slicing the lens from this
structure~\cite{Yan:2014,Yan:2010,Bionta:1994}.  To date, imaging resolutions of about
\SI{5}{\nano\meter} have been achieved~\cite{Doering:2013,Bajt:2018}.  Nanometer
resolution requires layers of comparable or smaller period than the diffraction-limited
spot size, positioned with high enough
accuracy to avoid wavefront aberrations. At the wavelengths considered here---of about
\SI{0.1}{\nano\meter} and below---reasonable
diffraction efficiency demands structures that are hundreds or thousands of times thicker
(in the propagation direction) than the layer period and tilted to fulfil the Bragg
condition of diffraction~\cite{Yan:2014}.  Such diffractive optical elements can be used as an objective lenses to
construct various kinds of X-ray microscopes, including a transmission microscope where
the lens forms a magnified image of an object on an area detector or a scanning
transmission microscope where the lens creates a focused probe through which the object is
scanned while mapping the transmission or emission of the sample.  In all cases, an
understanding of the imaging characteristics of the lens helps to optimise the design of
the microscope and to specify its tolerances.

X-ray diffractive optical elements such as Fresnel zone plates and MLLs are usually
designed according to ray-optics principles and often additionally analysed in the
framework of wave optics calculations, dynamical diffraction, or coupled-wave theories to
model the performance at one or several field
points~\cite{Maser:1992,Yan:2007,Hu:2015,Yan:2017,Ali:2020}.  Ray-tracing allows rapid and
accurate analysis of complex geometries over extensive parameter spaces, and can be used
as a computational engine for optimisation of particular parameters in a design.  Such an
analysis is used in the optical industry for the design of instruments such as telescopes,
microscopes, and micro-lithography systems, where optimal optical performance is paramount.
Ray tracing has long been applied to model Fresnel zone plates, giving insights into
aplanatic designs, for example~\cite{Murty:1960,Young:1972,Welford:1973,Hetirick:1986}.  The modelling
of diffractive optical elements in complex optical systems has been well established, with
the development of holographic optical elements to produce arbitrary wavefronts abetted by
the rise of computational ray-tracing methods and
software~\cite{Latta:1971,Close:1975,Welford:1975,Welford:1986,Fairchild:1982,Lindlein:1992,Klein:2008}.
Some of these principles have been recently rediscovered for the analysis of
meta-lenses~\cite{Yu:2011,Aieta:2013}, and we use them here for a comprehensive study of MLLs for
nanometer focusing.  We consider both axi-symmetric lenses formed by depositing layers
onto a cylindrical substrate, for example, as well as lenses that focus only in one
direction, created by deposition onto a flat substrate.

Achieving a resolution of \SI{1}{\nano\meter} is certainly a challenge from the points of
view of lens fabrication and instrumentation stability, but it appears that the
concepts and technologies to achieve this are in hand.  Layer periods below
\SI{1}{\nano\meter} have been demonstrated~\cite{Windt:2003} and methods to produce lenses with
tilted layers---required for rays to satisfy the Bragg condition throughout the lens
pupil---have been developed~\cite{Prasciolu:2015}.  A recent theoretical and experimental analysis
of crossed 1D MLLs has pointed out the stringent requirements to align the lenses with
respect to each other~\cite{Yan:2017}.  Here we characterise the field-dependent
aberrations of axi-symmetric and 1D MLLs formed on flat and curved surfaces, using
ray-tracing methods established for thick holographic optical elements (HOEs) in which the phase
profile (caused by additional wavelength of path of each ``fringe'' or period in the
structure) is combined with a ``modified Snell's law''~\cite{Welford:1986} that redirects rays by Bragg
diffraction.  We find that curving the lenses provides useful extra degrees of freedom in
the optical design, but for small focal lengths (giving diameters of lenses that can
be reasonably fabricated) the fields of view are limited by the rocking-curve
width of Bragg diffraction.  The ray-tracing approach is reviewed in
\sref{sec:holographic-elements} and further expanded in \sref{sec:ray-trace} for
X-ray MLLs.  Fourth-order expressions of the wavefront aberrations are derived in
sections~\ref{sec:off-axis} and \ref{sec:1D-MLL} and exact computations presented for lens
systems with \SI{1}{\nano\meter} resolution.  Finally, some tolerances for the fabrication
of MLLs are given in \sref{sec:imperfect}.  The symbols used in this
paper are listed in \tref{tab:symbols} for convenience.

\begin{table*}
  \centering
   \caption{List of symbols and their meanings.}
 \begin{scriptsize}
  \begin{tabulary}{\linewidth}{l>{\hangindent=1em}L}
    \br 
    $x$, $y$, $z$ & Cartesian coordinates of ray positions in the lens with the origin at
                   the lens vertex \\
    $r$, $\psi$, $z$ & Cylindrical coordinates of ray positions \\
    $x_1$, $y_1$; $x_2$, $y_2$ & Coordinates of rays in the first and second lenses (for
    a crossed pair of 1D lenses) \\
    $\rho$, $\rho_x$, $\rho_y$ & Angular pupil coordinates, $=r/f$, $x/f$, $y/f$ \\
    $x_i$, $y_i$ & Image-plane coordinates \\
    $\lambda$ & X-ray wavelength as assumed in the design of the lens\\
    $\lambda_m$; $\Delta \lambda$ & X-ray wavelength of a measurement; $\Delta \lambda = \lambda_m-\lambda$ \\
    $k$ & Wavenumber, $=\lambda/(2 \pi)$ \\
    $f$; $f_1$, $f_2$ & Designed focal length of the lens; focal lengths of the first and
    second lenses \\
    $f_m$ & Focal length of the lens at the measurement wavelength, $=f \lambda/\lambda_m$ \\
    $a$, $b$ & Distances from the lens to the object and image, both positive for a real image \\
   $\alpha_x$, $\alpha_y$ & Field angles in the $x$ and $y$ directions \\
    NA & Numerical aperture \\
    $\delta$ & Imaging resolution, $= 0.66 \lambda/\text{NA}$ for a circular pupil and
    $0.5 \lambda/\text{NA}$ for a square pupil \\
    $D$ & Lens diameter, $=2\text{NA} f$\\
    $\tau$ & Thickness of the lens (in the direction parallel to the optic axis) \\
    $R$; $R_1$, $R_2$ & Radius of curvature of the lens surface; radii of curvature of the
    first and second lenses.  Positive for surfaces that are convex as seen from the field
    at infinity. \\
    $s(r)$ & Sag of the lens surface in the $z$ direction \\
    $\vec{s}$ & Vector from the vertex to the ray intersection with the surface of the lens \\
    $\hat{\vec{n}}$ & Normal of the surface of the lens \\
    $\phi_r$, $\phi_o$ & Reference-wave and object-wave phases used in the construction of
    a HOE \\
    $\bar{\phi}$ & Holographic phase field, $=\phi_o-\phi_r$ \\
    $\phi$; $\phi_1$, $\phi_2$ & Phase imparted by the lens onto the reference-wave beam
    ($2\pi$ per layer period);
    phases for the first and second lenses \\
    $\phi_m$ & Phase  imparted by the as-manufactured lens onto the reference-wave beam ($2\pi$ per layer period)\\
    $\theta$ & Bragg angle, with a deflection angle given by $2\theta$ \\
    $\Delta \theta$ & Deviation of a ray from the Bragg angle \\
    $r_n$; $r_{n(m)}$ & Radial position of the $n^\text{th}$ layer pair from the optic
    axis; ditto for the as-manufactured lens \\
    $d(r)$; $d_n$ & Layer period at a position $r$ in the lens; period of the
    $n^\text{th}$ layer pair \\
    $C$ & Intersection point of planar or conical layers that approximate the holographic phase field $\bar{\phi}$ \\
    $\vec{q}$ & Reciprocal-space vector with a magnitude $1/d$ and direction normal to the
    layers \\
    $\vec{k}_\text{in}$, $\vec{k}_\text{out}$ & Wave-vectors of the incident and diffracted
    rays \\
    $\vec{r}$, $\vec{r}'$ & Normal vectors of the incident and diffracted rays\\
    $l_1$, $l_2$, $l_{12}$ & Lengths of rays traced from the incident wavefront to the
    (first) lens, from the (second) lens to the image plane, and between the lenses \\
    OPL & Optical path length \\
    OPD & Optical path difference \\
    $\text{OPD}^{(4)}$; $\text{OPD}_\text{SA}$ & Fourth-order series expansion of the OPD;
    OPD due to spherical aberration \\
    $I_L$ & Efficiency of Bragg diffraction as a function of the deviation from the Bragg
    condition \\ 
    $\Gamma_L$ & Pendellosung period for dynamical diffraction.  Maximum diffraction
    efficiency is obtained for $\tau = \Gamma_L/2$. \\
    $\epsilon$; $\epsilon_m$ & Deviation parameter for rays (equal to zero when the Bragg
    condition is satisfied); deviation parameter for the as-manufactured lens. \\
    $w_\epsilon$, $w_\theta$ & Width of the diffraction rocking curve in terms of
    $\epsilon$; in terms of $\Delta \theta$.\\
    $\eta$ & Dimensionless parameter, $=\epsilon/w_\epsilon = \Delta \theta/w_\theta$ \\
    $\delta_1-\delta_2$ & Difference of the optical constants of the two materials of the
    multilayer, $=\num{6.7e-6}$ for SiC/WC at a wavelength of \SI{0.075}{\nano\meter} \\
    $\chi_x$, $\chi_y$, $\chi_z$ & Rotations of the second lens relative to the
    first, about the $x$, $y$, and $z$ axis \\
    $h$ & Scale factor of the lens \\
    $p$ &  Deposition rate of material in the fabrication of the MLL\\
    $c(t)$; $c_1$ & Drift of the deposition rate with time; coefficient of linear drift \\
    $\beta$ & Relative change in deposition rate per unit length of material deposited, $=c_1/p$ \\
    $\Delta d$ & Offset error in the bi-layer period $d$ \\
    $\gamma$ & Inclination of the surface normal of a MLL relative to the optic axis\\
    \br 
  \end{tabulary}
  \end{scriptsize}
  \label{tab:symbols}
\end{table*}

\section{MLLs as holographic optical elements}
\label{sec:holographic-elements}

A perfect focusing lens transforms the wavefront $\phi(\vec{r})$ of an incoming plane wave into
a converging spherical wave such that at each point a ray, defined by the normal of
$\phi$, is deflected towards a common point on the optic axis.
The deflection $\sin 2\theta (\vec{r}) = -\nicefrac{\lambda}{2 \pi} \nabla \phi (\vec{r})$ must
therefore increase with distance $r=|\vec{r}|$ from the optic axis, for a wavelength $\lambda$. 
In a
refractive lens, deflection is caused by refraction and the change in deflection is
achieved with a variation of the direction of the surface normal with position---that is,
with a curved surface.  In diffractive optics, deflection is caused by diffraction.
Rays no longer traverse the path of least time to arrive at the focus, but instead accrue
extra path-lengths of integral multiple wavelengths to constructively interfere.

\begin{figure}
  \centering
  \includegraphics[width=\linewidth]{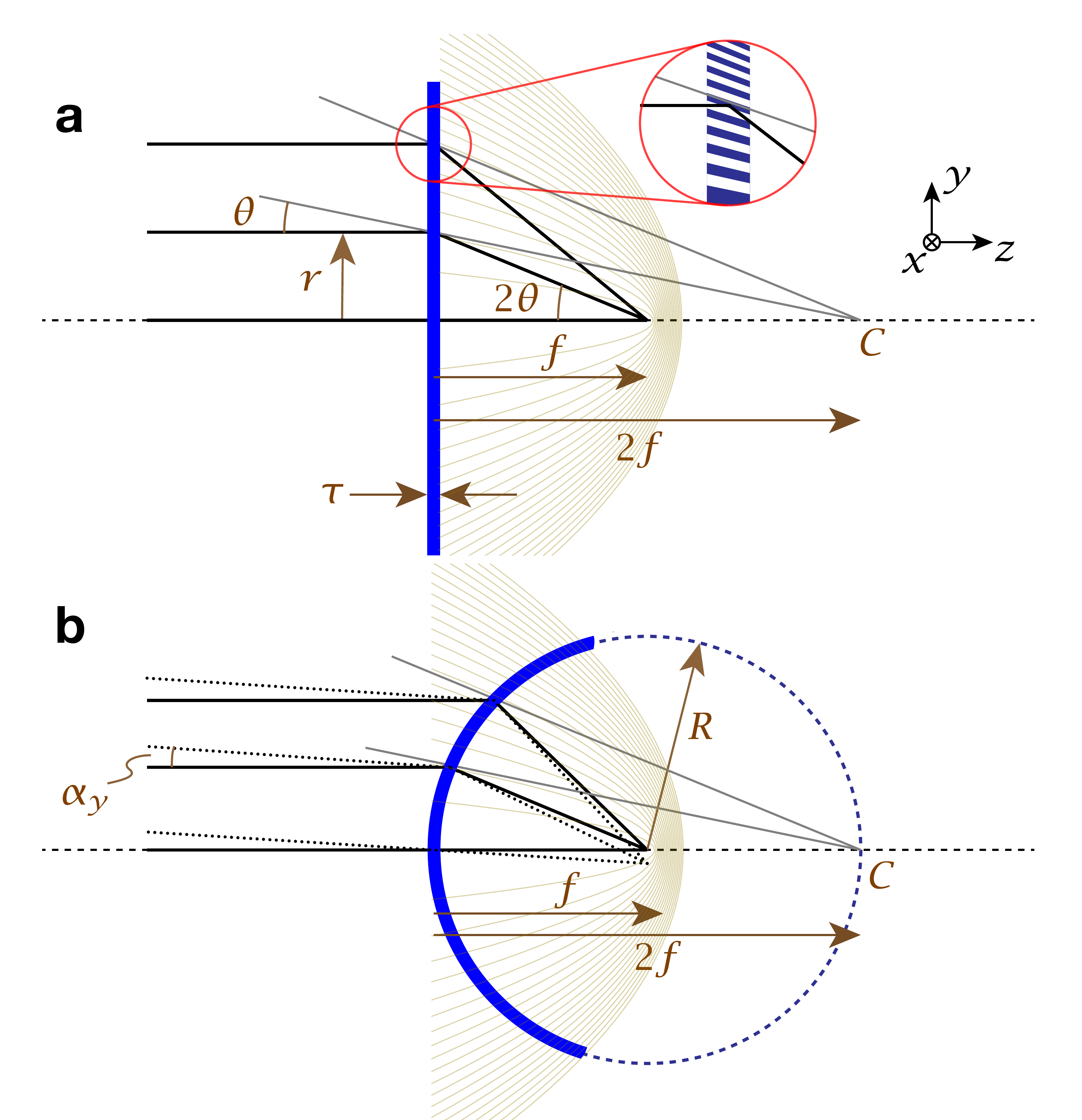}
  \caption{Schematic diagrams of focusing by a flat (a) and a curved (b) multilayer Laue
    lens.  The lens of thickness $\tau$ is depicted in blue and consists of layers whose surfaces (seen in the
    inset) follow a family of paraboloids shown in brown and which can be approximated as
    planes or cones that intersect at the point $C$, located a distance $2f$ from
    the lens.  The radius of the lens is $R$, equal to infinity in (a) and to $f$ in (b).
    Off-axis rays at an angle $\alpha_y$  are shown as dotted lines. 
     }
  \label{fig:MLL-schematic}
\end{figure}

Conceptually, a diffractive lens that focuses an incoming plane wave (i.e. rays from a
point source located at infinity) to a point a distance $f$ from the
lens can be formed holographically by interfering that plane wave with a
spherical wave emanating back towards the lens from the focus. For a cylindrical coordinate system $(r,z)$, the
phase of the resultant interfering field is thus the difference of the ``reference'' plane wave
phase $\phi_r = -k z$ and the ``object'' spherical wave phase,
$\phi_o = k[r^2+(z-f)^2]^{1/2}$, with $k = 2\pi/\lambda$. The
phase field $\bar{\phi}(r,z) = \phi_o-\phi_r$ gives surfaces at constant $\bar{\phi}(r,z)$ on which
the incident and deflected rays sum to a constant path length.  The family of surfaces at
$\bar{\phi}(r,z) =k f + 2\pi n$, where $n$ is an integer, provide the form of a diffracting
structure that can be fabricated on a surface specified by $z = s(r)$.  Rays deflected
by such structures will constructively interfere at the focus.  The family of curves may
be written as
\begin{equation}
  \label{eq:curve-family}
  z + \sqrt{r^2+(z-f)^2} = f+n\lambda,
\end{equation}
which can be expanded and simplified to an expression describing a set of paraboloids
\begin{equation}
  \label{eq:zp-thick}
  r^2_n = 2 n \lambda \left(f + \frac{n\lambda}{2} - z \right)
\end{equation}
that are illustrated in figure~\ref{fig:MLL-schematic}.  For example, the flat diffractive
lens formed at $z =s(r) = 0$ reproduces the Fresnel zone plate
formula with bi-layer zones separated at radii $r_n$.

 The phase imparted on the incident plane wave (replicating the reference wave) by the diffracting structure fabricated on
the surface $s(r)$ is given by
\begin{eqnarray}
  \label{eq:phase-s}
 \phi(r) &= \bar{\phi}(r,s(r)) - \frac{2\pi f}{\lambda} \nonumber\\
  &= \frac{2 \pi}{\lambda} \left( s(r) + \sqrt{r^2+\left(f-s(r)\right)^2} -f \right),   
\end{eqnarray}
such that $\phi(0) = 0$.  Since each period in a diffracting structure adds an additional
$2\pi$ of phase to the wave, the number of periods from the optic axis to the point $r$ on
the surface is $n = \phi(r)/(2 \pi)$. The change in ray direction caused by the diffracting structure
is given by the vector phase gradient $-\nicefrac{\lambda}{2 \pi} \nabla \phi$, which can
be computed from equation (\ref{eq:phase-s}) as~\cite{Fairchild:1982}
\begin{eqnarray}
  \label{eq:phase-gradient}
    \frac{\lambda}{2\pi} \frac{\partial \phi}{\partial r} &= \frac{\lambda}{2\pi}  \left( \frac{\partial
      \bar{\phi}}{\partial r}+\frac{\partial \bar{\phi}}{\partial s}\frac{\partial
      s}{\partial r} \right) \nonumber\\
& = \frac{r}{\sqrt{r^2+(f-s)^2}} + \frac{\partial s}{\partial r} \left(1
  -\frac{f-s}{\sqrt{r^2+(f-s)^2}}\right) \nonumber\\
& = \sin 2\theta + 2 \frac{\partial s}{\partial r}\, {\sin^2} \theta,
\end{eqnarray}
where $2\theta$ is the deflection angle as shown in figure~\ref{fig:MLL-schematic}.
For a flat lens ($s(r) = 0$), we find that
\begin{equation}
  \label{eq:phase-gradient-flat}
  \frac{\lambda}{2\pi} \frac{\partial \phi}{\partial r} = \sin
  2\theta  = \frac{\lambda \cos \theta}{d(r)}.
\end{equation}
Here we have considered X-rays interacting with three-dimensional diffracting structures, where a
deflection by $2\theta$ is attained by a structure that modulates in density in one
direction by a period $d$ satisfying Bragg's law, $\lambda = 2d\sin \theta$. This gives
the equality to the final term in (\ref{eq:phase-gradient-flat}), from which an
expression for $d(r)$ can be obtained.  Efficient diffraction requires that the direction
of the modulation be parallel to the momentum transfer, equal to the difference of the
wave-vectors of the rays, $\mathbf{k}_\mathrm{out} - \mathbf{k}_\mathrm{in}$, for a
wave-vector of the incident ray given by $\mathbf{k}_\mathrm{in}$ and the outgoing
wave-vector $\mathbf{k}_\mathrm{out}$.  This is equivalent to the rays reflecting from the
``layers'' of constant density in the structure, oriented at an angle $\theta$ to the $z$
axis.  Indeed, (\ref{eq:phase-gradient-flat}) indicates that the gradient of phase (in
the transverse direction $r$) is inversely proportional to the separation of layers in the
$r$ direction, $d(r)/\cos\theta$, for their minimal separation $d(r)$ in the direction
normal to the layers.  This would be equal to the deposited height of layers (ie. their
thickness in the direction normal to the substrate they are deposited
onto) which should thus follow the recipe of
\begin{equation}
  \label{eq:d-plate}
  \frac{d(r)}{\cos\theta} =\frac{\lambda f}{r} \sqrt{1+\frac{r^2}{f^2}} \approx
  \frac{\lambda f}{r} \left(1 + \frac{r^2}{2f^2}\right).
\end{equation}
Here, we have assumed that the period $d$ is well defined at any point $r$ along the
height of the lens, which is to say that the period changes slowly compared to $r$, or
$|\partial d / \partial r| \ll 1$.  Taking the leading term in the approximation in
(\ref{eq:d-plate}), this is equivalent to $d \ll r$. 

The optical performance of diffractive optics have long been modelled in this fashion, in
which a phase function $\phi(r)$ on a surface $s(r)$ defines a structure with periods that
vary as $d(r)$. These functions can then be used in a computational ray-tracing procedure
to deflect rays according to $\nabla \phi(r)$, as a modified Snell's law~\cite{Welford:1986}.
When the structure is fabricated not through interference of waves as a traditional HOE,
but rather as a ``computer-generated hologram'' or ``meta-lens'' such as by lithography or
multilayer deposition, we must contend with the fact that the structure might differ from
the ideal.  Furthermore, we are interested in understanding off-axis aberrations and
conditions where Bragg diffraction, for example, is not strictly satisfied.  An
established method to treat these cases is to use the structural properties of the lens
such as the local layer period $d(r)$ to prescribe the path of a ray, taking into account
the accrued phase $\phi(r)$ due to the number of diffracting periods in the structure.
The wavefront aberration is calculated by tracing many such rays throughout the pupil of
the optical system, from which a wave-optical calculation can be made to determine imaging
characteristics such as the point spread function, or to express the form and magnitude of
particular aberrations.
There are some approximations in this approach, such as the
assumption that locally, the diffracting structure acts as a grating of a single period.  More accurate treatments utilise
dynamic diffraction of strained crystals~\cite{Yan:2007}, the multislice algorithm~\cite{Li:2017c}, or coupled-wave
numerical modelling~\cite{Hu:2015}.

\section{Analysis of multilayer Laue lenses}
\label{sec:ray-trace}
\subsection{Design of a multilayer Laue lens}
\label{sec:design}
X-ray multilayer Laue lenses (MLLs) can be considered as thick diffractive optical
elements, in which diffraction occurs as a volume effect.  Efficient diffraction is
obtained by satisfying the Bragg condition as mentioned above, which requires the layers
to be tilted by $\theta$, half the deflection angle.  MLLs are fabricated by the
alternating deposition of two (or more) materials onto a substrate followed by slicing to
produce a structure similar to that illustrated in the inset of
figure~\ref{fig:MLL-schematic} (a).  The variation of period $d(r)$ throughout the lens is
controlled by varying the amount of material deposited in each layer (e.g.\ by changing
the time the substrate is exposed to each sputtering target) and the required tilt can be
obtained by placing a mask edge between the sputtering target and the
substrate~\cite{Prasciolu:2015}.  MLLs provide high efficiencies for X-ray energies above
about \SI{10}{\kilo\eV} (wavelengths less than about \SI{0.1}{\nano\meter}) with periods typically in the range of
\SIrange{1}{100}{\nano\meter}. The optimum thickness $\tau$ of the lens that the X-rays transmit
through is half the so-called Pendell\"{o}sung period $\Lambda_L$, which can vary from several
micrometers at lower X-ray energies to many tens of micrometers at harder energies,
depending on the materials and (weakly) on the period.

X-ray MLLs are usually fabricated as flat lenses with bi-layers (periods) positioned
according to the Fresnel zone plate condition of (\ref{eq:zp-thick}) with $z=0$. Given
the large thickness
$\tau$ of an MLL, rays can be
thought to reflect not only on the front surface of the structure but at points some
distance $z$ into the interior of the MLL.  In order that these rays are directed to a
common focal point the layers should follow the paraboloids given by
(\ref{eq:zp-thick}).  The approximation of the paraboloids to cones (or
tilted wedged layers for one-dimensional focusing) can be found from a Taylor series of
$r_n(z)$, showing the gradient of each layer is approximately
$r_n/(2f)$, which is indeed equivalent to the angle $\theta(r)$ of layers mentioned above
and shown in figure~\ref{fig:MLL-schematic}.
Also, given a bi-layer thickness $d_n = r_{n+1}-r_n$ it can be shown that
$d_n^2 \approx f \lambda/(4(n+\nicefrac{1}{2}))$, and from (\ref{eq:zp-thick}) it
follows that $d_n \approx f\lambda/r_n$.  For the purpose of ray tracing, the phase
imparted to the ray that intersects the lens at a position $r$, equal to $2 \pi n$, is given by $\phi(r)$ of
(\ref{eq:phase-s}).  

\subsection{Ray tracing of MLLs}
\label{sec:ray-tracing}
\begin{figure}
  \centering
  \setlength{\unitlength}{\linewidth}
  \begin{picture}(0.75,1.0)(0,0) 
    \put(0.05,0.52){
      \includegraphics[width=0.75\linewidth]{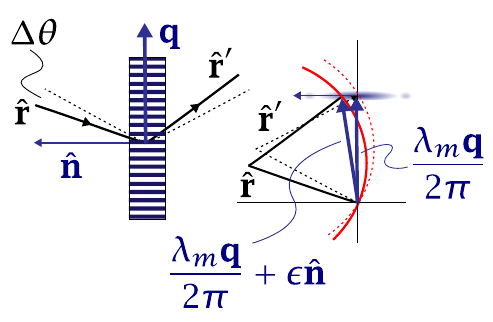}}
    \put(0.125,0){
      \includegraphics[width=0.5\linewidth]{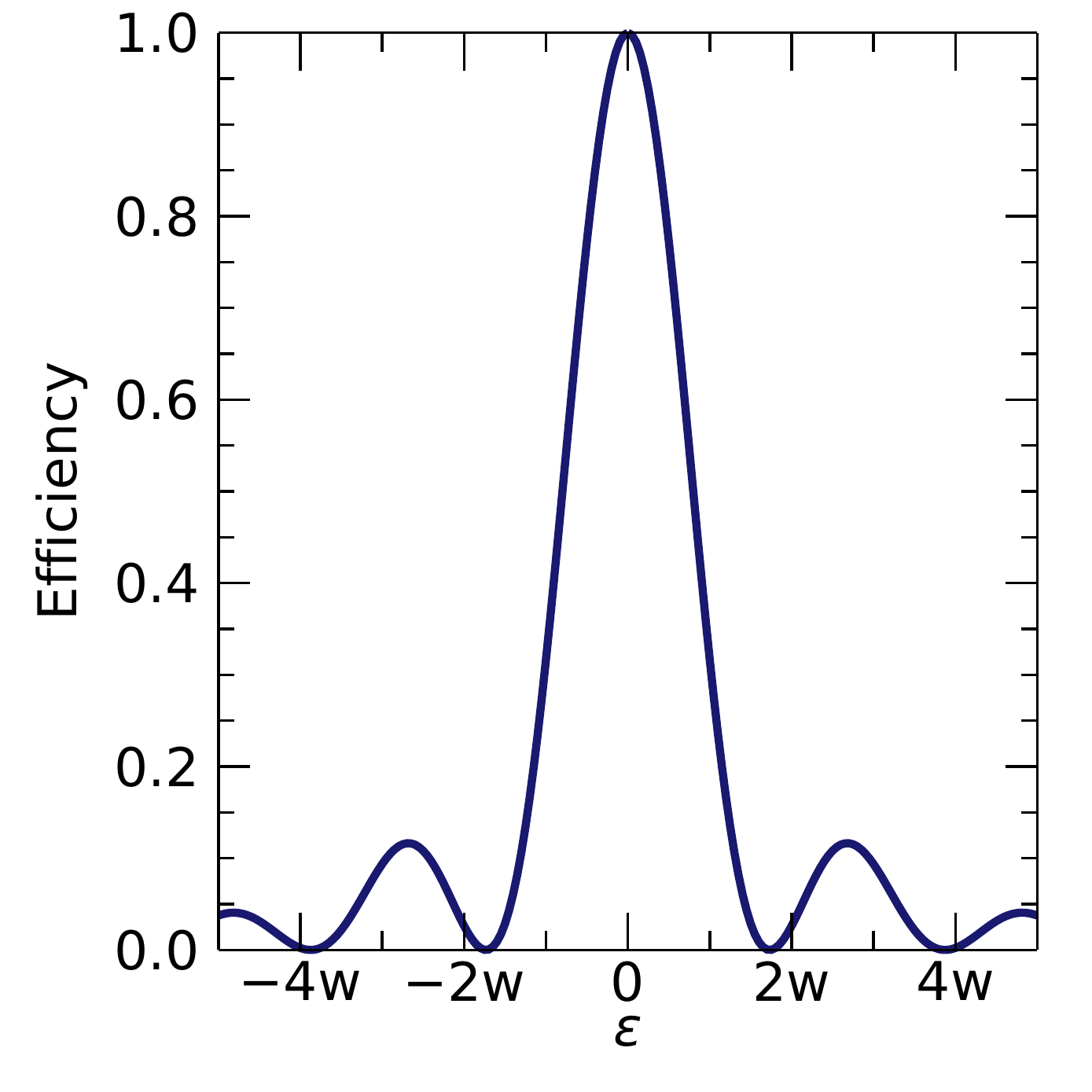}}
    \put(0.0,0.95){\fontfamily{phv}\normalsize\selectfont\text{(a)}}
    \put(0.08,0.45){\fontfamily{phv}\normalsize\selectfont\text{(b)}}
  \end{picture}
   \caption{(a) The off-Bragg reflection from a periodic multilayer in Laue geometry shown
     in real space to to left is
     found from the reciprocal-space Ewald sphere construction at right, considering that the diffraction peak
     lies along a truncation rod normal to the surface. (b) Rocking curve of a multilayer
     reflecting in a Laue geometry with an optimum thickness and neglecting absorption, as
     a function of the deviation parameter
     $\epsilon \approx \Delta\theta \,\lambda_m |\vec{q}|/(2\pi)$ and for a width
     $w_\epsilon$.}  
  \label{fig:rocking}
\end{figure}

We first consider the ray-trace analysis of perfect MLLs that are designed to deflect
incident parallel rays to the focal point.
In \sref{sec:imperfect} we consider lenses with imperfections (due to manufacturing processes, for
example) where $\phi(r)$ must be modified due to deviations in $d(r)$ from the ideal.     
Generally, ray tracing tracks a ray from a particular point in an object,
via straight-line propagation through homogeneous space to an interface where it is
refracted (according to Snell's law) or diffracted (according to a grating equation or
Bragg's law) before propagating to the next interface, and so on, until the plane of
interest is reached~\cite{Welford:1986}.  Unlike a more complete calculation that may utilise Huygens wavelets
formed at each scattering point of the structure (such as used in a multislice
calculation~\cite{Li:2017c}), a ray described by the unit vector
$\hat{\vec{r}}$ incident upon a diffractive optic is taken to follow a single
trajectory only. As mentioned above, the direction of the diffracted ray
$\hat{\vec{r}}'$ is set by Bragg's law, cast here as
$(2\pi/\lambda_m)\hat{\vec{r}}' = \vec{k}_\text{out} = \vec{k}_\text{in}+\vec{q}$
for an incident wave-vector $\vec{k}_\text{in} = (2 \pi/\lambda_m)\hat{\vec{r}}$ (see
figure~\ref{fig:rocking}~(a)). That is,
the ray has a ``measurement'' wavelength $\lambda_m$ which may differ from the parameter $\lambda$ used in
the design of the diffracting structure that modulates with a period $d = 2\pi /|\vec{q}|$ in the
direction of $\vec{q}$. This presumes that the diffracting structure will indeed be
oriented in the Bragg condition.  For thin structures, or to account for diffraction of a
ray in an off-Bragg condition (whether that be due to an angular deviation or a change in
wavelength), it can be assumed that the termination of the periodic
structure by the surface leads to truncation rods in reciprocal space that extend from
$\vec{q}$ along the direction of the surface normal $\hat{\vec{n}}$, in which case the
``refraction'' of the single ray can be described as
\begin{equation}
  \label{eq:off-Bragg}
  \hat{\vec{r}}' = \hat{\vec{r}}+\frac{\lambda_m}{2 \pi} \vec{q}+\epsilon \hat{\vec{n}},
\end{equation}
where the deviation parameter $\epsilon$ is chosen to ensure that $|\hat{\vec{r}}'| = 1$, and 
$\vec{q}$ points towards the optical axis (rather than away from it) for a focussing
optic.  Although (\ref{eq:off-Bragg}) was derived in the context of the geometric
theory of diffraction (see Chapter 3.3 of \cite{Authier:2001}) the expression also holds
in the framework of dynamical diffraction theory (see Chapter 4.8.5 of
\cite{Authier:2001}).  A similar approach used in the ray-tracing of thick HOEs is
referred to as the blurred grating vector approach~\cite{Goodman:1996} 

As the ray is traced sequentially from interface to interface, the optical path length
(OPL) of the ray is accumulated.
In a perfect imaging system based on refractive optics, rays originating from a point in the object plane
should all accumulate the same optical path (proportional to the flight time) upon their
intersection of the image plane at the image point.  A map of the differences of the optical
path lengths of the rays to that of a reference ray (optical path difference, OPD), as a function of the pupil
coordinates of the rays, gives the map of the wavefront aberration.  This aberration can
be related to the point spread function of the lens through a wave optic calculation,
treating the aberration as a phase error ~\cite{Born:2002}.  In a diffractive optic such as an MLL, however, rays
reflecting from each subsequent bi-layer accrue an extra wavelength of path-length, for a total path difference of
$n\,\lambda_m$ for $n$ periods, or a phase of $2\pi n$. For a wavelength equal to the
design wavelength, $\lambda_m = \lambda$, this phase is exactly $-\phi(r)$ as provided
by the diffracting structure. The formation of the image
is dictated by the constructive interference of waves at the measurement plane rather than
the principle of least time.  The lag of rays by a wavelength per period of the structure is
thus not apparent and it is accounted for by forming the OPD as~\cite{Welford:1986}
\begin{equation}
  \label{eq:OPD}
  \text{OPD}(r) = \text{OPL}(r)-\text{OPL}(r_0) + \frac{\lambda_m}{2\pi}\phi(r),
\end{equation}
where $r_0$ is the coordinate of the reference ray (such as the chief ray).  It is
not necessary to discretise $\phi(r)$ into the effects of individual layers since
the coordinates $r$ in the pupil need not be determined with a precision better than
the period of the diffracting structure.  Chromatic aberrations can be computed when the
measurement wavelength $\lambda_m$ is different to the design wavelength $\lambda$.

\subsection{Off-Bragg reflections}
\label{sec:off-Bragg}
Equations~(\ref{eq:off-Bragg}) and (\ref{eq:OPD}) 
can be used to trace rays through the optical system and analyse its wavefront
aberrations.  These equations do not account for any variation in diffraction efficiency,
which drops precipitously for rays outside of the so-called Darwin width of the reflection profile
(also referred to as the rocking-curve width), which will occur for rays originating from
a source point off the optic axis, for example, as well as for rays incident at the Bragg
angle but which have the wrong wavelength.  No matter whether a deviation from the Bragg
condition is caused by a change in wavelength $\lambda_m$ or a change in incident ray
direction $\hat{\vec{r}}$, or a combination of both, solving
$|\hat{\vec{r}}'|=1$ using (\ref{eq:off-Bragg}) will generate
a non-zero error term $\epsilon$. If the normal $\hat{\vec{n}}$ is perpendicular to
$\vec{q}$ (which is the case for symmetric Laue diffraction) then the equivalent angular deviation from the
Bragg condition can be estimated as
\begin{equation}
  \label{eq:deviation}
  \Delta \theta \approx \frac{2 \pi \epsilon}{\lambda_m |\vec{q}|} = \frac{\epsilon d}{\lambda_m}.
\end{equation}
The diffraction efficiency of a ray as a function of $\Delta \theta$ can be calculated by dynamical diffraction
theory~\cite{Authier:2001} or, equivalently, coupled waved theory~\cite{Kogelnik:1969}.
Bajt \emph{et al.} \cite{Bajt:2012} and Yan \emph{et al.} \cite{Yan:2007} give expressions
for dynamical diffraction of multilayer structures.  In particular, for a thick multilayer grating constructed
of equal layer heights of the two materials per period, a thickness of the lens $\tau$, and ignoring absorption,  the normalised rocking curve of the
symmetric Laue reflection is given by
\begin{equation}
  \label{eq:rocking}
  I_L(\eta) = {\sin}^2 \left(\frac{\pi\,\tau}{\Lambda_L}\sqrt{1+\eta^2} \right)\,\frac{1}{1+\eta^2}
\end{equation}
for a  Pendell\"{o}sung period $\Lambda_L$ and the normalised deviation parameter $\eta$
equal to
\begin{equation}
  \label{eq:eta}
 \eta = \frac{\Delta\theta}{w_\theta} = \frac{\Delta\theta \, \pi\,\sin
   2\theta}{2|\delta_1-\delta_2|} \approx \frac{\epsilon\,\pi}{2|\delta_1-\delta_2|} =
 \frac{\epsilon}{w_\epsilon} 
\end{equation}
for a Bragg angle $\theta$, and where $\delta_1$ and $\delta_2$ are the real parts of
the optical constants of the layer materials at the particular photon energy used. 
The expression of $\eta$ in terms of $\epsilon$ was obtained from $\sin 2\theta \approx
2\theta \approx  \lambda_m/d$.  
 In the
following we assume that the MLL is cut at a thickness $\tau=\Lambda_L/2$ to give the maximum
diffraction efficiency  such that
$I_L(0) = 1$.  The width of the Laue rocking curve is given by the Darwin
width $w_\theta$, at which the efficiency drops to $0.317$.  The efficiency
$I_L(\epsilon/w_\epsilon)$ is plotted in figure~\ref{fig:rocking}~(b) and can be included in the ray tracing
procedure~\cite{Lindlein:1992} by multiplying the ray intensity (initially unity) by the
diffraction efficiency at each interface.  As seen from equations~(\ref{eq:rocking}) and (\ref{eq:eta}),
this calculation does not require any further specification of the $d$ spacing of the
multilayer or the wavelength, which are implicitly accounted for in the prescription of
$\vec{q}$, as demonstrated in the next section.

\section{Off-axis aberrations of axi-symmetric MLLs}
\label{sec:off-axis}
Diffractive optical elements provide extra degrees of freedom for optical design as
compared with refractive elements since the phase profile, as determined by the
diffracting structure, can be decoupled from the profile of the surface.  For example, it
can be shown that the Seidel aberration of coma (varying linearly with the field angle) is
eliminated for particular imaging conjugates when a zone plate is constructed on a
spherical rather than a flat surface~\cite{Murty:1960,Young:1972,Welford:1973,Bokor:2001}. When
combined with the prescription of zones or layer periods $d(r)$ that result in zero spherical
aberration (as given in \sref{sec:design}), the
appropriately-curved zone plate is thus aplanatic (free of aberrations over a paraxial
field of view) and obeys the Abbe sine condition.  The required surface of the zone plate is given
by the circle of Apollonius, for which the radius $R$ satisfies $1/R=1/b-1/a$ for an
object distance $a$ and image distance $b$, and where positive $R$ represents a convex
surface as seen from the object plane.  For all points on this circle, the ratio of the
distances from a point on the circle to either the object or image points is a constant.
When $a=\infty$, $R=b=f$, which is the case illustrated in figure~\ref{fig:MLL-schematic}~(b).

Using the formalism laid out in the previous section we thus analyse the case of an
axi-symmetric MLL (as made by depositing materials onto a wire) with conical or
paraboloidal layers, imaging a source at $a=\infty$, to
determine if there is an advantage to polishing the MLL to a radius $R$. This serves as an
illustrative case to compare with the analysis of imaging with pairs of one-dimensional
MLLs and for the examination of MLLs with various imperfections in \sref{sec:imperfect}.

\begin{figure*}
  \centering
  \includegraphics[width=0.66\linewidth]{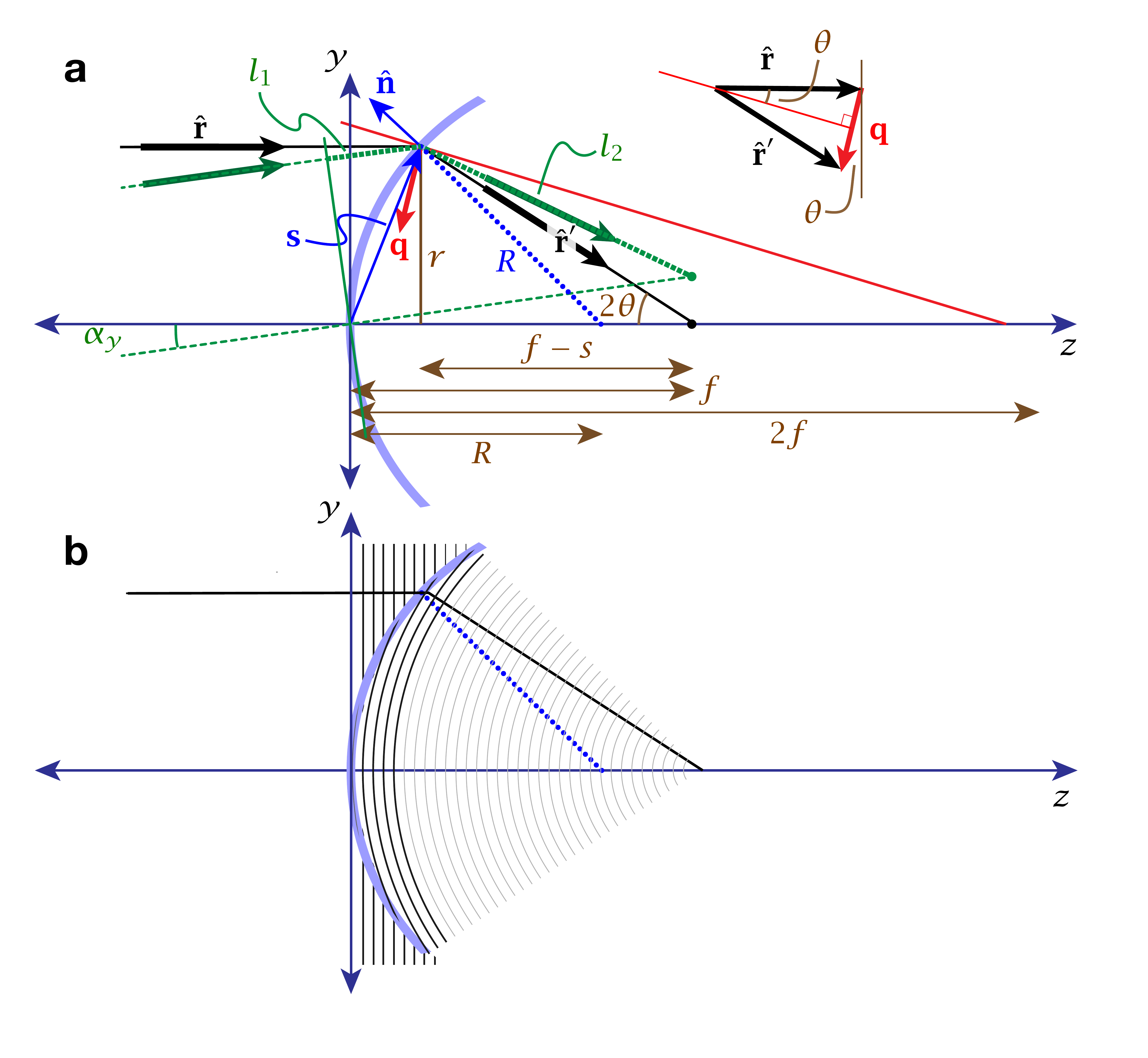}
  \caption{Geometry for ray trace calculations of a single MLL of focal length $f$. (a)
    The lens shape is the shell of a sphere or cylinder of radius $R$, shown in blue, with
    a surface described by $\vec{s}(r)$ for the radial coordinate $r$.  Rays are deflected
    by the diffracting structure described by $\lambda/(2\pi)\vec{q}(r)$ (in red) and the
    surface normal $\hat{\vec{n}}(r)$. This structure fulfils the Bragg condition (see
    inset) for the on-axis field point at infinity. The path of an incoming ray with
    direction $\hat{\vec{r}}$ is shown in black.  Paths for rays from a field angle
    $-\alpha_y$ are shown in green, and the path lengths $l_1$ and $l_2$ are indicated for
    that case by the thick dashed green lines. (b) Deflection of rays is also a
    consequence of the accumulation of $2\pi$ of phase from each layer pair, which
    themselves match the interference of the on-axis plane wave with a converging
    spherical wave. }
  \label{fig:ray-trace}
\end{figure*}

Consider the axi-symmetric MLL shown in figure~\ref{fig:ray-trace} which is polished to a
sphere of radius $R$ (which may differ from the focal length, $f$). The surface of the MLL is given by the ``sag'',
equal to the distance $z=s(r)$ from the plane at $z=0$ to the surface, which can be
parameterised as
\begin{equation}
  \label{eq:sag}
  \vec{s}(r,\psi) = \left(r \sin \psi, r \cos \psi, R-\sqrt{R^2-r^2}\right).
\end{equation}
Here we use cylindrical polar coordinates $(r,\psi,z)$, and a positive $R$ corresponds to
a centre of curvature towards $+z$. The definition of $\psi=0$ along the cartesian $y$
axis follows the convention of optical ray tracing. The outward-facing normal to this
surface is parallel to the cross product of the gradients of $\vec{s}$ in the $x$ and $y$
directions, giving a unit normal
\begin{equation}
  \label{eq:normal}
  \hat{\vec{n}}(r,\psi) = \left(\frac{r}{R}\sin\psi, \frac{r}{R} \cos\psi,-\sqrt{1-\frac{r^2}{R^2}}\right).
\end{equation}

The MLL is constructed with layers of varying
period and tilt designed
to perfectly focus incident rays of a particular wavelength $\lambda$ and parallel to the optic axis (taken to be the $z$ axis) to a point
on that axis a distance $f$ from the MLL.  Therefore, at a distance $r$ from the optic axis, layers of period
$d(r)=\lambda/(2\sin\theta(r))$ are tilted by an angle $\theta(r)$ to reflect rays by $2\theta(r)$ as
needed to direct them to the focus. This requires
\begin{equation}
  \label{eq:sin-2theta}
  \sin 2\theta(r) = \frac{r}{\sqrt{r^2+(f-s)^2}}
\end{equation}
so that the layer periods (not equivalent to the deposited layer thicknesses) are given by
\begin{equation}
  \label{eq:d-R}
  d(r) = \frac{\lambda}{2\sin\theta} \approx \frac{\lambda f}{r}\,\left(1+\left(\frac{3}{8}-\frac{f}{2R}\right)\frac{r^2}{f^2} \right).
\end{equation}
More completely, the momentum transfer vector caused by diffraction from the
tilted layers is given by
\begin{eqnarray}
  \label{eq:q}
 \fl   \frac{\lambda}{2\pi}\,\vec{q}(r,\psi) = (-\sin 2\theta(r) \,\sin \psi, -\sin 2 \theta(r)\,\cos\psi,
    \cos 2\theta(r)-1) \nonumber\\
    =\left( \frac{-r\sin\psi}{\sqrt{r^2+(f-s)^2}}, \frac{-r\cos\psi}{\sqrt{r^2+(f-s)^2}},\frac{f-s}{\sqrt{r^2+(f-s)^2}}-1\right).
\end{eqnarray}
This prescription of the layer period is consistent with (\ref{eq:phase-gradient})
because the structures described in equations~(\ref{eq:d-R}) and (\ref{eq:q}) are located on the
surface $z=s(r)$. 
The phase induced by the diffracting structure that achieves this common focus is given by
(\ref{eq:phase-s}) which can be expressed as
\begin{eqnarray}
  \label{eq:phase-R}
    \frac{\lambda}{2\pi}\,\phi(r) &= s+\sqrt{r^2+(f-s)^2} - f \\
    &\approx \frac{r^2}{2f}+\left(\frac{2f}{R}-1\right)\frac{r^4}{8 f^3} \nonumber.
\end{eqnarray}
The functions $\vec{s}(r,\psi)$, $\hat{\vec{n}}(r,\psi)$, $\vec{q}(r,\psi)$, and $\phi(r)$ fully define the
structure and optical
properties of the MLL. They can be used, with equations~(\ref{eq:off-Bragg}) and (\ref{eq:OPD}), to determine the
wavefront aberrations, with the simplifying assumption that the thickness of the MLL in the $z$ direction is negligible.  

Now consider an off-axis source point at infinity but located at a field angle $-\alpha_y$ in
the direction along $\psi=0$.  Incident rays are all parallel to
$\hat{\vec{r}} = (0,\sin\alpha_y,\cos\alpha_y)$. We must compute the
$\text{OPL}=l_1+l_2$  for rays
from a plane with a normal $\hat{\vec{r}}$ (such as the one passing through the
origin, defined by $\vec{x}\cdot \hat{\vec{r}} = 0$) and the image point at a
height $y_i=f\tan\alpha_y$ in the plane a distance $z=f$ from the MLL vertex at the origin.
An incident ray intersects the spherical surface of the MLL at the point
$\vec{s}(r,\psi) = (r\sin\psi,r\cos\psi,s)$, and the path-length $l_1$ from
the inclined plane to this point is given by
\begin{equation}
  \label{eq:l1}
  l_1= \vec{s}\cdot \hat{\vec{r}} = r \cos \psi \sin \alpha_y + s \cos\alpha_y.
\end{equation}
The length $l_2$ of the ray from $\vec{s}$ to the image plane depends on its direction
$\hat{\vec{r}}'$ which can be found from (\ref{eq:off-Bragg}), utilising
equations~(\ref{eq:q}) and (\ref{eq:normal}) and substituting $\sin 2\theta$ for the expression in
(\ref{eq:sin-2theta}).  The incident ray no longer satisfies the Bragg condition, so
the error length $\epsilon$ must be solved for, which can be done by constraining the
magnitude of $\hat{\vec{r}} + \lambda_m/(2\pi)\,\vec{q}+\epsilon \hat{\vec{n}}$ to be
unity (for rays of the ``measurement'' wavelength $\lambda_m$).
The
optical path length of the ray travelling from the MLL to the image plane can be obtained
by noting that $l_2\,\hat{\vec{r}}' \cdot \hat{\vec{z}} = f-s$.

The analysis can be carried out numerically or 
using a symbolic mathematics program, which can yield an exact result for the OPD given by
\begin{equation}
  \label{eq:OPD-R}
  \text{OPD}(r) = l_1+l_2-\frac{f}{\cos\alpha_y} -\frac{\lambda_m}{2\pi}\phi(r),
\end{equation}
using the expression for $\phi(r)$ from (\ref{eq:phase-R}). As noted above, $\phi(r)$
accounts for the number of bi-layers $n$ in the structure from the reference point to $r$,
which give rise to an effective optical path of $n\,\lambda_m$ when measured at a
wavelength $\lambda_m$. 
For imaging at the wavelength that the
lens was designed, $\lambda_m = \lambda$, the series
expansion of the exact expression is found to be
\begin{equation}
  \label{eq:OPD-R4}
\fl \text{OPD}^{(4)}(\rho) = -\frac{3f}{2}\alpha_y^3 \rho \cos\psi
  +f\left(\frac{3}{4}-\frac{f}{2R}\right)(2+\cos 2\psi) \alpha_y^2 \rho^2 +
  f\left(\frac{f}{R}-1\right) \alpha_y \rho^3 \cos \psi,
\end{equation}
where $\rho = r/f$ is the the normalised pupil coordinate equal to the tangent of the
angle of rays converging onto the focus.  At the edge of the pupil, $\rho \approx
\text{NA}$. 
Only terms to fourth order in powers of $\alpha_y$ and $\rho$ are
retained in this series expression, corresponding to the five Seidel (or primary) aberrations of an axi-symmetric system.  In
(\ref{eq:OPD-R4}) there are three summands.  The first varies linearly with the pupil
coordinate $\rho$ and indicates a wavefront tilt that grows with the third power of the field
angle $\alpha_y$.  This tilt corresponds to a shift of the image point and hence a distortion to
the image.  The second term consists of astigmatism and field curvature.  The dependence
on $\rho^2$ indicates a defocus which is different in orthogonal planes $\psi=0$ and $\psi=\pi/2$.  The last
term is coma, an aberration that depends on the third power of the pupil coordinate.  The
expression of (\ref{eq:OPD-R4}) does not contain the Seidel aberration
dependent only on $\rho^4$, which is spherical aberration.  This is the only Seidel
aberration present for a source point on axis ($\alpha_y=0$) and since the MLL was designed
to have zero aberrations on axis at the design wavelength, this term is also absent in this series approximation.
As anticipated above, the case where $R=f$ gives zero coma.
It is also interesting to note that the condition $R=(2/3)f$
eliminates astigmatism and field curvature, at the cost of finite coma.

In this analysis we find that changing the condition for off-Bragg reflection modifies the terms
in (\ref{eq:OPD-R4}), and thus the formulation of (\ref{eq:off-Bragg}) does matter.
For example, setting $\hat{\vec{n}}$ to be parallel to the optic axis rather than normal
to the spherical surface modifies the
distortion and field curvature coefficients.  

\subsection{Transmission of an axi-symmetric MLL}
\label{sec:transmission-2D}
Since all rays from the on-axis field point exactly match the Bragg condition in the lens,
they will be fully transmitted (under the ideal assumptions discussed in
\sref{sec:off-Bragg}) and the transmission of the lens will be uniform across the
pupil.  Rays from off-axis field points will deviate from the Bragg condition with an
efficiency $I_L(\epsilon)$ where $\epsilon$ is the deviation parameter. An off-axis field angle leads to
an angular deviation $\Delta \theta \approx \alpha_y$, so we expect from
(\ref{eq:deviation}) that $|\epsilon| \approx \lambda \alpha_y/d$.  Since the smallest
periods $d$ occur at the edge of the lens, the lens will be apodised---it will lose transmission at the
outermost regions of the pupil as the field angle is increased from zero. Off-axis
field points therefore become vignetted (have lower transmission)~\cite{Murray:2019} with reduced NA and degraded
resolution.  The expression for $\epsilon$ is
obtained in the analysis that leads to (\ref{eq:OPD-R4}) by solving for
$|\hat{\vec{r}}'| = 1$. The series expansion of this expression  to
first-order in $\alpha_y$ is given by
\begin{equation}
  \label{eq:epsilon-2D}
 \epsilon^{(1)}(\rho) = -\alpha_y \rho \cos\psi + \frac{f}{2R}\left(1-\frac{f}{R}
  \right)\,\alpha_y\,\rho^3 \cos\psi.
\end{equation}
That is, the magnitude of the deviation parameter does indeed increase linearly with field
angle and also depends linearly on the pupil coordinate in the direction of the field
angle, $\rho\cos\psi$.  In this direction, lens apodisation occurs when the deviation parameter
at the edge of the lens ($\rho = \text{NA}$) exceeds the rocking-curve width, or
\begin{equation}
  \label{eq:vignette-2D}
  \alpha_y < \frac{w_\epsilon}{\text{NA}} = \frac{2 |\delta_1-\delta_2|}{\pi\,\text{NA}} =
  w_\theta.
\end{equation}
This restriction is independent of the focal length of the lens.

\subsection{Chromatic aberrations of axi-symmetric MLLs}
\label{sec:chromatic}
The paraxial focal length of a diffractive lens scales inversely with the wavelength, and
so such lenses exhibit strong chromatic aberrations dominated by defocus.  The design of
the layers according to (\ref{eq:q}) will give zero on-axis aberrations (such as
spherical aberration) only for the wavelength at which it was designed.  Off-axis
aberrations will also be modified with a change in wavelength, since the image of an off-axis source point at infinity
will be focused to a plane a distance $f_m=f\,\lambda/\lambda_m$ from the lens with a
transverse displacement of $f_m \,\alpha$ from the optic axis instead
of $f \,\alpha$. The ray tracing procedure can be carried out using the governing
equations of (\ref{eq:off-Bragg}) and (\ref{eq:OPD}) and solving the path length
$l_2=(f_m-s)/(\hat{\vec{r}}'\cdot \hat{\vec{z}})$ to determine the wavefront aberrations
encountered when utilising the lens at a different wavelength to its design.  In this
case it is found that the spherical aberration term is given by
\begin{equation}
  \label{eq:OPD-SA}
    \text{OPD}_\text{SA}(\rho;\lambda_m) = \frac{3 f}{8}
  \frac{\lambda_m\, \Delta \lambda}{\lambda^2}\left( 2 -\frac{2f}{R}+\frac{\Delta \lambda}{\lambda}\right)\rho^4,
\end{equation}
where $\Delta \lambda = \lambda_m-\lambda$. This expression indicates that in addition to the design wavelength
($\Delta \lambda=0$), spherical aberration is nulled at the wavelength
$\lambda_m = \lambda (2 f/R -1)$.  That is, by setting the radius to
$R = 2f/(1+\lambda_m/\lambda)$, the spherical aberration is zero for both $\lambda$ and
$\lambda_m$.  Of course, images at each of these wavelengths will be located at different
planes, so such a design could have utility in scanning transmission or fluorescence
microscopy of thick objects at two discrete wavelengths (above and below an absorption edge, for example)
with a detector able to discriminate photon energies.
Equation~(\ref{eq:OPD-SA}) also shows that in the limit $\lambda_m
\rightarrow \lambda$, the optimum radius for minimising the spherical aberration is
$R=f$.  In this case $\text{OPD}_\text{SA}$ varies quadratically with 
$\Delta\lambda$ instead of linearly.

As with a change in the field angle, rays will no longer obey the Bragg condition as the
wavelength is moved from the design.  The series expansion of the deviation parameter for
the on-axis field point, as determined in the analysis used to derive (\ref{eq:OPD-SA})
is given by
\begin{equation}
  \label{eq:epsilon-chromatic}
  \epsilon^{(2)} (\rho) = \frac{\lambda_m \,\Delta \lambda}{2 \lambda^2} \rho^2.
\end{equation}
For this on-axis field point, apodisation will therefore occur radially with a tolerance
given by
\begin{equation}
  \label{eq:vignette-chromatic}
  \frac{\Delta \lambda}{\lambda} < \frac{2 w_\epsilon}{\text{NA}^2} = \frac{4|\delta_1-\delta_2|}{\pi\, \text{NA}^2}.
\end{equation}

Both apodisation and spherical aberration may scale approximately linearly with
$\Delta\lambda$ (as is the case for a flat lens, for example), but the degree of
apodisation increases with the square of the NA whereas spherical aberration grows with the
fourth power of NA and linearly with the focal length.  Thus for focal lengths shorter
than about $0.15 \lambda / (\text{NA}^2 w_\epsilon)$, the
limitation to wavelength changes will be the reduction of the active pupil diameter.  (Here the spherical aberration
tolerance from the Marechal condition is $\text{OPD}_\text{SA} (\text{NA}) /\sqrt{10} < \lambda/14$.)

\subsection{Ray tracing an axi-symmetric MLL for 1\,nm focusing}
\label{sec:2D-1nm}
%
%
As an example, we consider a lens designed to achieve a resolution of about
\SI{1}{\nano\meter} at a wavelength $\lambda = \SI{0.075}{\nano\meter}$ (\SI{17}{\kilo\eV}
photon energy) and $\text{NA}=0.0375$. The Rayleigh resolution of this lens for incoherent
imaging (such as the case in a scanning transmission microscope) is
$0.61 \lambda / \text{NA} = \SI{1.22}{\nano\meter}$. The focal length of the lens is
chosen to be $f = \SI{1}{\milli\meter}$, and thus the diameter of the lens is
\SI{75}{\micro\meter}.  The ray-tracing procedure detailed above was used to numerically
compute the wavefront aberrations exactly---that is, without approximating to a
fourth-order series expansion as for (\ref{eq:OPD-R4}).  The procedure, as formulated above, applies to the
case of field points at infinity, displaced from the optical axis by a field angle
$\alpha_y$ as set by the ray direction $\hat{\vec{r}}$. This situation represents, for
example, the formation of a focused beam for a scanning microscope.  In this case the
field angle also represents the angular misalignment of the optic axis of the lens relative to say a
``beam axis'' set by the direction of a far-off source.  These calculations also correspond
to the aberrations expected in full-field imaging at high magnification (where the image
plane is far from the lenses and the sample is near to the focal plane).  Finite
conjugates can be simulated too, with an appropriate re-definition of $\hat{\vec{r}}$ and
$l_1$. 

\begin{figure*}[t]
  \centering
  \setlength{\unitlength}{\linewidth}
   \begin{picture}(1.,0.7)(0,0)
    \put(0,0.25){\makebox(1,0.55){
        \includegraphics[height=0.33\linewidth]{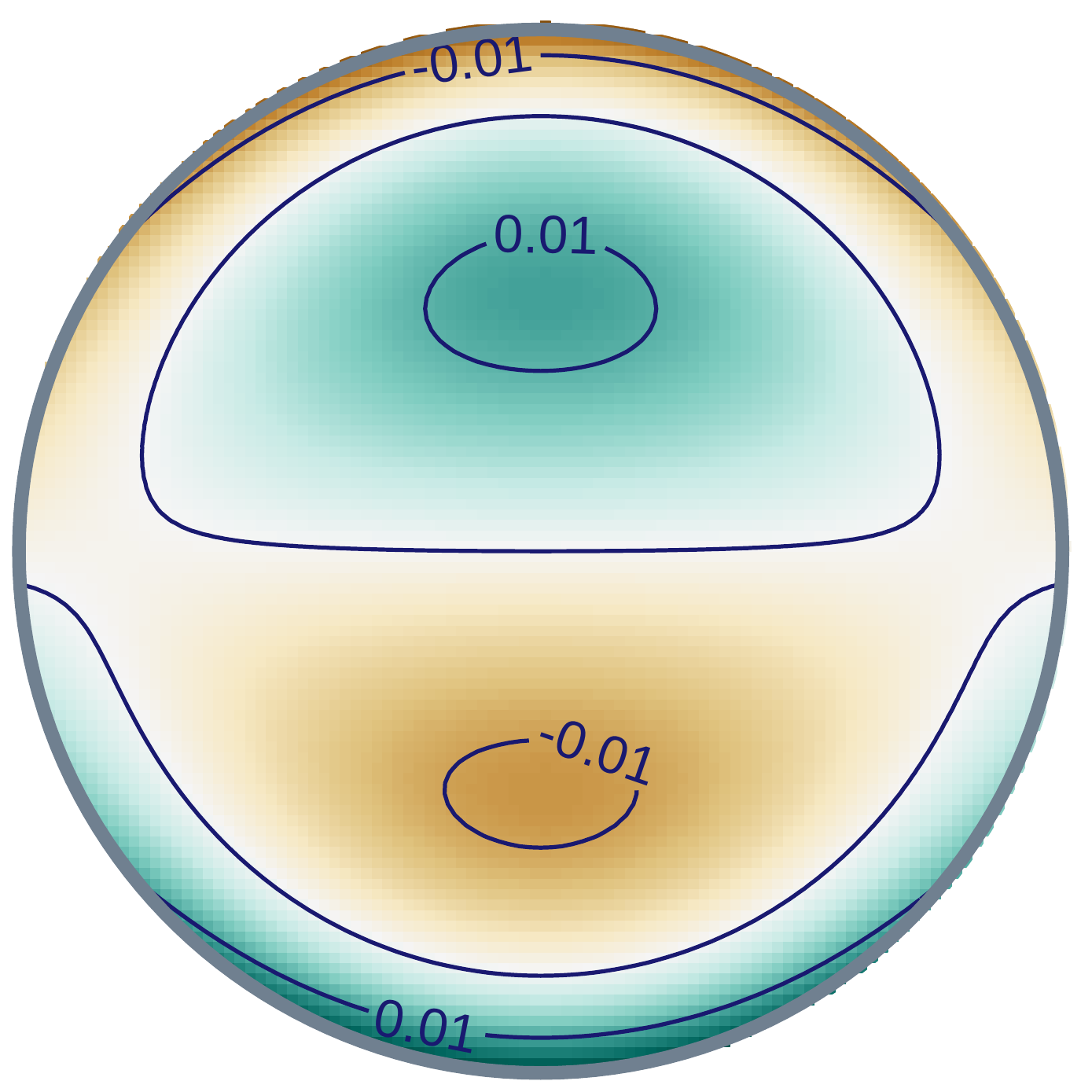}\hskip 1mm
        \includegraphics[height=0.33\linewidth]{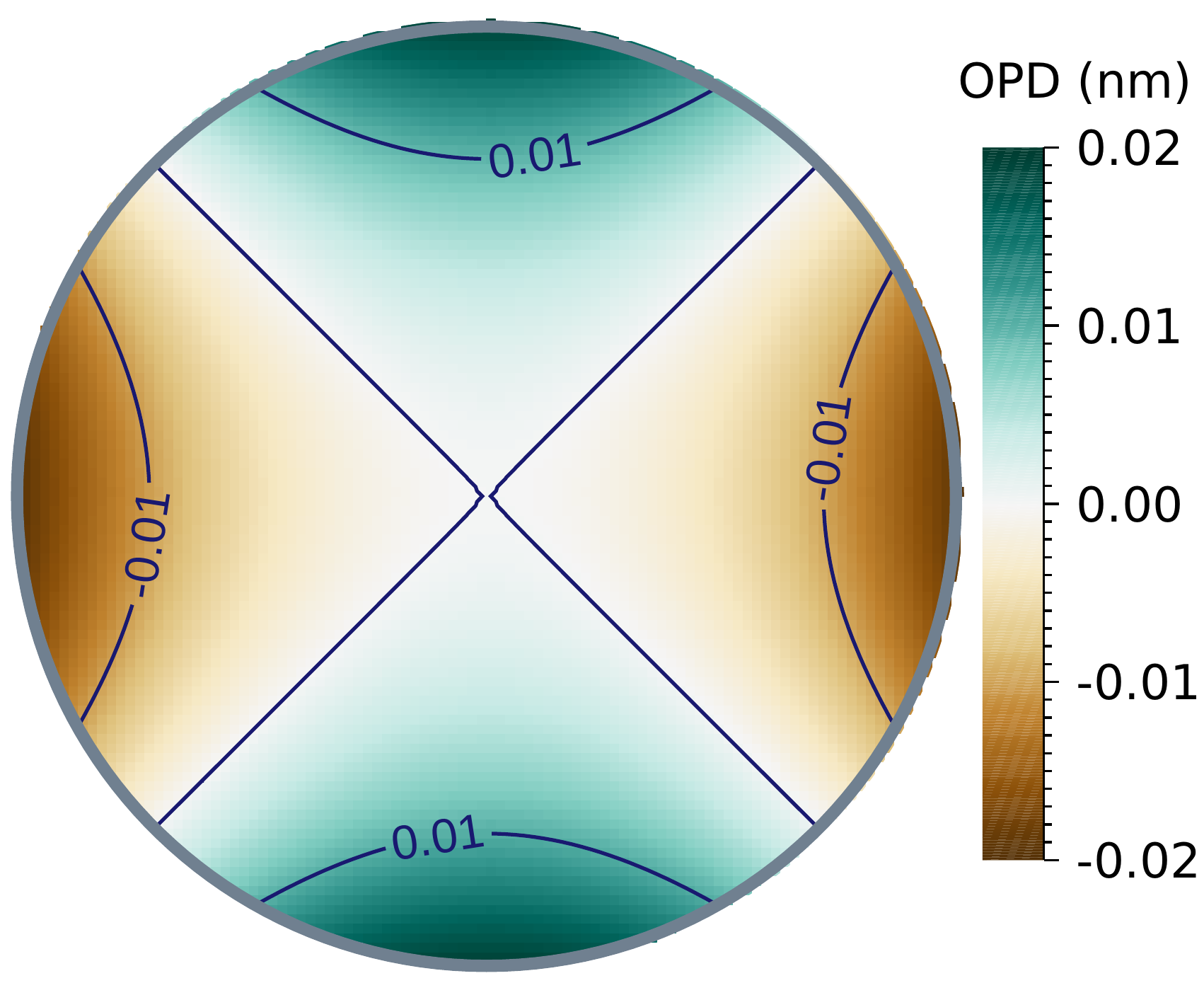}}}
    \put(0,0){\makebox(1,0.3){
        \includegraphics[width=0.5\linewidth]{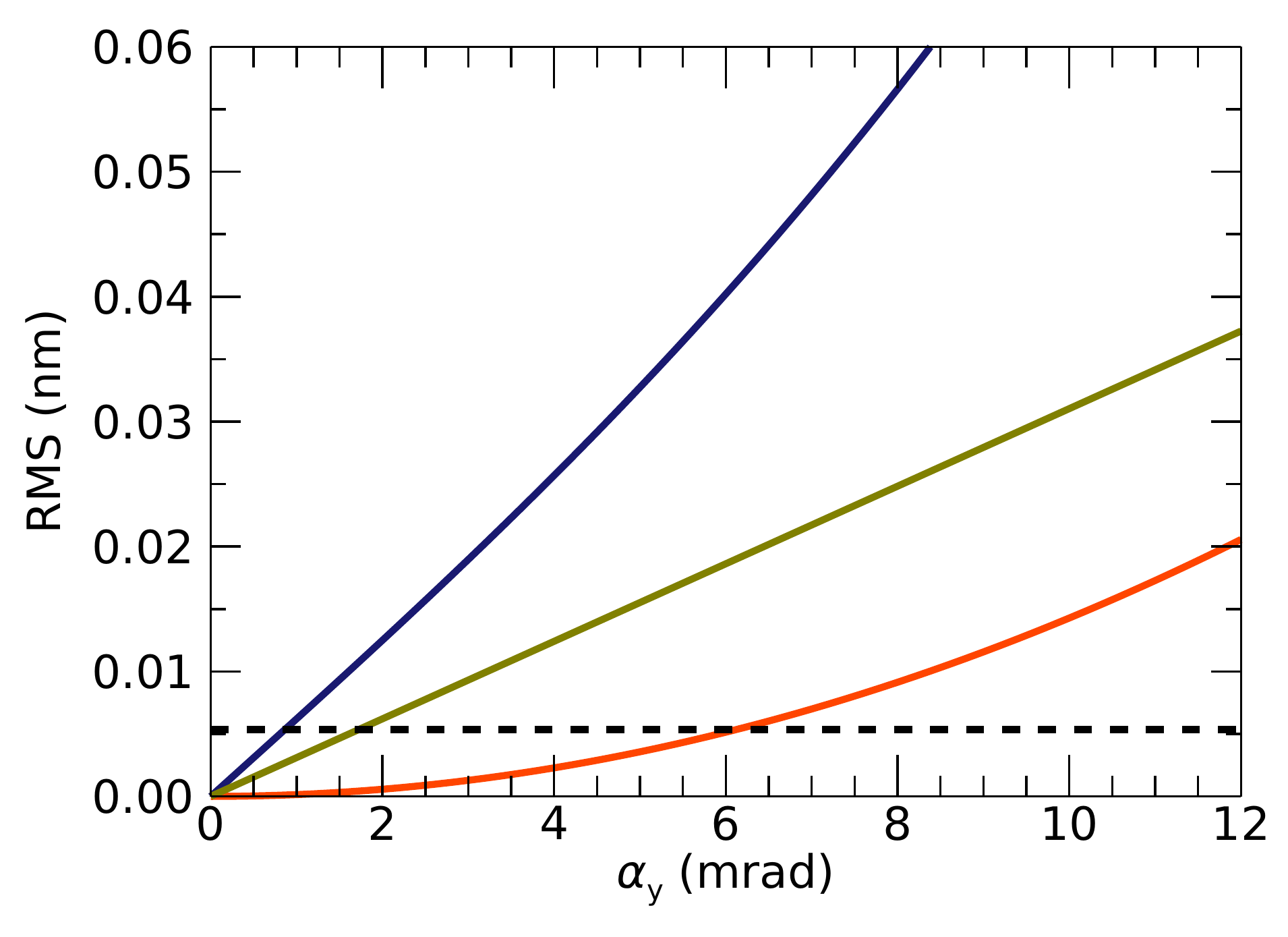}
        \includegraphics[width=0.5\linewidth]{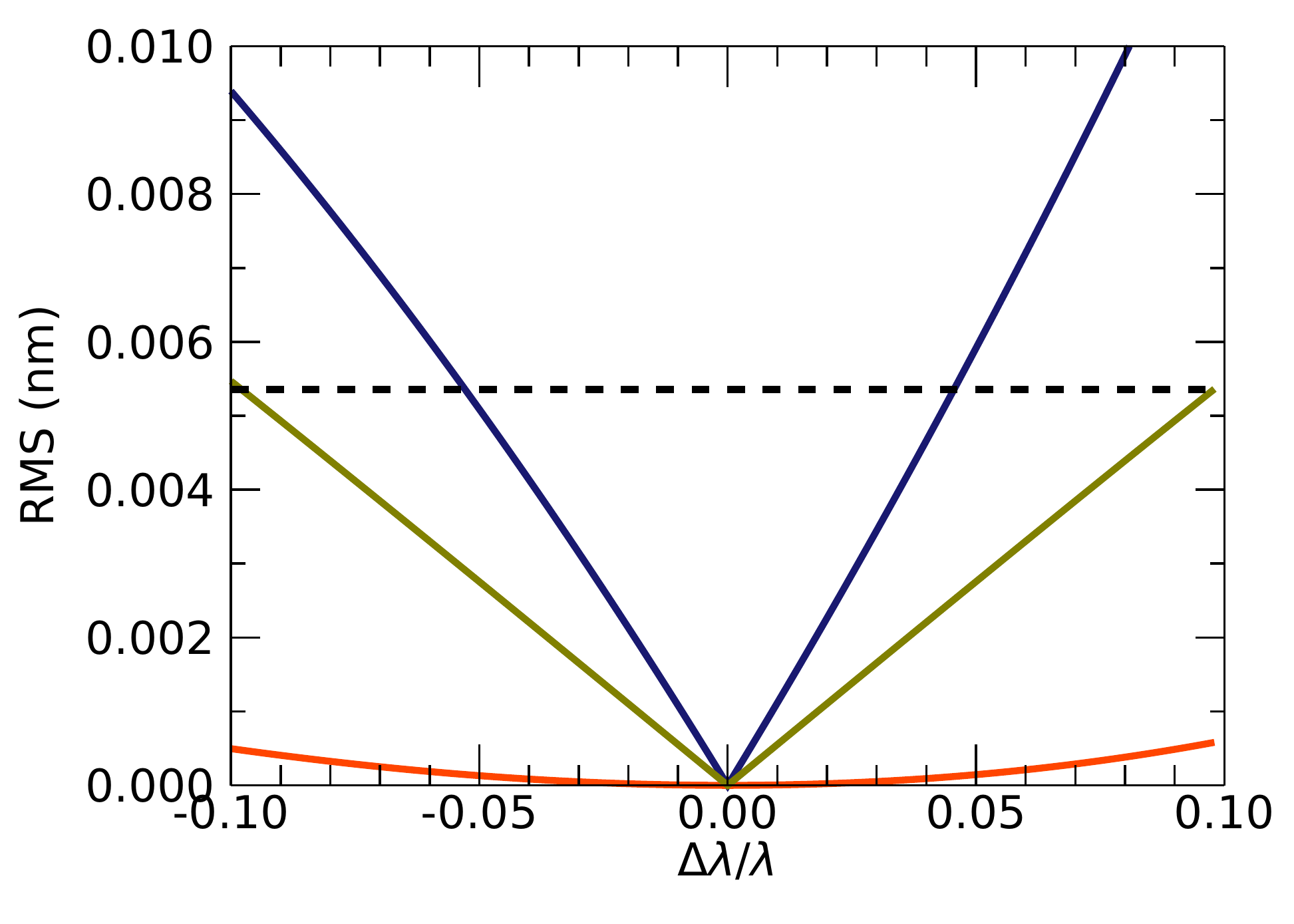}}}
    \put(0.15,0.66){\fontfamily{phv}\normalsize\selectfont\text{(a)}}
    \put(0.49,0.66){\fontfamily{phv}\normalsize\selectfont\text{(b)}}
    \put(0.11,0.27){\fontfamily{phv}\normalsize\selectfont\text{(c)}}
    \put(0.65,0.27){\fontfamily{phv}\normalsize\selectfont\text{(d)}}
  \end{picture}
     \caption{Wavefront errors (excluding defocus) of axisymmetric lenses with $\text{NA} = 0.0375$ and a focal
      length $f=\SI{1}{\milli\meter}$ and (a) a flat surface $R=\infty$ at a field angle
      $\alpha_y=\SI{1}{\milli\radian}$ and (b) a spherical surface with $R=f$ at a field
      angle $\alpha_y=\SI{7.5}{\milli\radian}$. (c) Plot of the RMS errors as a function
      of field angle for $R=\infty$ (blue), $R=(2/3)f$ (olive green) and $R=f$ (red). (d)
      RMS error as a function of the relative wavelength deviation from the design, where
      $\Delta\lambda=\lambda_m-\lambda$, and for $R=\infty$ (blue), $R=(2/3)f$ (olive
      green) and $R=f$ (red).  The Marechal condition of $\lambda/14$ is shown by the
      dashed line for a wavelength of \SI{0.075}{\nano\meter}.  The wavefront magnitude
      scales proportionally with the focal length $f$.}
  \label{fig:wavefronts}
\end{figure*}

At the design wavelength the calculated map of the OPD as a function of pupil coordinates for the on-axis field
point ($\alpha_y=0$) is zero to numerical precision, for both the flat MLL ($R=\infty$) and for
$R=f$. Thus, the prescription of the layer periods and tilts according to (\ref{eq:q})
avoid spherical aberration and all other aberrations as desired.  The wavefront map
for an off-axis field point at $\alpha_y = \SI{1}{\milli\radian}$ for a flat axisymmetric MLL is given in
figure~\ref{fig:wavefronts} (a).  This wavefront error, with an RMS value of
\SI{0.0062}{\nano\meter} (0.082 wavelengths), is dominated by coma and astigmatism in agreement with
(\ref{eq:OPD-R4}). Further calculations show that as the field angle increases, coma
increases linearly and becomes the dominant term.  However, by curving the lens surface
such that $R=f$, this term is essentially eliminated, leaving astigmatism as the dominant
aberration.  For the same field angle of \SI{1}{\milli\radian} the RMS wavefront for the
curved lens is only
\SI{0.0001}{\nano\meter} and so the OPD map is not visible on the same scale as for
figure~\ref{fig:wavefronts} (a). Instead, the map is shown for $R=f$ at a field angle of
\SI{7.5}{\milli\radian} in figure~\ref{fig:wavefronts} (b), after subtracting the best-fit
tilt and focus, with an RMS error of
\SI{0.0080}{\nano\meter}.   

According to the Marechal condition~\cite{Born:2002}, diffraction-limited imaging with
a Strehl ratio above \SI{80}{\percent} requires an RMS wavefront aberration less than
$\lambda/14$, or \SI{0.005}{\nano\meter} in this example. The quarter-wave rule of
Rayleigh corresponds to an absolute deviation of the OPD by \SI{0.019}{\nano\meter}.  Thus
it is seen that the field angles chosen for the calculations of figure~\ref{fig:wavefronts}
(a) and (b) just slightly exceed the aplanatic region that provides diffraction-limited
imaging.  The RMS error of both flat and curved lenses is plotted as a function of the
field angle in figure~\ref{fig:wavefronts} (c).  It can be seen that by curving the surface
of the lens to match the focal length, the radius of the aplanatic field increases from
\SI{0.8}{\milli\radian} to \SI{6.0}{\milli\radian}.  The quadratic dependence of the RMS
on field angle for the curved lens indicates this is dominated by the Seidel term for
astigmatism in (\ref{eq:OPD-R4}).  Indeed, after subtracting defocus and tilt,
$\text{OPD}^{(4)} (\rho_x,\rho_y)= (f/4) \alpha_y^2 (\rho_y^2 - \rho_x^2)$ for $R=f$.  The RMS of this
aberration over the circular pupil is equal to $f\,\alpha_y^2\,\text{NA}^2/(4\sqrt{6})$.
A plot of this curve cannot be distinguished from that shown in figure~\ref{fig:wavefronts}
(c), and the residual RMS aberration after subtracting this term from the numeric
computation of OPD is less than \SI{1e-3}{\nano\meter}.  Indeed, the contribution of other
aberrations besides astigmatism only reach an RMS value equal to $\lambda/14$ at a field
angle of \SI{100}{\milli\radian}.  The RMS error is also plotted in
figure~\ref{fig:wavefronts} (c) for the zero-astigmatism condition of $R=(2/3)f$.  As
mentioned above, coma is still present with $\text{OPD}^{(4)} = (f/2) \alpha_y \,\rho_y^3$,
giving rise to the linear dependence of the RMS wavefront error on field angle for lenses
with this surface radius. 

The aplanatic field of view was calculated here for the angular positions of objects at
infinity, but this also corresponds to forming a magnified image in the limit of high
magnification. In this case the field of view at the object plane is the field angle
multiplied by $f = \SI{1}{\milli\meter}$, or a diameter of \SI{12}{\micro\meter} for the
curved lens, which would be increased to \SI{200}{\micro\meter} if the astigmatism were
corrected.  A well-known approach to do this in lens design is to position the stop at a
different plane.  For a single refractive surface of radius $R$, astigmatism can be
eliminated by placing the stop a distance $R/2$ from that surface.  Unfortunately, this
remedy does not work well here since the reduction of astigmatism requires a stop to be
placed close to the focal plane.  It may be possible, instead, to use a refractive
meniscus lens near to the focal plane to change the curvature of the wave-field in the
direction of the field displacement and by an amount that increases with that
displacement.

\begin{figure}
  \centering
  \setlength{\unitlength}{\linewidth}
  \begin{picture}(1,1.29)(0,0) 
    \put(0.025,0.63){
      \includegraphics[width=0.95\linewidth]{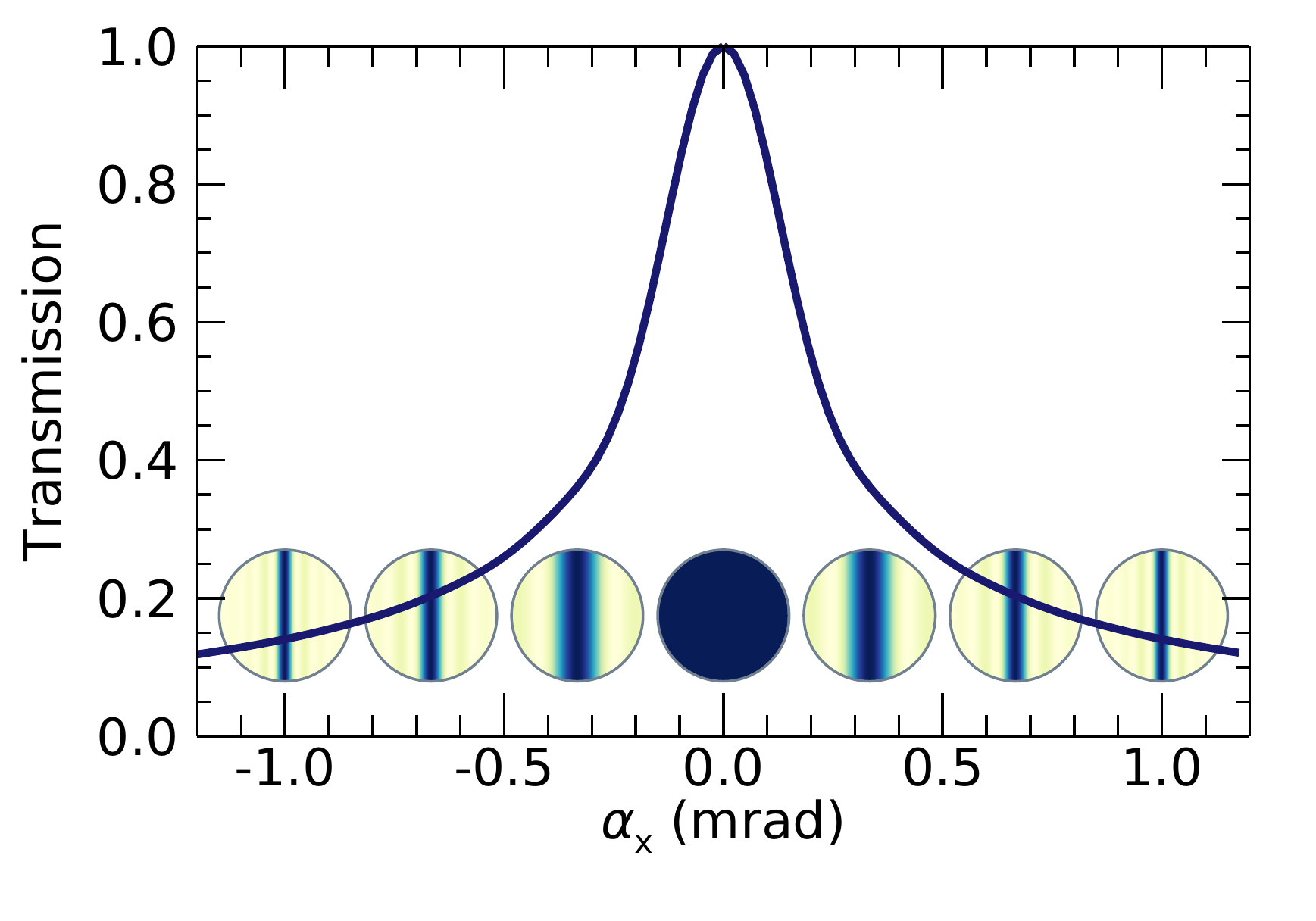}}
    \put(0.025,0){
      \includegraphics[width=0.95\linewidth]{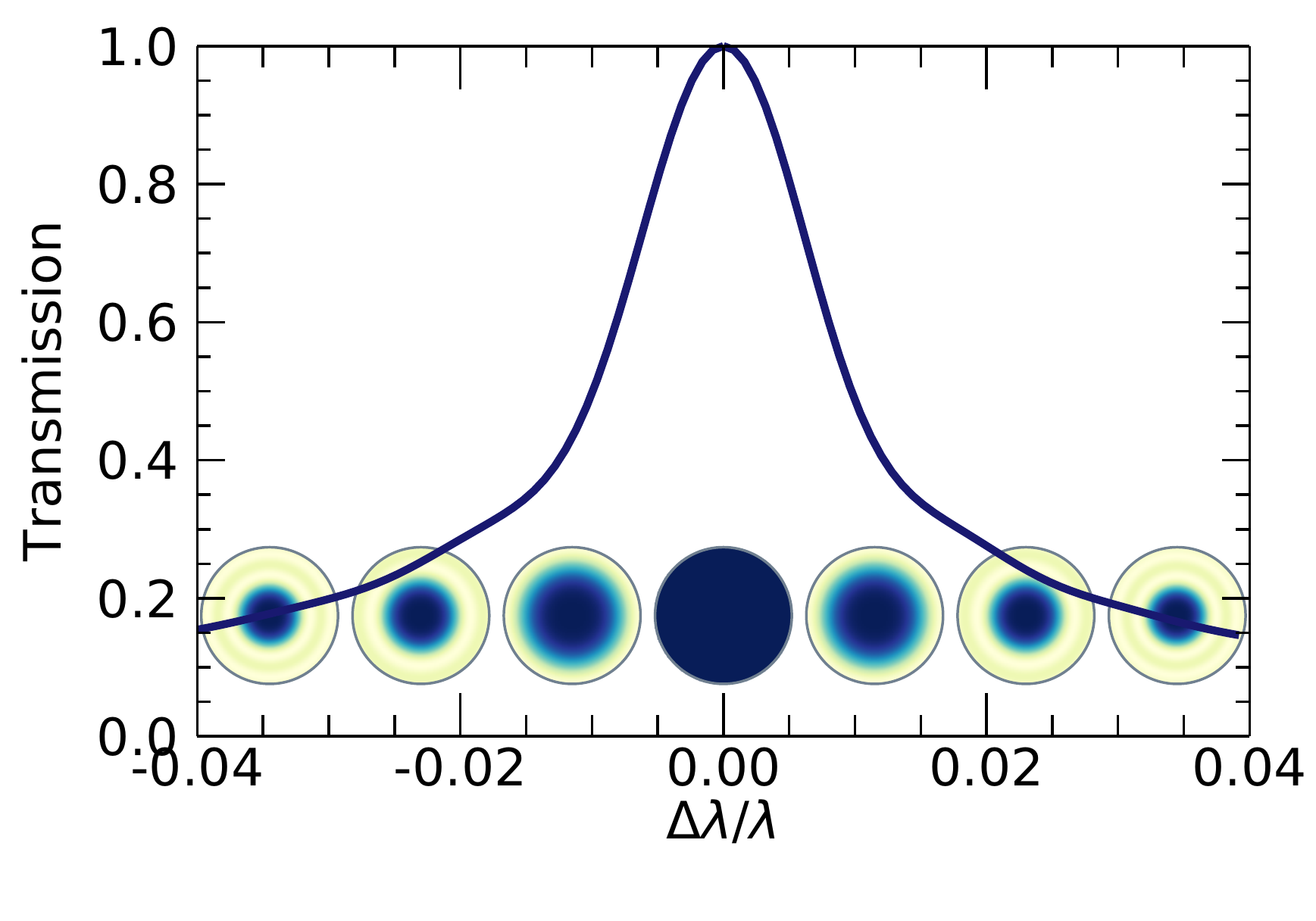}}
    \put(0,1.26){\fontfamily{phv}\normalsize\selectfont\text{(a)}}
    \put(0,0.62){\fontfamily{phv}\normalsize\selectfont\text{(b)}}
  \end{picture}
  \caption[Transmission of the MLL]{Transmission of an
    axisymmetric MLL with $\text{NA}=0.0375$ and a focal length $f=\SI{1}{\milli\meter}$ 
    as a function of (a) the field angle, and (b) the relative change in wavelength. In
    (a) the transmission maps of the lens pupil are
    shown for field angles $\alpha_x=-1$, $-0.6$, $-0.3$, $0$, $0.3$, $0.6$, and
    \SI{1}{\milli\radian}, and in (b) for $\Delta\lambda/\lambda = -0.03$, $-0.02$,
    $-0.01$, $0$, $0.01$, $0.02$, and $0.03$.}
  \label{fig:transmission}
\end{figure}

As discussed in \sref{sec:transmission-2D}, the field of view of a MLL may be limited not by the field dependence
of the aberrations, but rather by the Darwin width of the Laue reflection, proportional to the difference of the optical constants
of the materials that make up the layers of the MLL, as given by (\ref{eq:vignette-2D}).  For SiC and WC layers with densities
of \SI{2.47}{\gram\per\centi\meter\cubed} and \SI{14.6}{\gram\per\centi\meter\cubed},
respectively, $|\delta_1-\delta_2| = \num{6.7e-6}$ at a wavelength of
\SI{0.075}{\nano\meter}, giving a half width of
$w_\theta = 2|\delta_1-\delta_2|/(\pi\,\text{NA}) = \SI{0.11}{\milli\radian}$ for the
layers at the boundary of the lens considered here.  
A plot of the total lens transmission obtained from the numerical analysis of this lens 
is given in figure~\ref{fig:transmission} as a function of the field angle.  Maps of the
transmission of the lens pupil are also shown for particular field angles, showing the
loss of transmission in the thinner layers first.  As seen in the maps, this transmission
loss depends only on the projection of the layer period in the direction of the transverse
displacement of the field, giving a band of the lens that transmits.  The reduction of
the aperture leads to a corresponding reduction in resolution in the direction of the
object or image point (for example, vertical lines will be less resolved than horizontal
lines when located at a horizontal position in the field).  The lens transmission does not
depend noticeably on the radius $R$. 

\begin{figure*}
  \centering
  \includegraphics[width=0.95\linewidth]{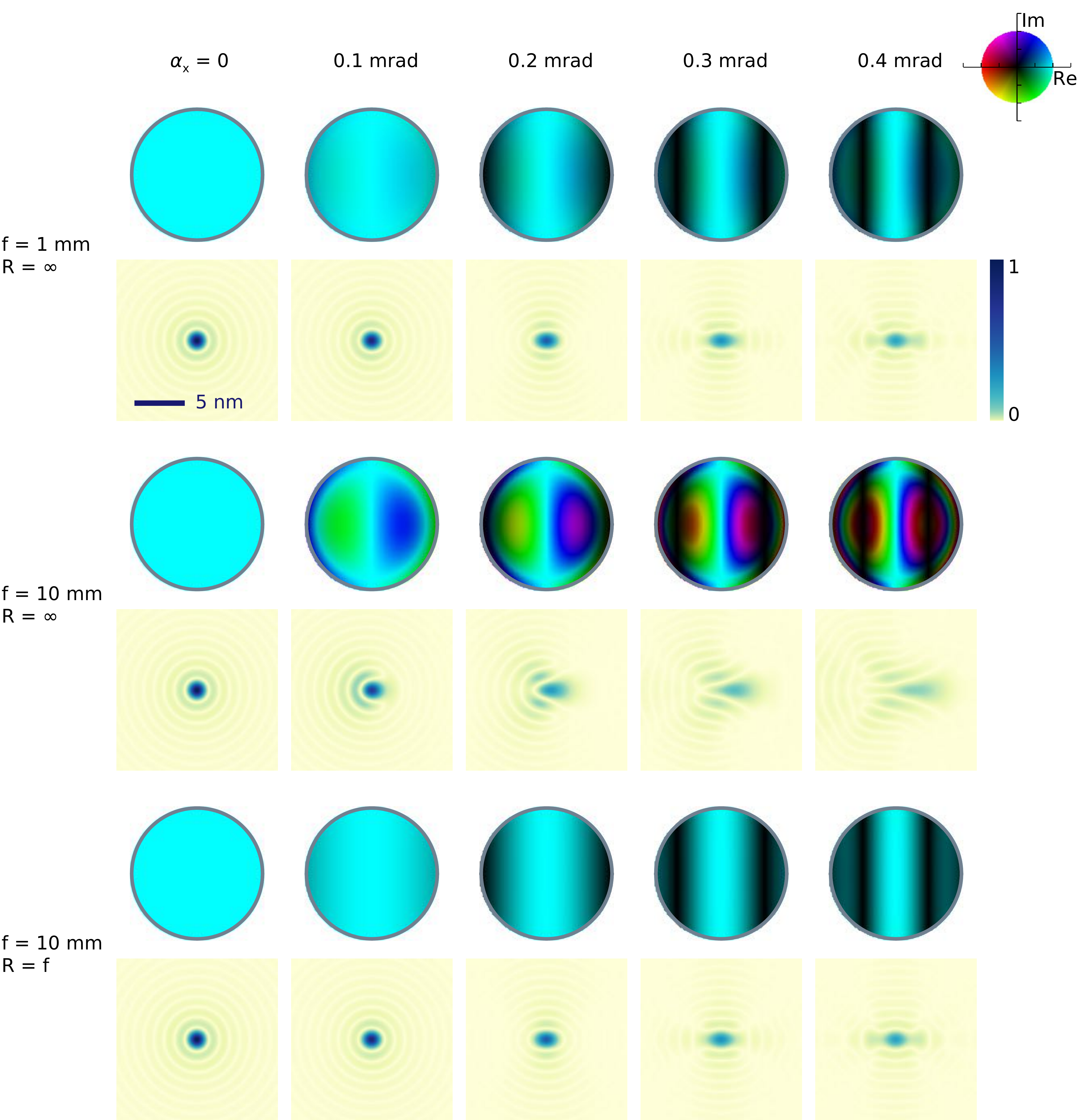}
  \caption[Field dependence of the complex-valued wavefront and PSF]{Complex-valued pupil
    functions of MLLs with $\text{NA} = 0.0375$ for five field angles $\alpha_x$ as
    indicated, along with their associated point spread functions.  Three lens
    configurations are shown as indicated by the labels to the right.  Complex values in
    the pupil function are visualised by hue (phase) and brightness (amplitude) according to the color-wheel
    shown at top right.}
  \label{fig:PSF-2D}
\end{figure*}

As seen in \sref{sec:transmission-2D}, the largest field angle that can be tolerated
by apodisation due to a loss of diffraction efficiency depends only on the multilayer
materials and inversely on numerical aperture, whereas the magnitude of the wavefront
aberrations are proportional to the focal length and inversely on powers of the numerical
aperture (depending on whether coma or astigmatism dominate).  The competition of the
effects of the wavefront aberration and apodisation of the pupil on the imaging properties
of the lens is illustrated in figure~\ref{fig:PSF-2D} where the combined complex-valued
pupil function is shown for various field angles.  The square modulus of the Fourier
transform of this yields the point spread function (PSF)~\cite{Born:2002,Goodman:1996,Cowley:1981}, also shown
in the figure.  The radius of the aplanatic field for the $f=\SI{1}{\milli\meter}$,
$\text{NA}=0.0375$ lens considered here is much larger than the limits due to diffraction
efficiency, as can be seen in the top row of figure~\ref{fig:PSF-2D}.  When the field angle
exceeds \SI{0.2}{\milli\radian} (or a image field radius of \SI{0.2}{\micro\meter})
efficiency is lost from two edges of the lens, leading to a broadening of the PSF in the
direction of the displacement of the object point.  Increasing the focal length of the
lens to \SI{10}{\milli\meter} for the same NA (implying a 10 times increase in the lens
diameter) leads to a corresponding 10-times increase in the OPD.  As seen in the second
pair of rows of figure~\ref{fig:PSF-2D}, coma causes a degradation of the PSF
at lower field angles, before apodisation becomes a problem.  Here, the field is limited
by the Marechal condition to \SI{0.1}{\milli\radian} or a field radius of
\SI{1}{\micro\meter}. The aberrations can be removed in principle by adopting an aplanatic
design by cutting the lens on a spherical surface with a radius $R=f$, giving a
performance similar to the \SI{1}{\milli\meter} lens (bottom pair of rows) and an image
field radius of about \SI{2}{\micro\meter} (or 2000 resolution elements).

Away from atomic resonances, the optical constants of materials,
and hence their difference $|\delta_1-\delta_2|$, vary with $\lambda^2$.  For lenses of a
given resolution (i.e. made with a minimum $d$ spacing) the NA increases in proportion
with $\lambda$ so that the angular width $w_\theta$ is linearly proportional to
wavelength.  Thus the largest ratio of field of view to resolution is achieved at longest
wavelength.

The dominant aberrations induced by a change in wavelength are tilt and defocus due to
the change in focal length (which varies inversely with wavelength) and the corresponding
change in the paraxial image position.  For an on-axis field point the OPD due to
defocus is given by $(f/2) (\Delta\lambda/\lambda) \rho^2$ with a RMS value of $(f/4)
(\Delta\lambda/\lambda) \text{NA}^2$.  Thus, applying the Marechal condition
(equivalent to a defocus of less than $(2/7)\lambda/\text{NA}^2$), a relative bandwidth of
$2\Delta\lambda/\lambda < (8/14) \lambda /(f \,\text{NA}^2) = \num{3.0e-5}$ can be tolerated
for \SI{1}{\nano\meter} imaging for the lens considered here.  A shorter focal length
allows a broader bandwidth since this gives fewer periods in the lens.

A MLL manufactured for a particular wavelength might be used at another, for example in
spectro-microscopy applications where the sample must be repositioned at the focal
distance of $f_m=f\,\lambda/\lambda_m$.  The range of wavelengths that can be used depends
both on the spherical aberration and the apodisation of the lens caused by the incorrect
tilt of the layers for the particular wavelength.  The RMS wavefront error is plotted in
figure~\ref{fig:wavefronts} (d) for the on-axis field point as a function of wavelength.  Here, the
ray-tracing was performed at the paraxial image plane at $f_m$. As expected from
(\ref{eq:OPD-SA}), the best compensation is achieved for $R=f$, where the dependence on
$\Delta\lambda$ is quadratic.  However, as seen from figure~\ref{fig:transmission} (b), and
expected from (\ref{eq:vignette-chromatic}), the
limitation on using the lens at different wavelengths arises from apodisation, not
spherical aberration.

\section{Off-axis aberrations of 1D MLLs}
\label{sec:1D-MLL}
Multilayer Laue lenses are typically fabricated by layer deposition onto a flat substrate,
to produce a structure that focuses only in one direction, similar to a cylindrical lens.
Two crossed cylindrical lenses can focus or form an image in two dimensions, as can two crossed
MLLs.  The imaging characteristics and off-axis aberrations of
such systems differ from axi-symmetric lenses examined above.  In particular, crossed MLLs
are anamorphic (giving a different magnification in each transverse direction due to the
different object and image distances to each lens), the
concept of entrance and exit pupils is not well defined, and there are 16 primary
aberrations instead of the five Seidel aberrations of an axi-symmetric
system~\cite{Yuan:2009}.
The surface $\vec{s}_1$, surface normal $\hat{\vec{n}}_1$, reciprocal vector $\vec{q}_1$, and accumulated phase
$\phi_1$ for a 1D MLL focusing in the $y$ direction (and polished to a cylindrical shape
of radius $R_1$ in that same direction) are listed in the
left-hand column of \tref{tab:crossed} as found by
substituting $r$ with $y$ in the expressions for the axi-symmetric case in \sref{sec:off-axis}, and setting to
zero the $x$ components of $\hat{\vec{n}}$ and $\vec{q}$.

The case of a single 1D MLL, with layers parallel to the $x$ axis, imaging an off-axis
source point at a field angle $\alpha_y$ in the $y$-$z$ plane is equivalent to the analysis of
\sref{sec:off-axis} for $\psi=0$. That is, the $y$-component of the rays obey the
expression $\text{OPD}(r)$ of (\ref{eq:OPD-R4}) with $\psi=0$:
\begin{equation}
  \label{eq:OPD-R1}
 \fl \text{OPD}^{(4)}(\rho_y;\alpha_y) = -\frac{3f_y}{2}\alpha_y^3 \rho_y 
  +3f_y\left(\frac{3}{4}-\frac{f_y}{2R_y}\right) \alpha_y^2 \rho_y^2 +
  f_y\left(\frac{f_y}{R_y}-1\right)\frac{1}{f_y} \alpha_y \rho_y^3.
\end{equation}
Thus, aplanatic focusing in one dimension can be achieved by cutting or polishing the MLL
to a cylindrical surface with a radius $R_y=f_y$.

The situation is slightly different for field points off the $y$-$z$ plane. Despite the symmetry
of a 1D MLL, we find that a source point at a field angle $\alpha_x$ in the $x$-$z$ plane
is not necessarily focused with zero aberration.  This is due to the tilt of the layers in
the MLL.  An incident ray at a field angle $\alpha_x$ can be obtained by rotating an axial ray
about the $y$ axis.  However, the $\vec{q}$ vectors that define the layers in the MLL are
not parallel to the $y$ axis, and as such these inclined rays no longer satisfy the Bragg
condition.  
The analysis can be carried out in the same way as in \sref{sec:off-axis},
starting with $\hat{\vec{r}} = (\sin \alpha_x,\sin\alpha_y,\{1-{\sin^2}\alpha_x-{\sin^2}\alpha_y\}^{1/2})$.
The series expansion of the resulting expression for the OPD to fourth order in products
of $\alpha_x$ and $y$ yields
\begin{equation}
  \label{eq:OPD-R-alphax}
\fl  \text{OPD}^{(4)}(\rho_x,\rho_y) = f_y\,\alpha_x \rho_x -\frac{f_y}{6}\alpha_x^3 \rho_x +
  f_y\left(\frac{3}{4}-\frac{f_y}{2R_y}\right) \alpha_x^2 \rho_y^2 + \text{OPD}^{(4)}(\rho_y;\alpha_y),
\end{equation}
where $\text{OPD}^{(4)}(\rho_y;\alpha_y)$ is given by
(\ref{eq:OPD-R1}). Equation~(\ref{eq:OPD-R-alphax}) shows that
in this fourth-order approximation
the only consequence of tilting the source
point out of the $y$-$z$ plane is distortion and curvature of field.  However, as we shall
see below, this leads to an aberrated wave-field when two such 1D lenses are crossed
orthogonally to image in two dimensions. 

\begin{figure}
  \centering
  \includegraphics[width=\linewidth]{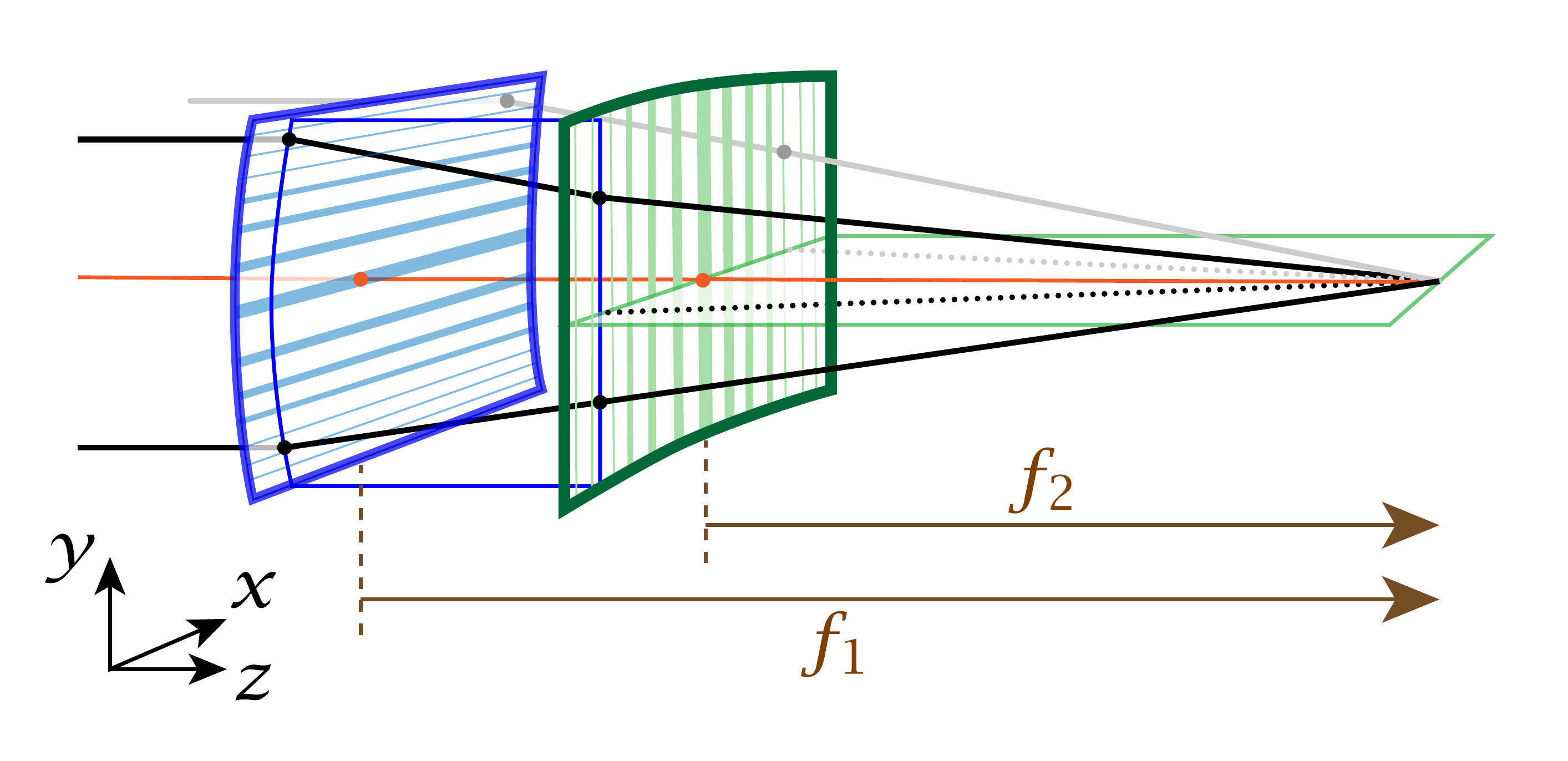}
  \caption{Schematic diagram of two 1D MLLs oriented to focus in orthogonal
    directions. The first lens focuses in the $y$ direction with a focal length $f_1$ and
    the second focuses in the $x$ direction with a focal length $f_2$, positioned a
    distance $f_1-f_2$ from the first.  The lenses are also cylindrically curved with radii of curvature
    $R_1$ and $R_2$, respectively.}
  \label{fig:crossed-lenses}
\end{figure}

\subsection{Off-axis aberrations of crossed 1D MLLs}
\label{sec:crossed}
To determine the performance of crossed 1D MLLs
we trace rays through an optical system of
a 1D MLL focusing in the $y$ direction with a focal length $f_1$ followed by another
focusing in the $x$ direction with a focal length $f_2$.  The distance between the two
lenses is set to $f_1-f_2$ to ensure they both focus to a common point, and for generality the lenses are polished to cylindrical
surfaces with radii $R_1$ and $R_2$, respectively (see figure~\ref{fig:crossed-lenses}).  The functions describing the geometry
and optical properties of the lenses are given in \tref{tab:crossed}.  The
calculation of the OPD proceeds as before, but now with a source point located at a
position given by the angles $\alpha_x$ and $\alpha_y$ such that incident rays on the
first MLL have a direction $\hat{\vec{r}}_1 = (\sin \alpha_x, \sin \alpha_y,
\{1-{\sin}^2\alpha_x -{\sin}^2\alpha_y\}^{1/2})$.  These rays are traced a distance $l_1$ from the plane
perpendicular to their direction to the first lens at the position $\vec{s}_1 =
(x_1,y_1,s_1)$, and then deflected to directions $\hat{\vec{r}}'_1$ as
described in sections~\ref{sec:off-axis} and \ref{sec:1D-MLL} (solving for the deviation
parameter $\epsilon_1$).  A ray intersects the
second lens at the point $\vec{s}_2
=\vec{s}_1 + l_{12}\,\hat{\vec{r}}'_1$ after propagating a distance $l_{12}$.   The $x$ component of this expression
yields $x_2 = x_1+l_{12}\,r'_{1,x}$, which can be substituted into
the expression for $s_2$ given in \tref{tab:crossed} to obtain, from the $z$
component, a quadratic equation that can be solved for $l_{12}$.  The ray at $\vec{s}_2$
incident at an angle $\hat{\vec{r}}_2 = \hat{\vec{r}}'_1$
can then be deflected into the direction $\hat{\vec{r}}'_2$ that is calculated using (\ref{eq:off-Bragg}) and the expressions for
$\hat{\vec{n}}_2$ and $\vec{q}_2$ in \tref{tab:crossed}, solving for the off-Bragg
deviation parameter $\epsilon_2$ by requiring that $|\hat{\vec{r}}'_2|=1$.   The distance $l_2$ from
$\vec{s}_2$ in the second lens to the image plane can be then obtained from
$l_2\,\hat{\vec{r}}'_2 \cdot \hat{\vec{z}} = f_2-s_2$. 

\begin{table*}
\caption{Expressions of the sag $s$, surface coordinates $\vec{s}$, surface normal $\hat{\vec{n}}$,
  reciprocal vector for diffraction $\vec{q}$, and phase lag due to the number of periods
  $\phi$, for the two crossed MLLs of focal lengths $f_1$ and $f_2$, and radii $R_1$ and
  $R_2$, respectively, separated by a distance $f_1-f_2$.}
\centering
\begin{scriptsize}
  \noindent\begin{tabular}{@{}l c l@{}}
             \br
             First lens (vertex at $0$) & & Second lens (vertex at $f_1-f_2$) \\
             \mr
            \begin{minipage}[t]{0.45\textwidth}
       
      \begin{eqnarray*}
 \fl      s_1 =R_1-\sqrt{R^2_1-y_1^2} \\
 \fl       \vec{s}_1 =(x_1,y_1,s_1) \\
 \fl       \hat{\vec{n}}_1 = \left(0,\frac{y_1}{R_1},-\sqrt{1-\frac{y_1^2}{R_1^2}}\right) \\
 \fl       \frac{\lambda}{2\pi} \vec{q}_1 = \left(0,
          \frac{-y_1}{\sqrt{y_1^2+(f_1-s_1)^2}},\frac{f_1-s_1}{\sqrt{y_1^2+(f_1-s_1)^2}}-1\right)\\
  \fl      \frac{\lambda}{2\pi} \phi_1= \sqrt{y_1^2+(f_1-s_1)^2}-f_1+s_1\\
      \end{eqnarray*}
    \end{minipage}
    & &
    \begin{minipage}[t]{0.5\textwidth}
      \begin{eqnarray*}
 \fl      s_2 =R_2-\sqrt{R^2_2-x_2^2} \\
 \fl       \vec{s}_2 =(x_2,y_2, f_1-f_2+s_2)\\
 \fl        \hat{\vec{n}}_2 = \left(\frac{x_2}{R_2},0,-\sqrt{1-\frac{x_2^2}{R_2^2}}\right) \\
 \fl        \frac{\lambda}{2\pi} \vec{q}_2 = \left(
           \frac{-x_2}{\sqrt{x_2^2+(f_2-s_2)^2}},0,\frac{f_2-s_2}{\sqrt{x_2^2+(f_2-s_2)^2}}-1\right)\\
 \fl        \frac{\lambda}{2\pi} \phi_2 = \sqrt{x_2^2+(f_2-s_2)^2}-f_2+s_2\\
       \end{eqnarray*}
     \end{minipage} \\
   \br
  \end{tabular}
\end{scriptsize}
\label{tab:crossed}
\end{table*}

As stated above, two crossed cylindrical lenses do not have a well-defined pupil (that is,
an aperture in a single plane that limits the angular extent of all rays) and therefore
this optical system does not have a well-defined chief ray.  Nevertheless we can express
the OPD as a function of coordinates $(x_1,y_1)$ of the first lens, as the difference
between the OPL of a ray that passes through those coordinates and the vertex of that
lens, as
\begin{equation}
  \label{eq:OPD-crossed}
\fl  \text{OPD}(x_1,y_1) = \text{OPL}(x_1,y_1) -
  \text{OPL}(0,0)-\frac{\lambda_m}{2\pi}\left[\phi_1(y_1)+\phi_2(x_1+l_{12}\,r'_{1,x})\right] 
\end{equation}
with $\text{OPL} = l_1 + l_{12}+l_2$ and $\lambda_m$ is the measurement wavelength.  The
phase lags $\phi_1$ and $\phi_2$ are proportional to the number of diffracting periods between the ray of
interest and the reference ray, and is thus determined by the coordinate $y_1$ or $x_2$ of
the ray on the particular lens.  Retaining terms to fourth order in pupil and field
coordinates, when $\lambda_m = \lambda$ the OPD is evaluated to
\begin{eqnarray}
  \label{eq:OPD4}
 \fl   \text{OPD}^{(4)}(\rho_x,\rho_y) &=  \frac{3}{4} f_2 \,\rho_x^2\,\rho_y^2 \nonumber\\
   & + f_2\left(\frac{f_2}{R_2}-1\right)  \alpha_x \, \rho_x^3 
     + f_1\left(\frac{f_1}{R_1}-1\right)  \alpha_y \, \rho_y^3 \nonumber\\
   & - \frac{3}{2} f_2 \,\alpha_y \, \rho_x^2 \, \rho_y
    + \frac{3}{2} (f_1-2f_2)\, \alpha_x \, \rho_x \, \rho_y^2 \nonumber\\
    & + \frac{3}{4} \left[ 4f_1\left(\frac{f_2}{R_2}-1\right)
      +f_2\left(7-\frac{6f_2}{R_2} \right)\right]\alpha_x^2 \,\rho_x^2
     +\frac{1}{4} f_1\left (3-\frac{2f_1}{R_1}\right)\alpha_y^2\,\rho_y^2 \nonumber\\
    & + \frac{3}{4}f_2\,\alpha_y^2\,\rho_x^2 
    +\frac{1}{4}\left[11(f_2-f_1)+\frac{3f_1^2}{f_2}\right] \alpha_x^2 \,\rho_y^2 \\
    &-3(f_1-2f_2)\,\alpha_x\,\alpha_y\,\rho_x\,\rho_y \nonumber\\
   & +\frac{3}{2}(f_1-2f_2)\,\alpha_x\,\alpha_y^2 \,\rho_x
    -\frac{1}{2}\left [11(f_2-f_1)-\frac{3
        f_1^2}{f_2}\right]\alpha_x^2\,\alpha_y\,\rho_y \nonumber\\
    & +3\left(\frac{f_1}{f_2}-2\right)\left[f_1\left(\frac{f_2}{R_2}-1\right)+
      f_2\left(\frac{3}{2}-\frac{f_2}{R_2}\right)\right]
     \alpha_x^3 \,\rho_x - \frac{3f_1}{2}\,\alpha_y^3\,\rho_y \nonumber
\end{eqnarray}
Here, $\rho_x = x_1/f_2$
and $\rho_y = y_1/f_1$.  That is, $(\rho_x,\rho_y)$ are equal to the tangents of the
angles of rays converging onto the focus, ignoring deflections of rays in the $x$
direction by the first lens.

There are 14 terms in (\ref{eq:OPD4}), out of a possible 16 allowed by the
symmetry of an anamorphic system~\cite{Yuan:2009}.  The two absent fourth-order aberrations
are the spherical aberration terms of the individual lenses, proportional to $\rho_x^4$
and $\rho_y^4$, which are zero by design (as given by the forms of $\phi$ and $\vec{q}$).
However there is a term in (\ref{eq:OPD4}) proportional to $\rho_x^2\,\rho_y^2$
which thus varies with the fourth power of the pupil coordinate along the diagonal and is
the only aberration for the on-axis field points $\alpha_x = \alpha_y =
0$.  
It is seen from (\ref{eq:OPD4}) that the coma terms (proportional to
$\alpha_y\,\rho_x^3$ and $\alpha_x\,\rho_y^3$) are zero under the same conditions as for
the individual lenses, namely that the curvatures of the lenses are set equal to the focal
lengths, $R_1 = f_1$ and $R_2=f_2$.  However, even then, there are some coma-like cross
terms (proportional to $\alpha_y \, \rho_x^2 \, \rho_y$ and
$\alpha_x \, \rho_x \, \rho_y^2$) that are independent of $R_1$ and $R_2$.  Indeed, the
only dependence of the OPD on the radii of curvature occur in terms that are proportional
to only $x$ or only $y$ coordinates, and we find through computational ray tracing that,
in general, errors in or modifications to the wavefronts of the individual lenses only
affect the wavefront in the corresponding coordinate of the system wavefront.  Thus, a
radius of curvature of the lenses equal to their focal lengths does reduce coma, and does
minimise off-axis aberrations, but it does not produce an aplanatic system as is
achievable for an axi-symmetric lens.  The remaining terms of (\ref{eq:OPD4})
are either quadratic in the pupil coordinates, representing curvature of field and
astigmatism, or linear in pupil coordinates, describing image distortion.  The
field-dependent \ang{45} astigmatism (or more correctly, oblique astigmatism),
proportional to $\rho_x\,\rho_y$, is zero for $f_2=f_1/2$, as are some of the other terms.
When $f_1 = f_2 = R_1 = R_2$ the expression for OPD is invariant in swapping the $x$ and
$y$ coordinates ($\rho_x \leftrightarrow \rho_y$ and $\alpha_x \leftrightarrow \alpha_y$)
as expected.

The minimum aberration occurs for a field point on axis, and unlike the case of the
individual lenses this aberration is non-zero, with
$\text{OPD} = (3/4)f_2\,\rho_x^2\,\rho_y^2$.  This non-zero aberration exists even though
each individual lens focuses an incident collimated beam without aberration. It can be termed ``oblique
spherical aberration'' since it varies with the fourth order of the pupil radial
coordinate along the diagonal ($\psi =\ang{45}$ for a square pupil), and is zero in the
$x$ and $y$ directions.  For on-axis cylindrical lenses where $\rho_x$ and $\rho_y$ range
from $-\text{NA}$ to $+\text{NA}$, the RMS wavefront error is evaluated to
$(3/20) f_2 \, \text{NA}^4$.  This can be compensated slightly by introducing defocus
proportional to $\rho_x^2+\rho_y^2$, reducing the RMS wavefront error to a minimum of
$\sqrt{2/7}(3/20) f_2 \,\text{NA}^4 = 0.11 f_2 \,\text{NA}^4$.  This error is significant
for high-NA lenses for imaging at \SI{1}{\nano\meter} resolution.  For example, for
$\text{NA} = 0.0375$, $\lambda = \SI{0.075}{\nano\meter}$, and $f_2=\SI{1}{\milli\meter}$,
the RMS error is equal to $\SI{0.16}{\nano\meter} = 2.1 \,\text{waves}$, excluding tilt,
defocus, and astigmatism.  The Marechal condition at this wavelength (i.e an RMS wavefront
less than $\lambda/14 = \SI{0.005}{\nano\meter}$) can be satisfied for lenses with a
\SI{1}{\milli\meter} focal length only for $\text{NA} < 0.016$, or a resolution of
\SI{2.3}{\nano\meter}.  Alternatively, a resolution of \SI{1}{\nano\meter} at this
wavelength requires a focal length $f < \SI{34}{\micro\meter}$.  Such a lens would have a
very small aperture, so a more practical approach would be to compensate the on-axis
aberration with an appropriately shaped refractive phase plate~\cite{Seiboth:2017}, akin
to a Schmidt corrector plate~\cite{Jenkins:1976}.  

The $\rho_x^2\,\rho_y^2$ error is not dependent on the curvatures of the surfaces of the
individual lenses, nor the distance between them, and is independent of the field angle.
The origin of this error can be traced to the curvature of field of the second MLL (here
focusing in the $x$ direction) due to rays incident out of the $x$--$z$ plane.  This is the
situation addressed by (\ref{eq:OPD-R-alphax}) (but with the $x$ and $y$ directions
reversed).  That is, consider rays from the on-axis field point with
$\hat{\vec{r}}_1 = (0,0,1)$ intersecting the first MLL at a particular height $y_1$ and a
range of values $x_1$.  These will be all be deflected in the $y$ direction by the same
angle, $\alpha_y \approx y_1/f_1$, and will impinge on the second lens with this angle,
leading to a defocus term that according to (\ref{eq:OPD-R-alphax}) is proportional to
$\alpha_y^2\,x_1^2$ and thus to $x_1^2 \,y_1^2$. One may expect that this could be
addressed by polishing 
the surface of the second lens to a cylinder of radius $f_2$ that curves in the $y$
direction so that the rays deflected by the first lens always impinge normal to the
surface (and hence seem to always be on axis).  Failing that, setting $R_2 = (2/3)f_2$ may
eliminate the curvature of field as suggested by (\ref{eq:OPD-R-alphax}).
The ray-trace analysis shows, however, that the
on-axis aberration 
remains unchanged for any choice of curvature, as given by the term in (\ref{eq:OPD4}).
The wavefront is dependent upon the tilts of the diffracting layers in the MLL, which
follow the cylindrical symmetry dictated by their deposition onto flat substrates. The
normals of the layers of the second lens are in the $x$-$z$ plane as described by $\vec{q}_2$ in
\tref{tab:crossed}.

The transmission of the lens system is also affected by the violation of the Bragg
condition in the second lens.  For the on-axis field point, the analysis leading to (\ref{eq:OPD4}) gives
the solution of the deviation parameter of the first lens as
$\epsilon_1(\rho_x,\rho_y) = 0$, showing that the Bragg condition is satisfied across the
entire pupil of the first lens, but in the second
lens
\begin{equation}
  \label{eq:epsilon-1D}
  \epsilon_2^{(4)}(\rho_x,\rho_y) =\rho_x^2 \rho_y^2/4.
\end{equation}
This equation suggests that the loss of transmission of the lens starts to occur at the corners of the
pupil ($\rho_x = \rho_y = \text{NA}$), even for the on-axis
field point, when $\text{NA}^4 > 4 w_\epsilon$, independent
of focal length.  For the
SiC/WC multilayer system considered above for a wavelength of \SI{0.075}{\nano\meter}
where $|\delta_1-\delta_2| = \num{6.7e-4}$,
(\ref{eq:epsilon-1D}) sets a limit of the NA of crossed 1D lenses to 0.064,
corresponding in this case to a resolution of \SI{0.59}{\nano\meter}. 

Equation~(\ref{eq:epsilon-1D}) confirms that the on-axis aberration stems from the fact that
the convergent beam incident on the second lens does not mirror the reference wavefront
that conceptually would produce the design of the second MLL as a computer-generated
hologram. An elimination of the on-axis aberration (and the apodisation) would thus require that the layers of
the second lens are fabricated to conform to the cylindrical symmetry introduced by the
first lens and the desired spherical convergent beam created by the second.  The use of a
phase plate, as suggested above, would address the aberration but not the lens
transmission.

\subsection{Relative alignment of the two lenses}
\label{sec:relative}
From the cylindrical symmetries of the two lenses, any transverse displacement of one
lens relative to the other either leaves the system invariant or simply shifts the optic axis,
defined by the $y$ coordinate of the first lens and the $x$ coordinate of the second lens
(see figure~\ref{fig:crossed-lenses}). A
tilt of one lens about either the $x$ or $y$ axes will cause an aberration of similar
magnitude as that from a field angle $\alpha_y$ or $\alpha_x$.  However, in that case
there no longer exists an axis of symmetry of the system, and instead one can consider the
coordinate system to be fixed to one of the lenses, such that a tilt of that lens is
equivalent to an opposite change in the field angle. Alignment tolerances are investigated
in \sref{sec:align} for lenses with \SI{1}{\nano\meter} resolution. 

As known from the analysis of crossed cylindrical lenses, an error in the orthogonality of
the two lenses, caused by a rotation of one relative to the other about the $z$ axis,
leads to oblique astigmatism with a wavefront error proportional to $\rho_x\,\rho_y$
\cite{Yan:2017}.  This can be seen in the simplifying approximation of the system
wavefront as the sum of the phase contributions of each lens, scaled to to the coordinates
of the second lens, given by $\lambda/(2\pi)\,\phi_1 = -y^2/(2f_2)$ and
$\lambda/(2\pi)\,\phi_2 = -x^2/(2f_2)$.  The combined wavefront of
$-f_2 (\rho_y^2 + \rho_x^2)/2$, represents (in the paraxial approximation) a spherical
wave converging onto the focus at $f_2$. However, if the second lens is rotated about
the optic axis by an angle $\chi_z$ then the second lens produces a phase proportional
to $-f_2(\rho_x \cos\chi_z + \rho_y \sin\chi_z)^2$. For small $\chi_z$ the wavefront error found
by subtracting the converging spherical wave is then equal to
$f_2 \,\chi_z\,\rho_x\,\rho_y$.  This is oblique (\ang{45}) astigmatism, and this
contribution is constant across the field. This term is orthogonal to the \ang{0}-\ang{90}
astigmatism and to defocus, and gives an RMS error of $f_2 \,\chi_z\, \text{NA}^2 /3$.
Given a tolerable degree of astigmatism of half a wave, the requirement on the alignment
of the two lenses is then $\chi_z < 3 \lambda/(2 f_2 \,\text{NA}^2)$.  For the example above
of $\lambda = \SI{0.075}{\nano\meter}$, $f_2=\SI{1}{\milli\meter}$, and
$\text{NA} = 0.0375$ we find a requirement of $\chi_z < \SI{80}{\micro\radian}$.  This
tolerance can in some cases be relaxed by finding the field location with an oblique
astigmatism that cancels this term due to the non-orthogonality.  As seen in
(\ref{eq:OPD4}), the coefficient of the field-dependent oblique astigmatism is
$-3(f_1-2f_2)\,\alpha_x\,\alpha_y$.  However, choosing the appropriate $\alpha_x$ and
$\alpha_y$ to zero the oblique astigmatism will introduce other off-axis aberrations.

\subsection{Chromatic aberrations of crossed MLLs}
\label{sec:chromatic-crossed}
Due to the anamorphism of crossed MLLs, and the fact that focal length scales inversely
with $\lambda_m$, a change in wavelength causes a different amount of defocus in each
focused direction.  That is, besides defocus, crossed MLLs exhibit a chromatic astigmatism
that is proportional to the difference in focal lengths $f_1-f_2$.  The magnitude of
off-axis aberrations (including distortion) will also depend on the field direction.  For
the on-axis field point, the chromatic defocus and
astigmatism can be compensated by repositioning the lenses so that their distances to the
image plane are $f_1\,\lambda/\lambda_m$ and $f_2\,\lambda/\lambda_m$ (respectively for
the first and second lenses).  As in \sref{sec:chromatic}, the ray tracing procedure
can be carried out for the repositioned lenses for $\lambda \neq \lambda_m$, in which case
the on-axis wavefront aberration is evaluated to 
\begin{eqnarray}
  \label{eq:OPD4-SA}
      \text{OPD}_\text{SA}(\rho;\lambda_m) =&
  \frac{3}{4}f_2\frac{\lambda_m^3}{\lambda^3}\,\rho_x^2\,\rho_y^2 + \frac{3 f_1}{8}
 \frac{\lambda_m\, \Delta \lambda}{\lambda^2}\left( 2 -\frac{2f_1}{R_1}+\frac{\Delta
     \lambda}{\lambda}\right)\rho_y^4 \nonumber\\
  & + \frac{3 f_2}{8}
 \frac{\lambda_m\, \Delta \lambda}{\lambda^2}\left( 2 -\frac{2f_2}{R_2}+\frac{\Delta \lambda}{\lambda}\right)\rho_x^4.
\end{eqnarray}
The coefficients of the $\rho_y^4$ and $\rho_x^4$ terms of (\ref{eq:OPD4-SA}) are
equivalent to the spherical aberration term for the axisymmetric lens of
(\ref{eq:OPD-SA}). These terms represent a distorted spherical aberration, which can be
zeroed independently in each direction, for a given $\lambda_m$, with the appropriate
radius of curvature $R_1$ and $R_2$.  As for the axisymmetric lens, setting $R_1 = f_1$
and $R_2=f_2$ gives a quadratic dependence of the spherical aberration on the wavelength
deviation $\Delta \lambda$ and therefore the lowest chromatic
aberration for positive and negative wavelength deviations from the design wavelength.
However, even though these terms can be minimised, the oblique spherical aberration
(proportional to $\rho_x^2\,\rho_y^2$) scales with $(\lambda_m/\lambda)^3 = (1 +\Delta
\lambda/\lambda)^3$ and therefore dominates. 

\subsection{Ray tracing of crossed MLLs for 1\,nm focusing}
\label{sec:1nm}

\begin{figure}[t]
  \centering
  \setlength{\unitlength}{\linewidth}
  \begin{picture}(1,1.15)(0,0)
    \put(0.1,0.5){
      \includegraphics[width=0.8\linewidth]{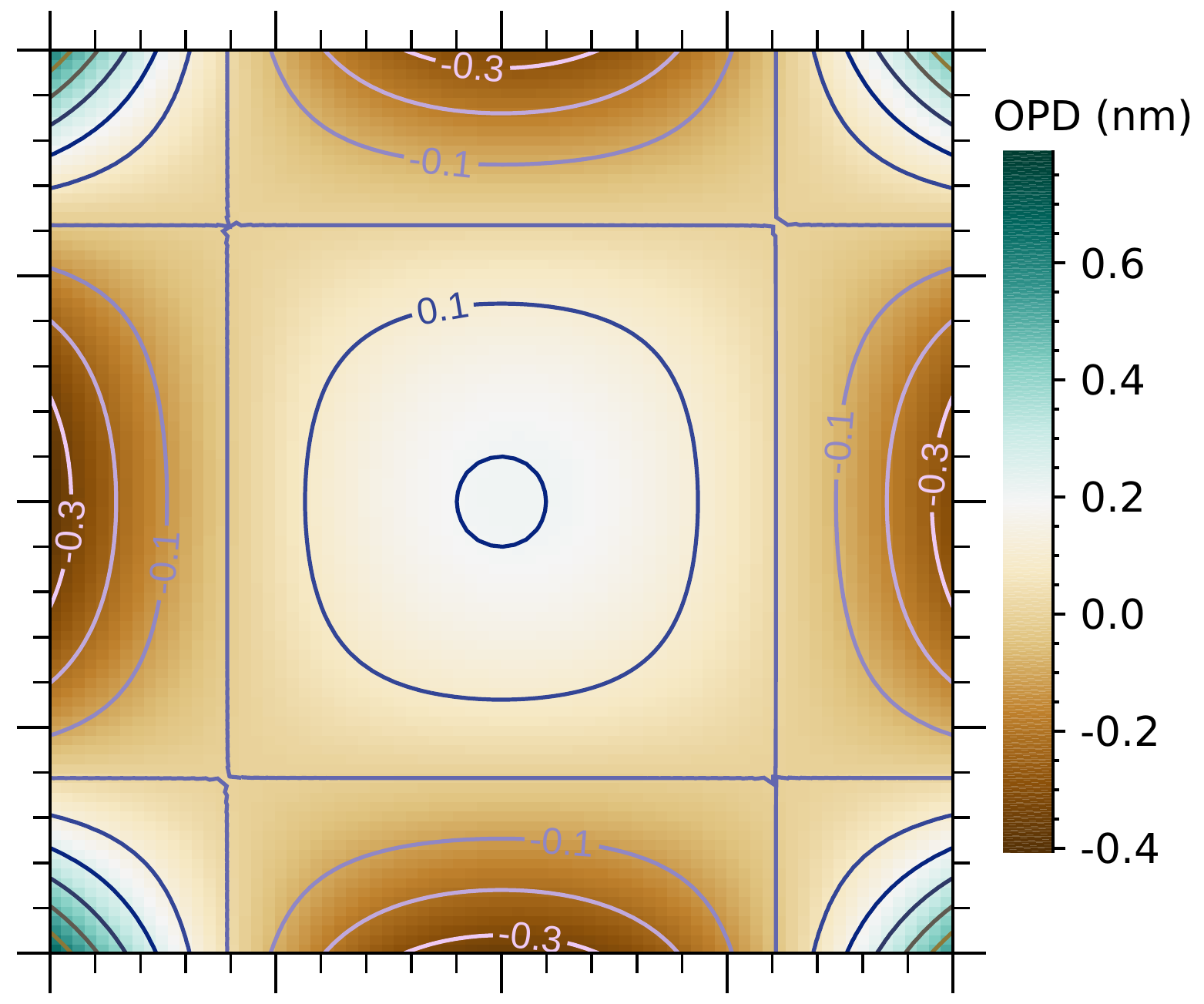}}
    \put(0,0){
      \includegraphics[height=0.49\linewidth]{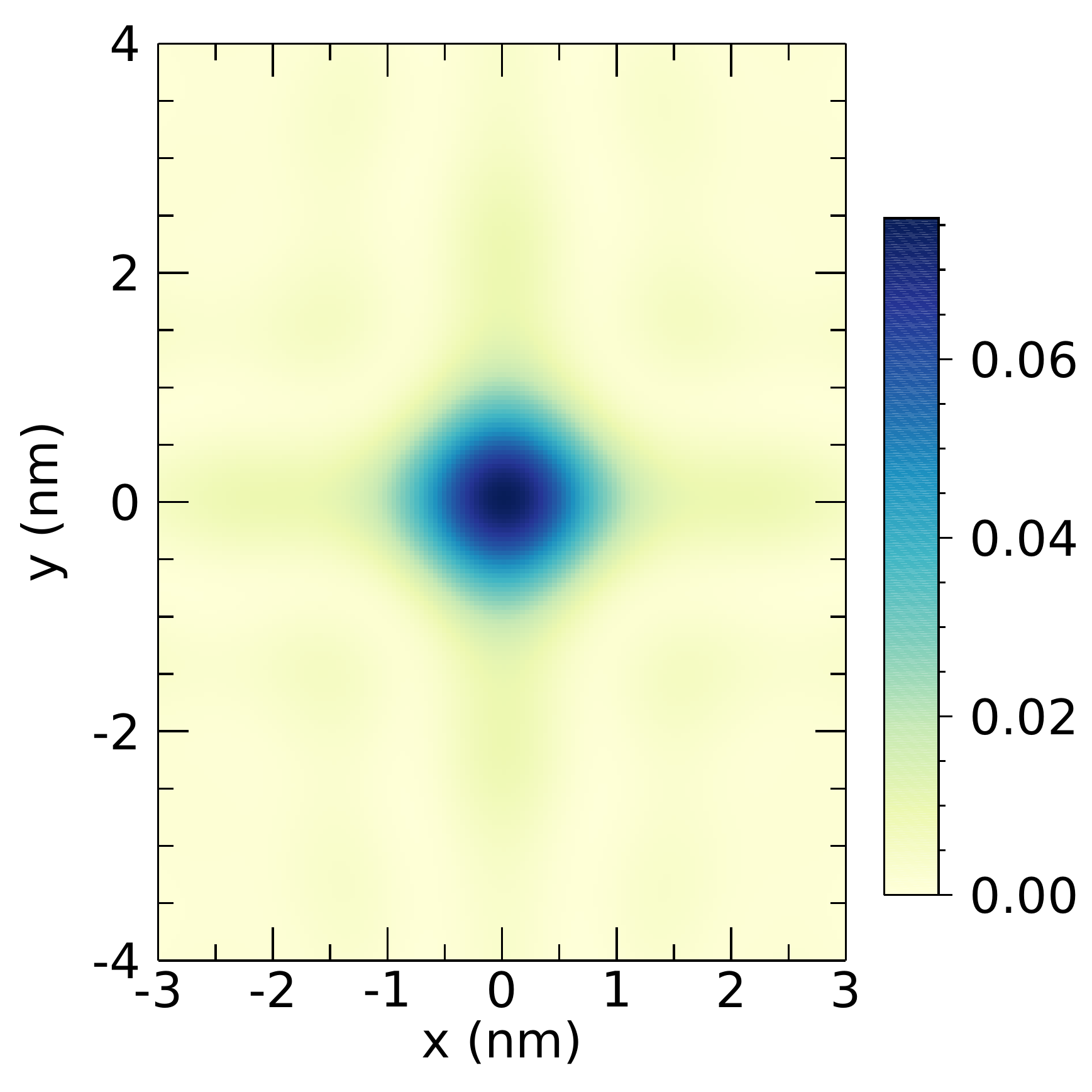}}
    \put(0.5,0){
      \includegraphics[height=0.49\linewidth]{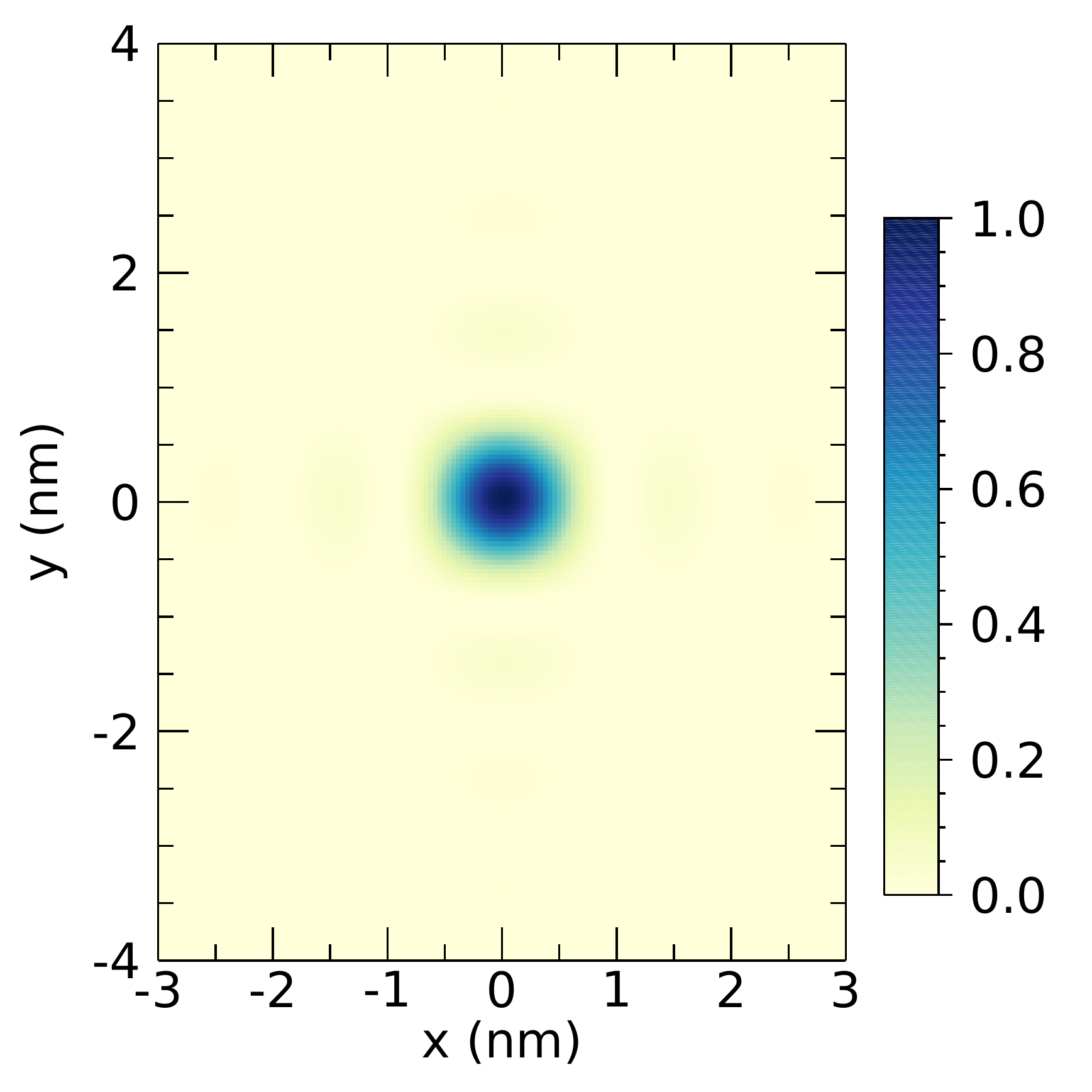}}
    \put(0.07,1.12){\fontfamily{phv}\normalsize\selectfont\text{(a)}}
    \put(0.0,0.45){\fontfamily{phv}\normalsize\selectfont\text{(b)}}    
    \put(0.5,0.45){\fontfamily{phv}\normalsize\selectfont\text{(c)}}
  \end{picture}
  \caption[Aberration maps]{(a) Map of the wavefront aberration for the on-axis field
    point for a system of crossed flat 1D MLLs of focal lengths of \SI{1.25}{\milli\meter} and
    \SI{1}{\milli\meter}, both with $\text{NA}= 0.0375$. The best-fit defocus
    has been subtracted. (b) The point spread function intensity computed from the wavefront
    aberration for the on-axis field and a wavelength of \SI{0.075}{\nano\meter}, and (c)
    the point spread function at the same wavelength and NA but zero wavefront aberration.}
  \label{fig:aberration-map}
\end{figure}
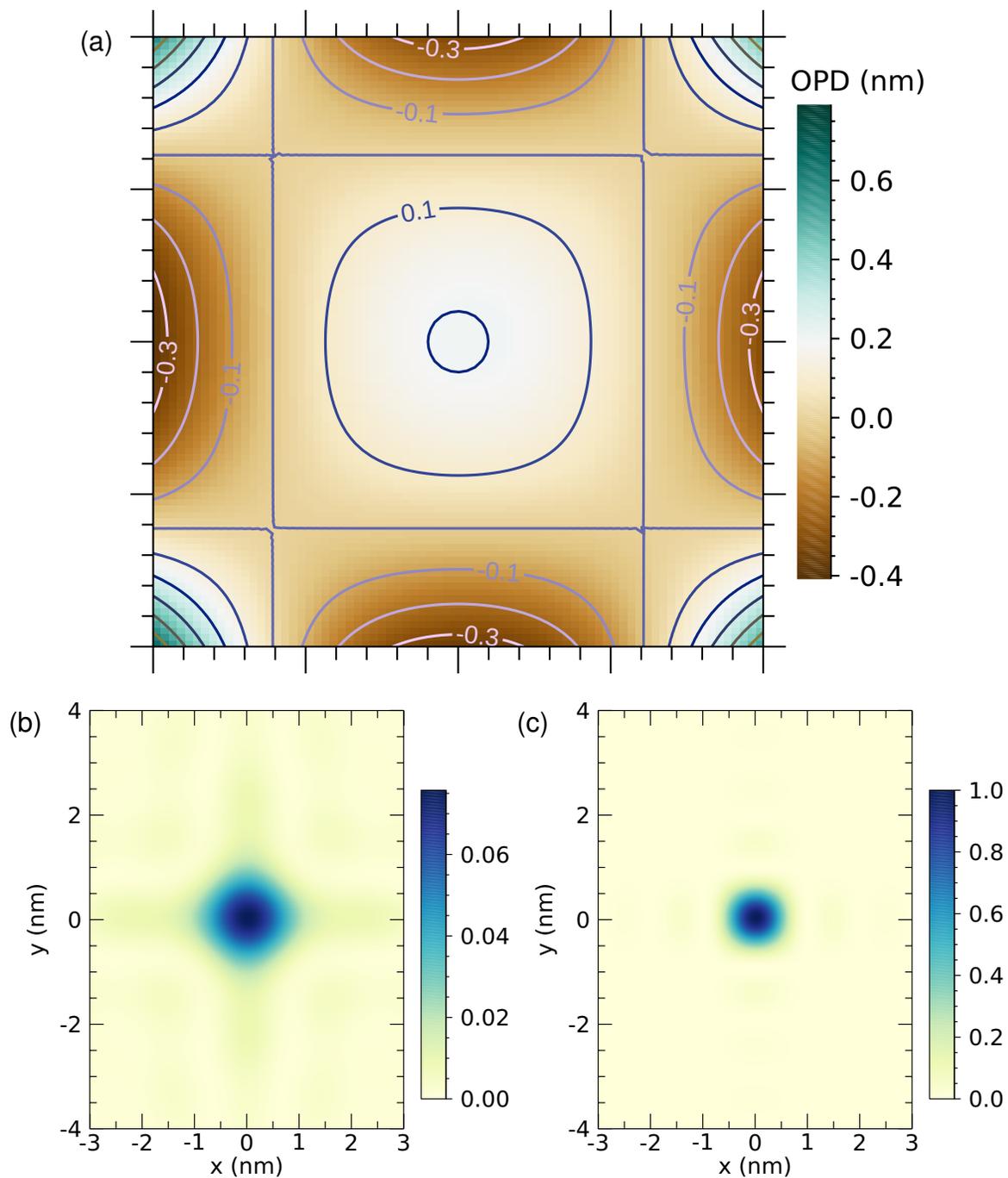

As an example, we consider the above-mentioned lens system with a resolution of
\SI{1}{\nano\meter} at a wavelength $\lambda = \SI{0.075}{\nano\meter}$ (\SI{17}{\kilo\eV}
photon energy) with a square exit pupil of $\text{NA}=0.0375$ in the $x$ and $y$
directions, (and $\text{NA} = 0.053$ along the diagonal). Here we define the resolution for a square-pupil optic to be
$0.5 \lambda / \text{NA}$ in the $x$ and $y$ directions. The focal lengths of the lenses
are taken to be
$f_1 = \SI{1.25}{\milli\meter}$ and $f_2 = \SI{1}{\milli\meter}$, and thus the heights of
the lenses are \SI{93}{\micro\meter} and \SI{75}{\micro\meter}, respectively.  The
ray-tracing procedure detailed above was used to numerically compute the wavefront
aberrations exactly---that is, without approximating to a fourth-order series expansion as
for (\ref{eq:OPD4}).  The particular procedure applies to the case of field points at
infinity, displaced from the optical axis by field angles $\alpha_x$ and $\alpha_y$ as set
by the ray direction $\hat{\vec{r}}_1$. This situation represents, for example, the
formation of a focused beam for a scanning microscope.  In this case the field angles also
represent the tilt of the optic axis of the lenses relative to say a ``beam axis'' set by
the direction of a far-off source.  The calculations also correspond to the aberrations
expected in full-field imaging at high magnification (where the image plane is far from
the lenses and the sample is near to the focal plane).

The aberrations are computed in terms of the points of intersection of rays on the two
lenses.  A map of the OPD as a function of the coordinates of the second lens is shown in
figure~\ref{fig:aberration-map} (a), for flat lenses ($R_1 = R_2 = \infty$).  The extent of
the plot is limited to the same numerical aperture of \num{0.0375} in the two orthogonal
directions.  The best-fit defocus has been subtracted and this map represents the
phase error of the exit wave converging onto the plane of best focus.  This map is
dominated by the fourth-order term $\rho_x^2\,\rho_y^2$, and the residual after
subtracting that term from the calculated OPD is proportional to higher order terms with
the same symmetry as displayed in figure~\ref{fig:aberration-map} (a).  The RMS wavefront
error of the OPD shown in figure~\ref{fig:aberration-map} (a) is \SI{0.134}{\nano\meter},
compared with the contribution of the fourth-order term
$0.08 f_2 \,\text{NA}^4 = \SI{0.158}{\nano\meter}$.  As previously, the point spread function (PSF) can
be computed from a Fourier transform of the complex-valued pupil function of the lens
system, constructed from the OPD and lens transmission.  The PSF for the on-axis
field point of the lens system is shown in figure~\ref{fig:aberration-map} (b) for a
wavelength of \SI{0.075}{\nano\meter}, compared with the PSF for a perfect system of the
same NA and wavelength in figure~\ref{fig:aberration-map} (c).  In addition to the increased
width of the PSF with a full-width at half maximum of \SI{1.64}{\nano\meter}, compared
with \SI{0.85}{\nano\meter} for the perfect system, the Strehl ratio of the PSF is 0.076.

\begin{figure*}
  \centering
  \setlength{\unitlength}{\linewidth}
  \begin{picture}(1.,0.35)(0,0)
    \put(0,0){\makebox(1,0.35){
        \includegraphics[height=0.3\linewidth]{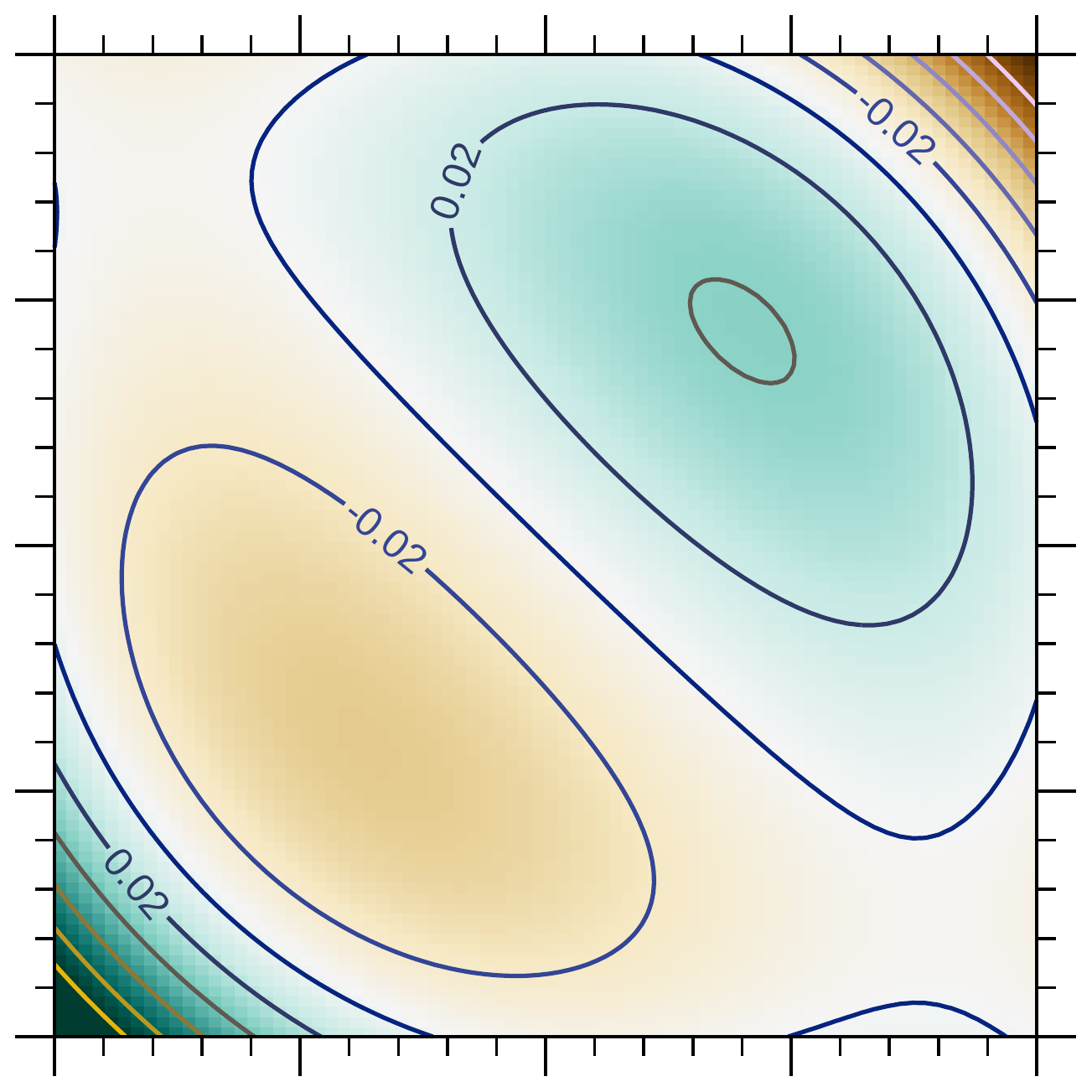}
        \includegraphics[height=0.3\linewidth]{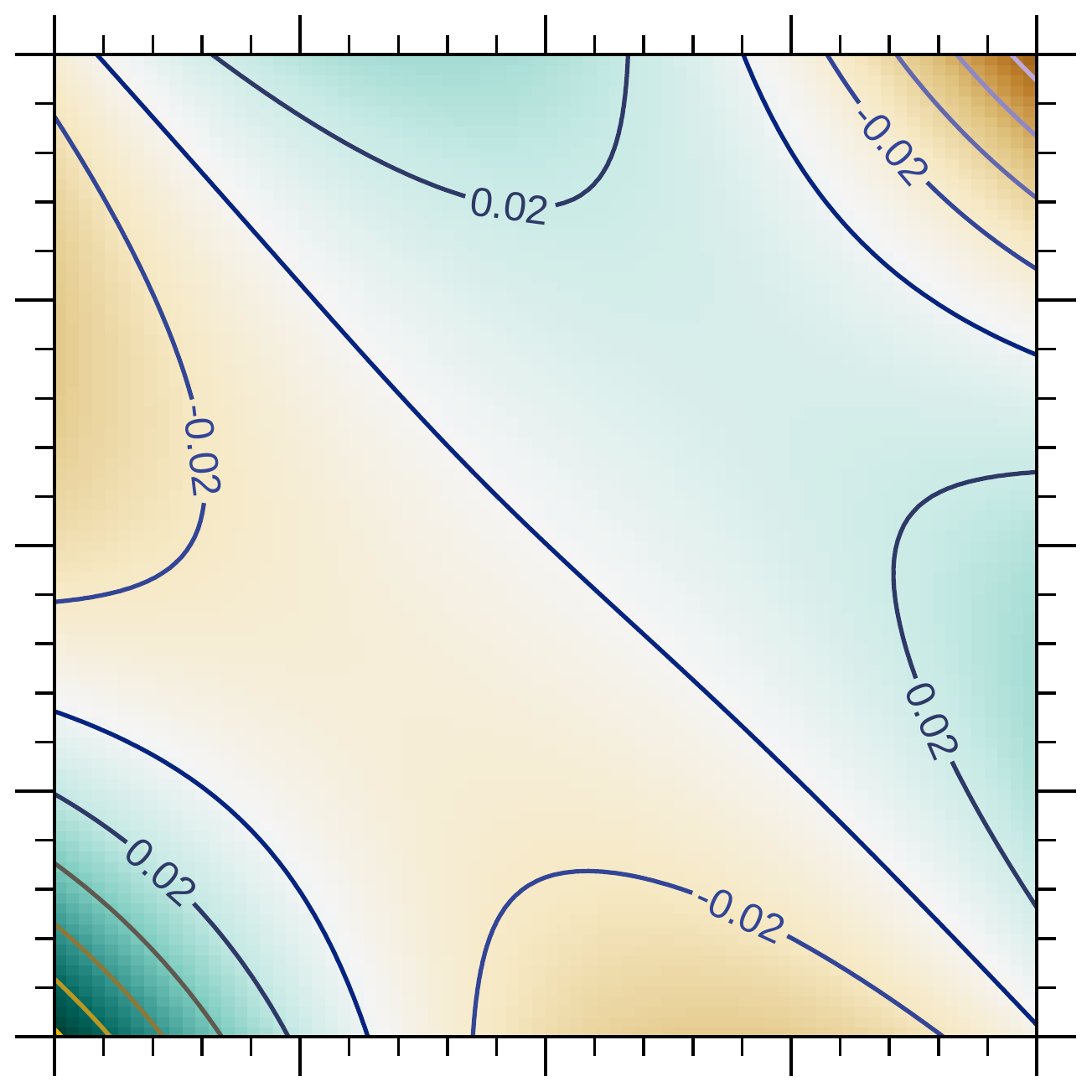} 
        \includegraphics[height=0.3\linewidth]{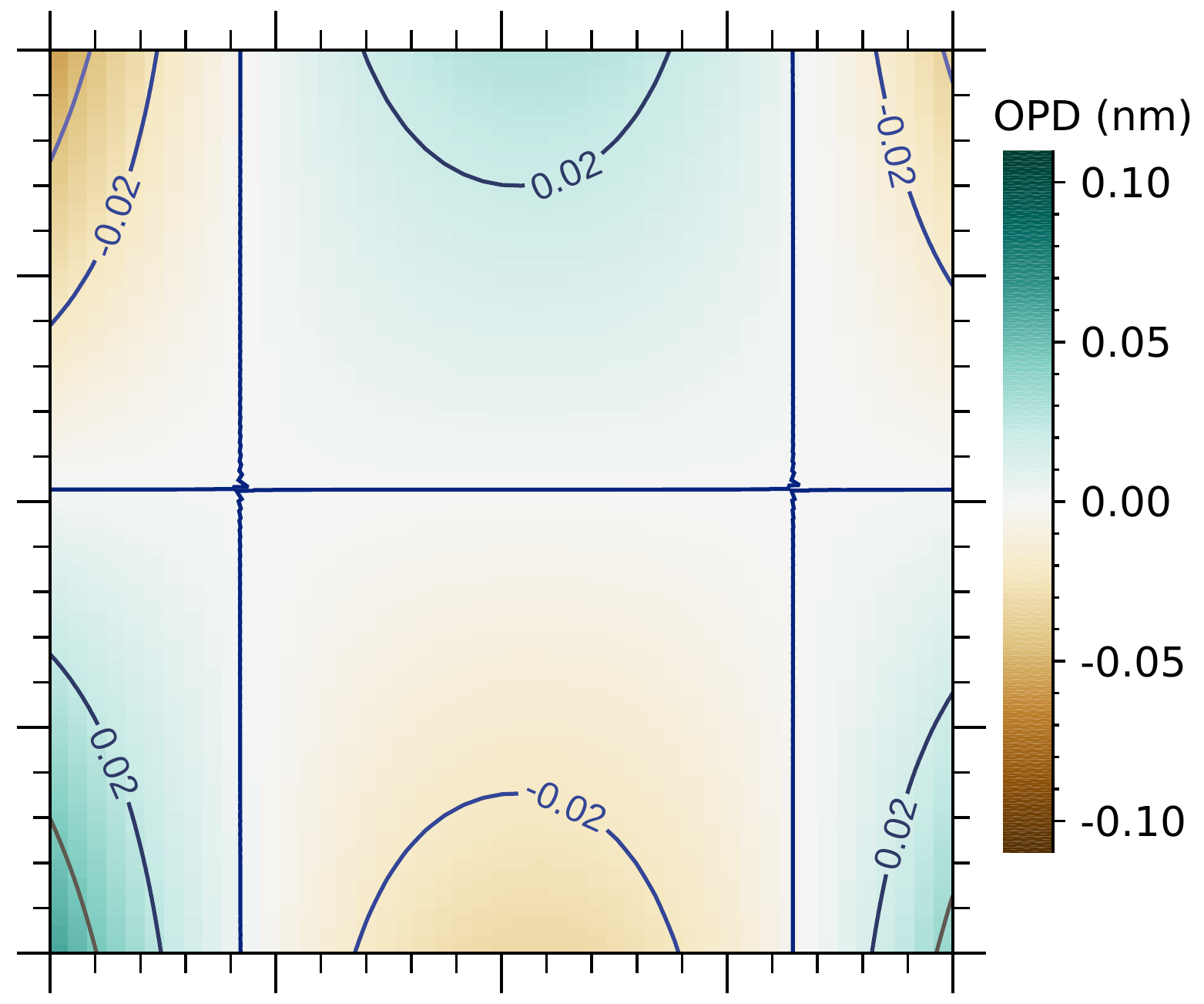}}}
    \put(0.04,0.33){\fontfamily{phv}\normalsize\selectfont\text{(a)}}
    \put(0.35,0.33){\fontfamily{phv}\normalsize\selectfont\text{(b)}}
    \put(0.66,0.33){\fontfamily{phv}\normalsize\selectfont\text{(c)}}
  \end{picture}
  \caption[Off-axis aberration maps]{Wavefront maps of systems of crossed 1D MLLs that are
    phase-compensated for the on-axis field point, calculated over the square pupil (set
    to be at the second lens) for the field point located
    at angles $\alpha_x = \SI{1.25}{\milli\radian}$ and
    $\alpha_y = \SI{1}{\milli\radian}$, for (a) flat MLLs, (b) MLLs with radii of
    curvature equal to the focal lengths ($R_1 = f_1$ and $R_2=f_2$), and (c) as for (b)
    but with $f_1 = 2 f_2$.  In all cases $f_2 = \SI{1}{\milli\meter}$.  In (a) and (b)
    $f_1=\SI{1.25}{\milli\meter}$, while in (c) $f_2 = \SI{2}{\milli\meter}$. In all cases the best-fit defocus
    and \ang{0}-\ang{90} astigmatism were subtracted.}
  \label{fig:aberration-map-offaxis}
\end{figure*}

The
calculated OPD for the on-axis field point far exceeds both the Marechal condition of the
RMS error being less than $\lambda/14$ or the Rayleigh condition of an absolute deviation
of the OPD by more than $\lambda/4$.  Further
computations show that the OPD at field angles less than about \SI{1}{\milli\radian}
appear similar to the on-axis aberration map, so we would expect to achieve aplanatic imaging
over a field of view of about this extent by compensating the wavefront equally at all
field points.  Such a compensation could be achieved by locating a phase plate close to
either of the lenses.

The aberrations of such a phase-compensated system were simulated simply by subtracting
the OPD computed for $\alpha_x = \alpha_y = 0$ from wavefront maps at other field
angles. This should be a valid assumption when the phase plate is close enough to the
second lens that its correction is the same for all field angles.  The wavefront map for
the field located at $\alpha_x = \SI{1.25}{\milli\radian}$ and
$\alpha_y = \SI{1}{\milli\radian}$ is given in figure~\ref{fig:aberration-map-offaxis} (a).
The RMS error is \SI{0.024}{\nano\meter}, which is certainly an improvement over the
non-compensated lens system, but still does not meet the Marechal condition.  It is
apparent from this figure that the wavefront is dominated by coma oriented along the same
direction as the field point ($\psi = \ang{45}$, accounting for the anamorphism).  That
is, the wavefront is proportional to $\rho_x^3 + \rho_y^3$.  These are the terms in the
second line of (\ref{eq:OPD4}), which can be eliminated by setting the radii of
curvature of the lenses equal to the focal lengths.  Figure~\ref{fig:aberration-map-offaxis}
(b) shows the calculated wavefront for that case. The wavefront aberration is
reduced slightly (especially along the $\rho_y = -\rho_x$ diagonal), and the RMS error is
also slightly improved with a value of \SI{0.019}{\nano\meter}. The aberration is still
dominated by coma, now given by the terms on the third line of (\ref{eq:OPD4}).
Equation~(\ref{eq:OPD4}) suggests one more optimisation to reduce the wavefront aberration,
which is to set the focal length of the second lens to half that of the first,
$f_2=f_1/2$, to eliminate one of the remaining two coma-like terms.  With this, and
cylindrical curvatures of the two lenses equal to the focal lengths, we find the lowest
RMS error of \SI{0.014}{\nano\meter} and the map of the wavefront given by
figure~\ref{fig:aberration-map-offaxis} (c). Here there is a remaining coma-like term that
only depends on the position of the field in the $y$ direction, which is the focusing
direction of the first lens.

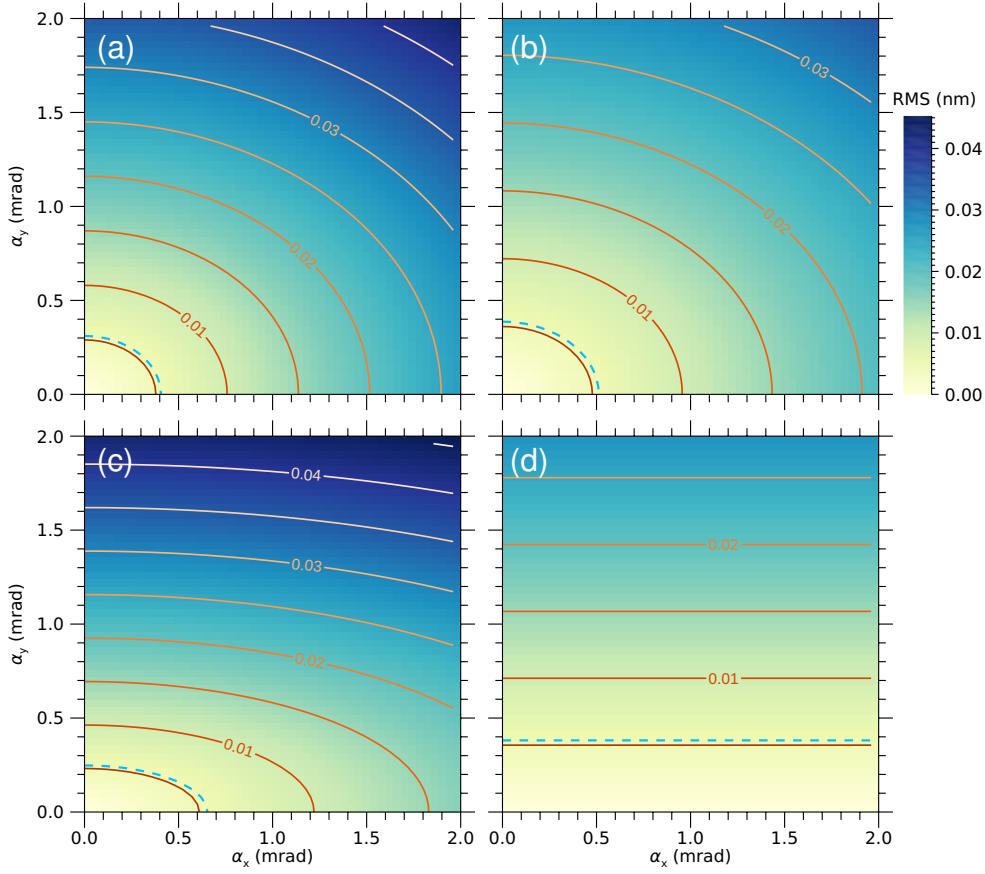
\begin{figure*}
  \centering
  \setlength{\unitlength}{\linewidth}
  \begin{picture}(0.85,0.8)(0,0)
    \put(0,0){
      \includegraphics[width=0.85\linewidth]{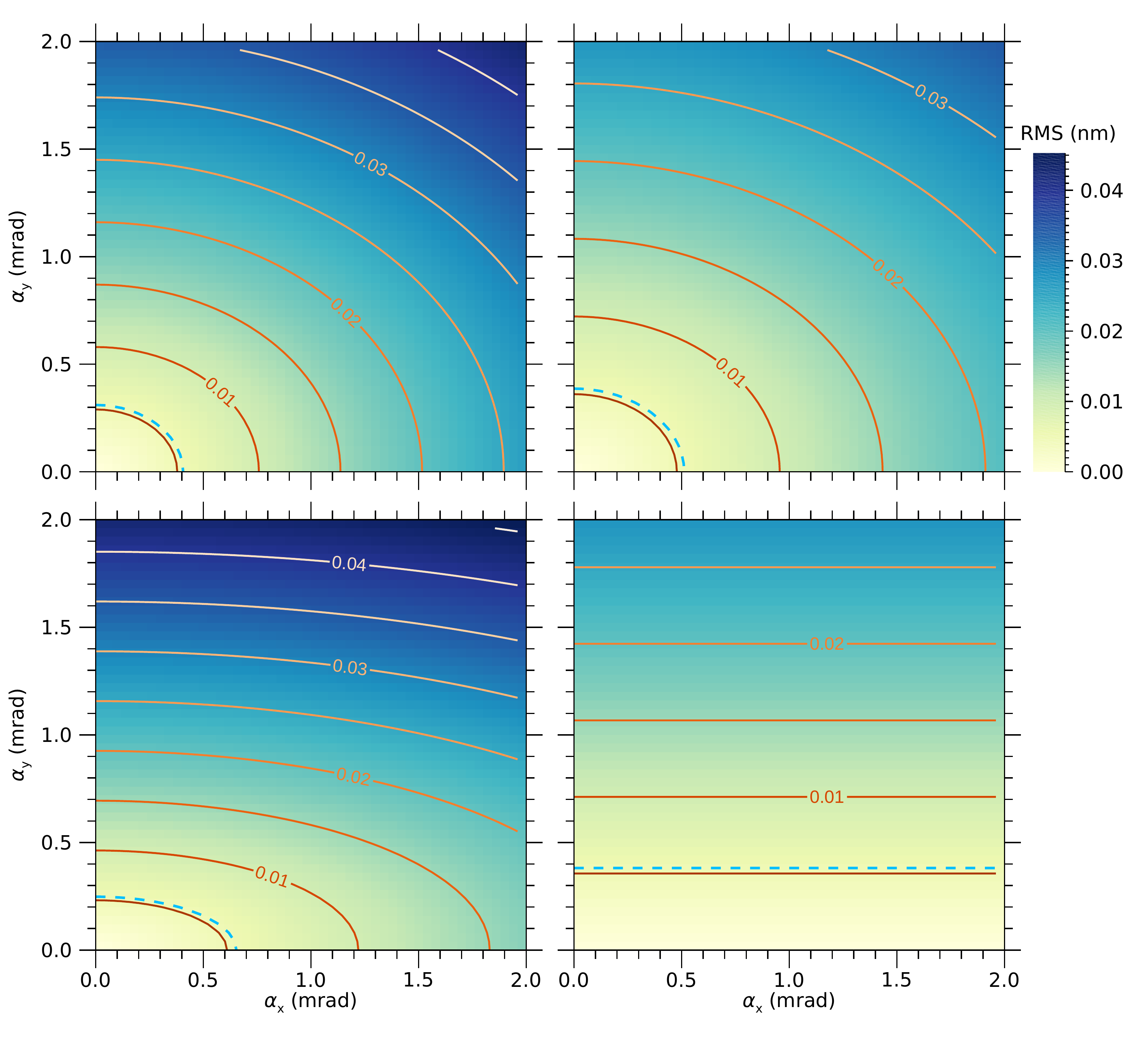}}
    \put(0.09,0.71){\fontfamily{phv}\normalsize\selectfont\color{white}\text{(a)}}
    \put(0.44,0.71){\fontfamily{phv}\normalsize\selectfont\color{white}\text{(b)}}
    \put(0.09,0.36){\fontfamily{phv}\normalsize\selectfont\color{white}\text{(c)}}
    \put(0.44,0.36){\fontfamily{phv}\normalsize\selectfont\color{white}\text{(d)}}
  \end{picture}
  \caption[Aberration maps]{Maps of the RMS wavefront aberration as a function of the
    field angles $\alpha_x$ and $\alpha_y$ for crossed MLLs with $\text{NA}=0.0375$,  a
    phase plate set to conjugate the aberration of the on-axis field point, and the focal
    length of the second lens $f_2 = \SI{1}{\milli\meter}$. (a)
    $f_1 = \SI{1.25}{\milli\meter}$, $R_1=R_2=\infty$; (b)
    $f_1 = \SI{1.25}{\milli\meter}$, $R_1 = f_1$, and
    $R_2=f_2$; (c) $f_1=2 f_2$, $R_1=R_2=\infty$; and (d)
    $f_1=2 f_2 = \SI{2}{\milli\meter}$, $R_1 = f_1$, and $R_2=f_2$. The blue
    dashed lines indicate the Marechal condition of $\lambda/14$ for
    $\lambda=\SI{0.075}{\nano\meter}$. These lines enclose the isoplanatic areas of the
    lens systems.}
  \label{fig:field-maps}
\end{figure*}

 Given a linear dependence on the field angle for coma, achieving the Marechal condition of
\SI{0.005}{\nano\meter} RMS aberration error therefore requires reducing the magnitude of the field angle to
about a quarter to a third of the value simulated in figure~\ref{fig:aberration-map-offaxis}
and thus we expect the aplanatic region to have a radius of about
\SI{0.3}{\milli\radian}.  Plots of the RMS error as a function of the field position
$(\alpha_x,\alpha_y)$ are given in figure~\ref{fig:field-maps} for the same lens
configurations as for figure~\ref{fig:aberration-map-offaxis}, with the additional case of
flat lenses with $f_1=2 f_2$.  The aplanatic area is indicated by a dashed line, which
gives the contour for an RMS of $\lambda/14 \approx \SI{0.005}{\nano\meter}$.  Comparing
Figs.~\ref{fig:field-maps} (a) and (b), it is seen that setting the radii of the lenses
to the focal lengths only provides a slight enlargement of the aplanatic area over the
case of flat lenses.  The RMS error increases linearly with the magnitude of the field
angle in these cases, indicating coma in accordance with the terms of (\ref{eq:OPD4}).
The largest aplanatic field (and hence the largest tolerance for misalignment of lenses)
is achieved by setting the focal length of the second lens to half that of the first, as
in figure~\ref{fig:field-maps} (c) and additionally curving each lens to a radius equal to
the focal length as depicted in Fig ~\ref{fig:field-maps} (d).

\begin{figure}
  \centering
  \includegraphics[width=0.95\linewidth]{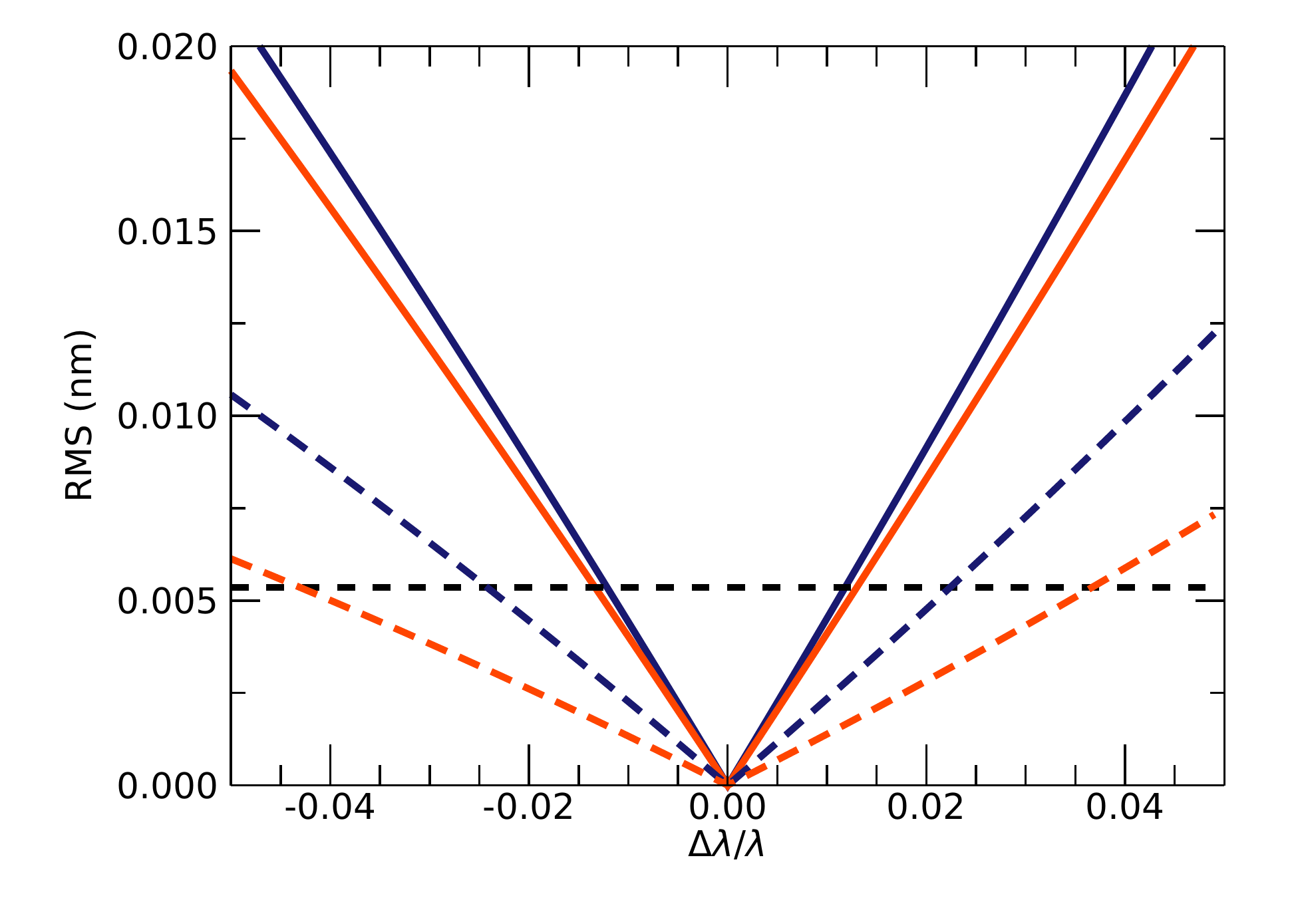}
  \caption{RMS wavefront aberration (excluding defocus and astigmatism) of crossed MLLs of
    $\text{NA} = 0.0375$, $f_1 = \SI{1.25}{\milli\meter}$ and $f_2 = \SI{1}{\milli\meter}$
    as a function of the relative deviation of the wavelength from the design for curved
    lenses ($R_1 = f_1$ and $R_2=f_2$) (orange) and flat lenses ($R_1 = \infty$ and
    $R_2 = \infty$) (blue).  The solid lines assume a corrector with a phase profile that
    is independent of wavelength and the dashed lines are for a phase corrector made of carbon.
    The Marechal condition of $\lambda/14$ is indicated by the black dashed line.}
  \label{fig:chromatic-crossed}
\end{figure}

Plots of the RMS wavefront error as a function of the relative wavelength deviation
$\Delta\lambda / \lambda = \lambda_m/\lambda-1$ for the on-axis field point are given in
figure~\ref{fig:chromatic-crossed}, for the cases of curved lenses ($R_1 = f_1$ and
$R_2=f_2$) and flat lenses ($R_1 = \infty$ and $R_2 = \infty$), as computed by ray tracing.  It is assumed in these
calculations that both lenses are refocused for each wavelength change.  As compared with
the aberrations for the axisymmetric lens, plotted in figure~\ref{fig:wavefronts} (d), the
RMS wavefront error is larger for the crossed lenses. It is found that the dominant
aberration is proportional to $\rho_x^2\,\rho_y^2$, as expected from
(\ref{eq:OPD4-SA}).   The calculations of the RMS errors given by the solid lines in
figure~\ref{fig:chromatic-crossed} furthermore assume that this error is corrected at
$\Delta\lambda = 0$, for example by a phase plate, but that the correction is independent
of wavelength.  However, the refractive index decrement $\delta$ of a phase plate constructed from a material
of low atomic number such as carbon (diamond) varies as $\lambda^2$ at wavelengths
comparable to the design wavelength of \SI{0.075}{\nano\meter}.  In this case, given that
the phase of the corrector is much less than unity, the correction scales as
$(\lambda_m/\lambda)^2$.  This somewhat compensates for the $(\lambda_m/\lambda)^3$
dependence of the $\rho_x^2\,\rho_y^2$ aberration of the crossed lenses, giving lower RMS
errors plotted as dashed lines in figure~\ref{fig:chromatic-crossed}.

\begin{figure}
  \centering
  \setlength{\unitlength}{\linewidth}
 \begin{picture}(1,1.35)(0,0)
    \put(0,0.66){
      \includegraphics[width=\linewidth]{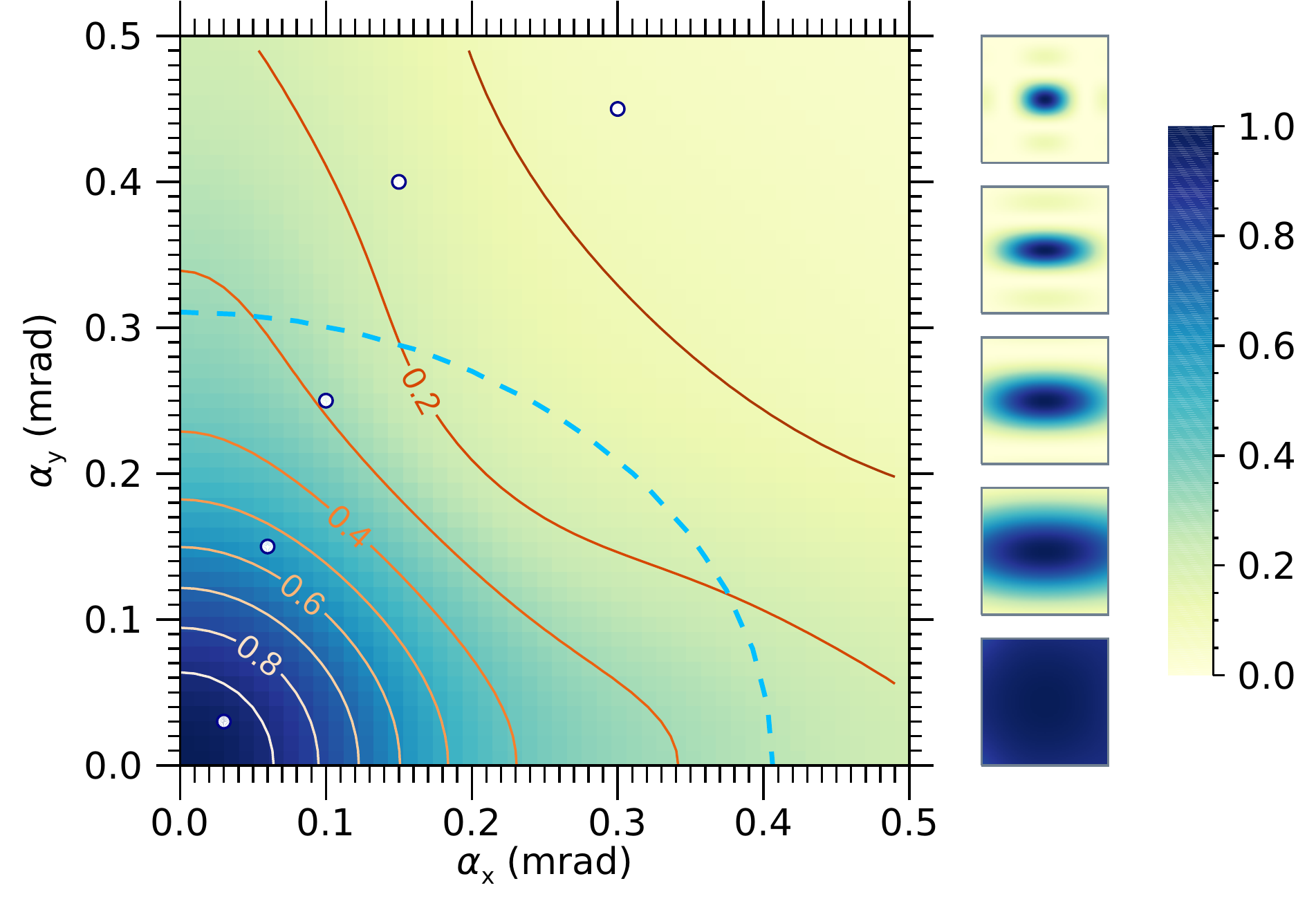}}
    \put(0,0){
      \includegraphics[width=0.95\linewidth]{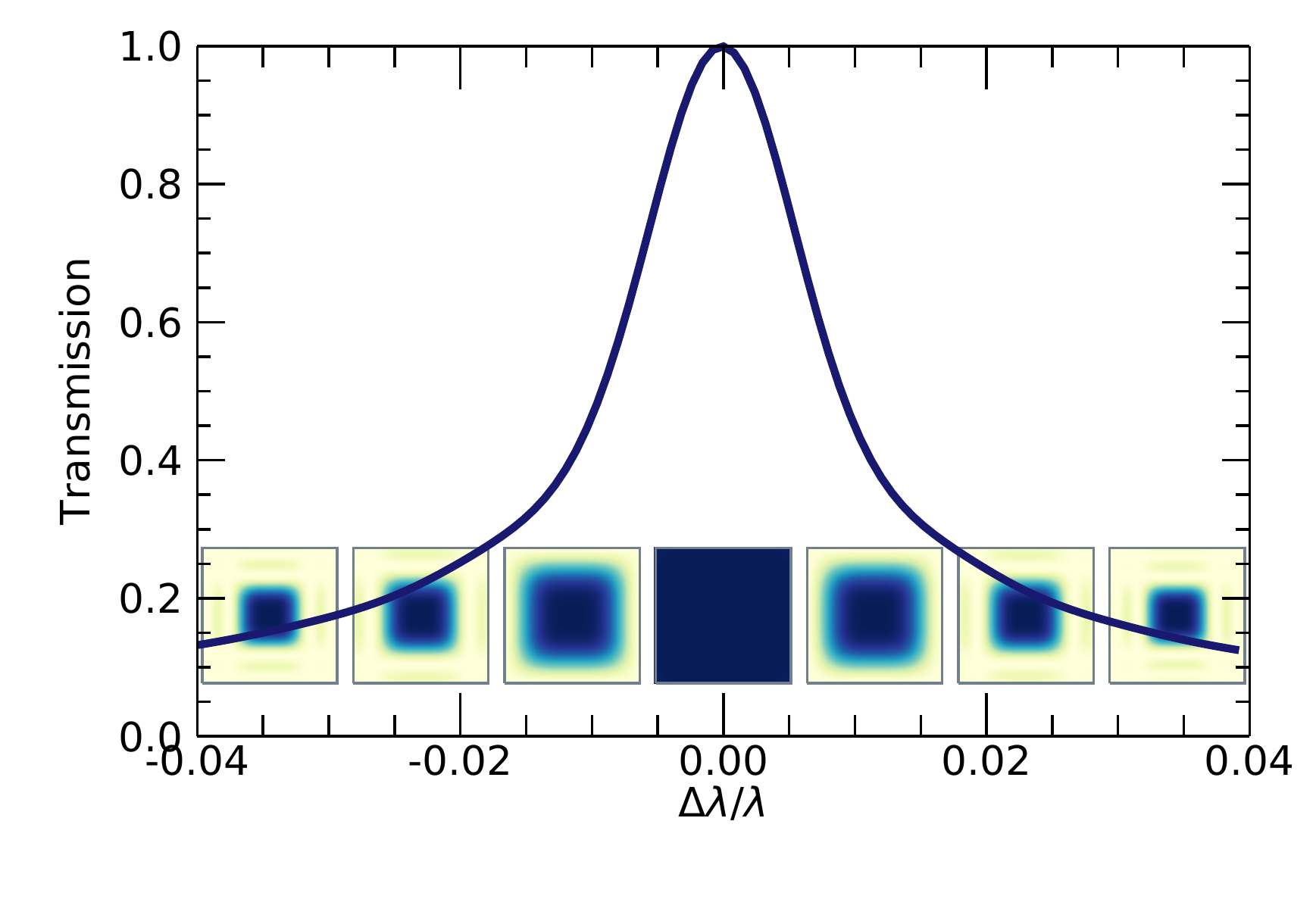}}
    \put(0.0,1.31){\fontfamily{phv}\normalsize\selectfont\text{(a)}}
    \put(0.0,0.62){\fontfamily{phv}\normalsize\selectfont\text{(b)}}
  \end{picture}
  \caption{Transmission of crossed MLLs with $\text{NA} = 0.0375$ and focal lengths $f_1 =
  \SI{1.25}{\milli\meter}$ and $f_2=\SI{1}{\milli\meter}$ as function of (a) the field
  position and (b) the relative change in wavelength. Transmission maps of the square lens-pair
  pupil are shown in (a) for the positions indicated by the white circles.  The Marechal
  condition for flat lenses and a wavelength of \SI{0.075}{\nano\meter} is indicated by the dashed sky-blue line in (a).}
  \label{fig:crossed-trans}
\end{figure}

As with the case for the axisymmetric lens, the Darwin width of the Bragg reflections of
these particular crossed lenses limits the field angles more stringently than the wavefront
aberrations.  A map of the total transmission of the lens pair is given in
figure~\ref{fig:crossed-trans} (a) as a function of field angle, for the same multilayer
properties as in the calculations of figure~\ref{fig:transmission}: $|\delta_1-\delta_2| =
\num{6.7e-6}$. For $\lambda = \SI{0.075}{\nano\meter}$ the half width $w_\theta =
\SI{0.11}{\milli\radian}$, as for the axisymmetric lens.  The transmission is found to be
independent of the radii of the lenses, as expected from (\ref{eq:epsilon-1D}).  Likewise, the apodisation caused by detuning the
wavelength has a similar behaviour to that found for the axisymmetric lens.  As seen in 
figure~\ref{fig:crossed-trans} (b), a deviation of the wavelength causes the effective
numerical aperture of the lens to decrease, and a transmission of at least \num{0.5} requires
$\Delta\lambda/\lambda < \SI{0.9}{\percent}$.

\begin{figure*}
  \centering
  \includegraphics[width=0.66\linewidth]{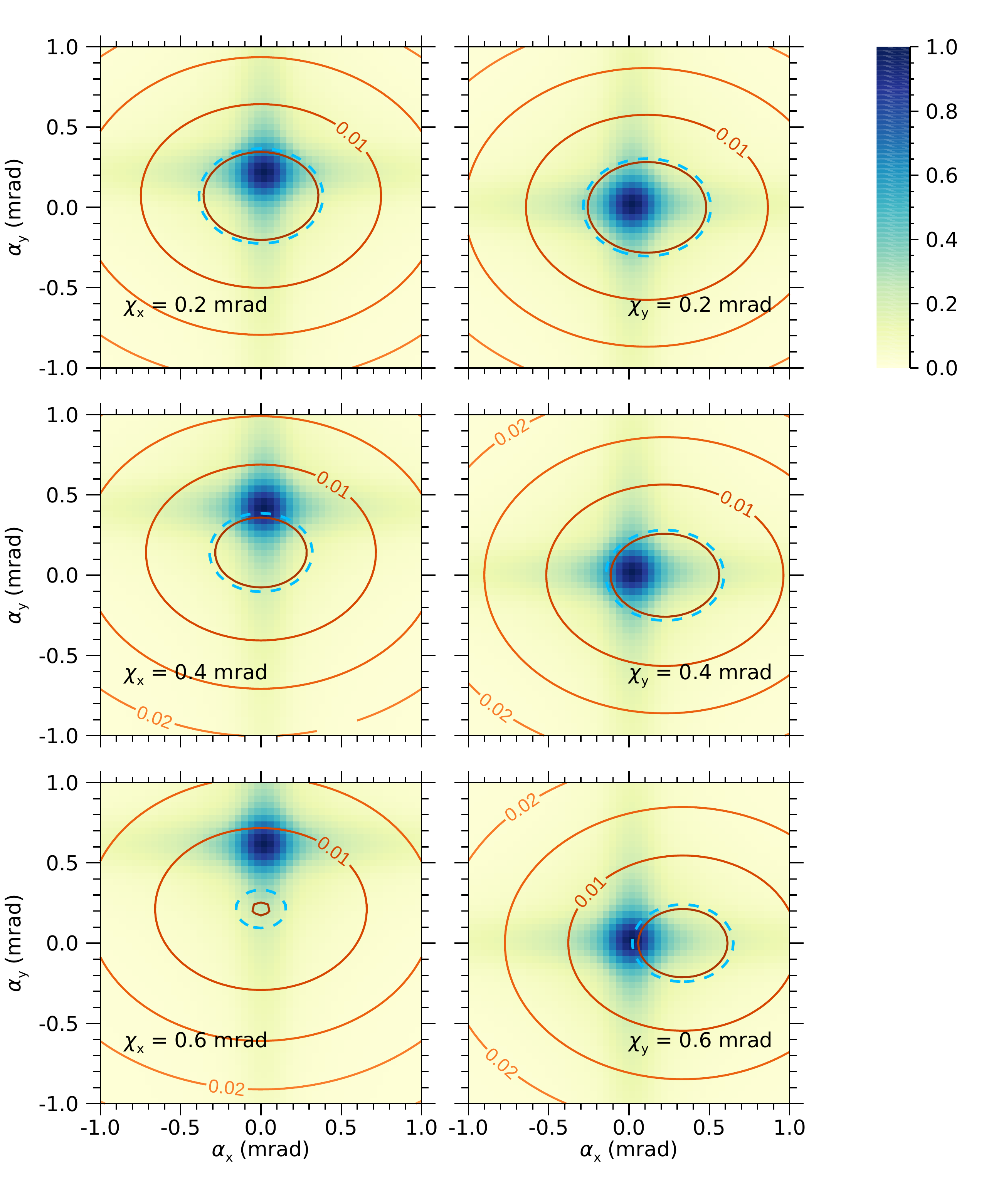}
  \caption{Field maps (in false colour) of the transmission of a system of  misaligned crossed flat MLLs of
    $\text{NA} = 0.0375$ and focal lengths $f_1 = \SI{1.25}{\milli\meter}$ and
    $f_2=\SI{1}{\milli\meter}$, for various rotations $\chi_x$ and $\chi_y$ of the first
    lens relative to the second. Contour maps of the RMS aberrations are superimposed,
    with the Marechal condition indicated by the dashed sky-blue line. }
  \label{fig:crossed-align}
\end{figure*}

\subsection{Relative alignment of crossed MLLs for  1\,nm focusing}
\label{sec:align}
The two lenses must be oriented with respect to each other by angles that are comparable
to the allowable field angles.  The tolerance can be explored by considering one lens to
be fixed (which sets the coordinate system) and setting the tilt of the other lens and the
field coordinates as free parameters.  Ignoring the orthogonality of the lenses (which was
discussed in \sref{sec:relative}) this four-dimensional space is illustrated in
figure~\ref{fig:crossed-align}. Here ray tracing calculations are presented for two flat
lenses ($R_1 = R_2 = \infty$). The second lens was held fixed and the first tilted by
$\chi_x$ about an axis passing through its vertex parallel to the $x$ axis or by an angle
$\chi_y$ about the $y$ axis.  The tilt by $\chi_x$ is given in the left-hand column of the
figure. This tips the lens in the direction that it focuses (refer to
figure~\ref{fig:crossed-lenses}) and it is seen that rays incident on the lens at that tilt
angle are efficiently transmitted through it and the second lens, as indicated by the
false-colour maps of the transmission as a function of field angle.  That the region of
high transmission tracks the tilt angle of the lens shows that the transmission of the
second lens (which focuses in the $x$ direction) is not sensitive to misalignment in the
$y$ direction.  The contours in the maps of figure~\ref{fig:crossed-align} show the RMS
wavefront aberration.  For the tilts of $\chi_x$  (the left column) the aplanatic area
shrinks and only tracks the tilt at about a third of the rate of the lens tilt.  Thus, the
region of the field with low aberrations moves out of the region of high transmission and
the tolerable magnitude of the lens tilt $\chi_x$ (where the system gives high transmission and low
aberrations) is about \SI{0.2}{\milli\radian}. 

The situation is somewhat more relaxed for tilts of the first lens by $\chi_y$ about the
$y$ axis, as displayed in the right-hand column of figure~\ref{fig:crossed-align}. Just as
the $\chi_x$ tilt had little effect on the transmission of the second lens, the tilt of
$\chi_y$ has little effect on the transmission of the first lens.  Therefore, the map of
the transmission of the lens system remains unchanged. However, the aplanatic region of the
field tracks with about half the rate of the lens tilt and does not diminish in area as
rapidly as for $\chi_x$ tilts.  The tolerable magnitude of the $\chi_y$ tilt is about
\SI{0.4}{\milli\radian}.  This tilt of the first lens by $\chi_y$ is equivalent to a
rotation by $-\chi_y$ of the second lens about the axis that passes through its vertex
parallel to $y$, combined with a rotation of the field angle.  Although the dependence of
the aberrations on field angle differs for different values of $R_1$, $R_2$, and the ratio
of focal lengths, it is found that the tolerances of the lens misalignment remain
approximately the same for the various cases that were investigated in
figure~\ref{fig:field-maps}.

\begin{figure*}
  \centering
    \includegraphics[width=0.9\linewidth]{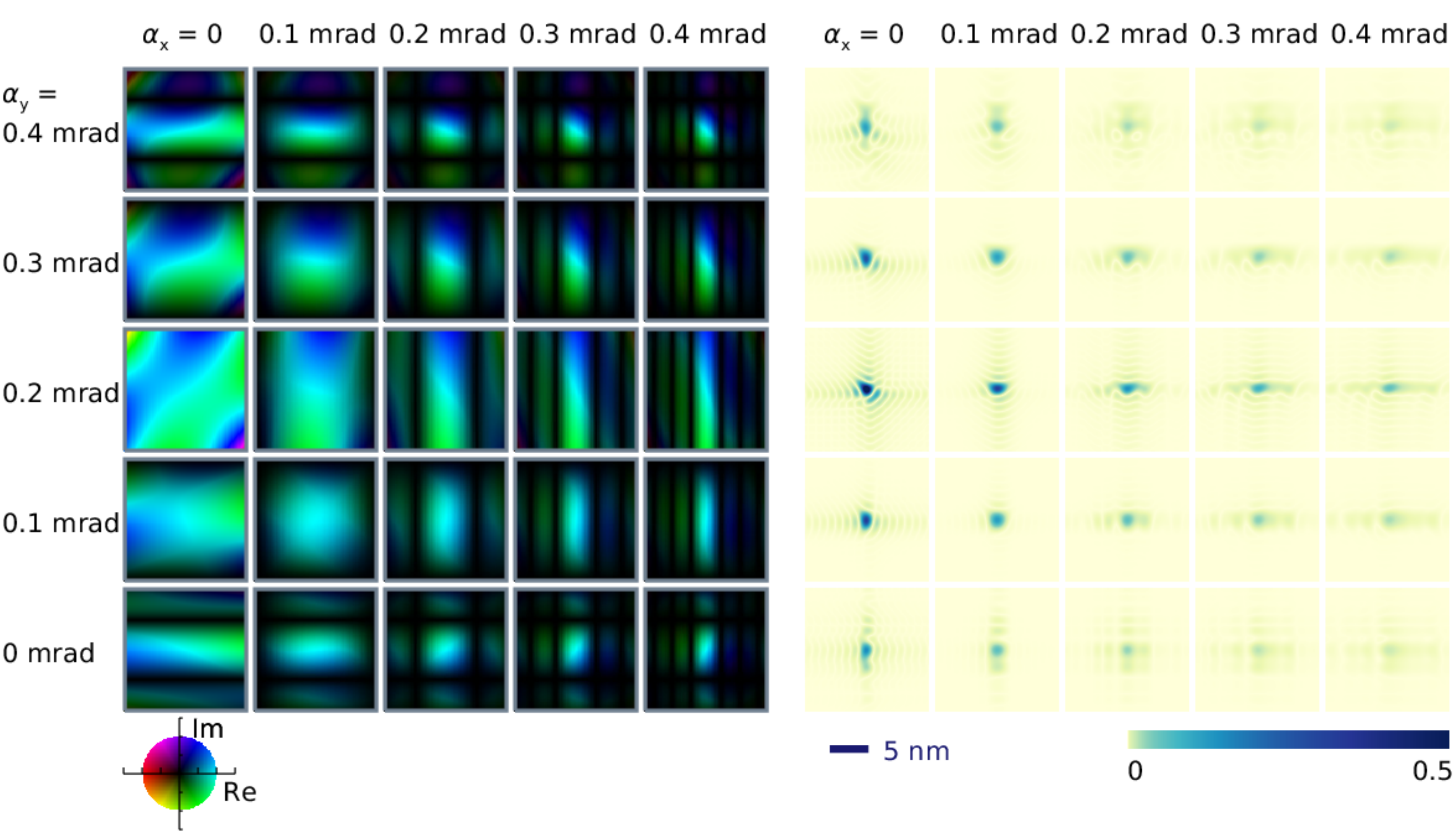}
    \caption{Complex-valued pupil functions and PSFs for a misaligned pair of 1D MLLs with
      $\text{NA} = 0.0375$ and focal lengths $f_1 = \SI{1.25}{\milli\meter}$ and
      $f_2 = \SI{1}{\milli\meter}$, for a range of field angles as indicated by $\alpha_x$
      and $\alpha_y$. The first lens is rotated by
      $\chi_x = \SI{0.4}{\milli\radian}$ and $\chi_y = \SI{0.4}{\milli\radian}$ relative
      to the second. The complex values of the pupil function are visualised by hue (phase)
      and brightness (amplitude) according to the colour wheel at bottom left. The
      corresponding PSFs are shown on a common intensity scale, which is truncated to a
      value of 0.5, compared with an intensity of 1 for a PSF of a lens with full
      amplitude and zero aberration. }
  \label{fig:PSF-1D}
\end{figure*}

The imaging performance depends both on the wavefront aberration and the apodisation, as
is visualised in figure~\ref{fig:PSF-1D} for a particular misalignment of two flat lenses
with $\chi_x = \SI{0.4}{\milli\radian}$ and $\chi_y = \SI{0.4}{\milli\radian}$,
compensated by a phase plate.  The phase plate is assumed to conjugate
the wavefront of perfectly-aligned lenses.  The
complex-valued pupil functions were constructed from the phase as given by the OPD and the
amplitude equal to the square root of the transmission, plotted in the figure 
over a range of field angles $\alpha_x$ and $\alpha_y$.  As expected from
figure~\ref{fig:crossed-align} the field position that gives maximum transmission, at
$(\alpha_x, \alpha_y) = (0, 0.2)\, \text{mrad}$ does not correspond to the location of
lowest aberration at $(\alpha_x, \alpha_y) = (0.1, 0.1)\, \text{mrad}$.  The wavefront at
the location with maximum transmission is dominated by a \SI{45}{\degree} coma.
The
width of the  apodised PSF at $(\alpha_x, \alpha_y) = (0.1, 0.1)\, \text{mrad}$ is similar
to that of the aberrated PSF at  $(\alpha_x, \alpha_y) = (0, 0.2)\, \text{mrad}$.  Further
from these field points, the PSF is dominated by the severe loss of diffraction efficiency
at the edges of the pupil.

\section{Aberrations of imperfect multilayer Laue lenses}
\label{sec:imperfect}
Imperfections of MLLs may include an error of the layer period $d$ at a particular
position of the lens, or an incorrect dependence of this period on position.  The most
trivial of these is a scaling of the lens, perhaps due to incorrect deposition rates, but
a more problematic case is the variation of the deposition rate over the course of its
fabrication.  Other errors may be introduced by slicing the lens with an angular mis-cut,
or by not achieving the necessary tilt of the layers in the deposition. Here we consider
some of these cases to understand the tolerances required to make diffraction-limited
MLLs. 

\subsection{Scale error of the MLL}
\label{sec:scaling}
An overall scaling of the MLL structure by a factor $h$ may occur due to an error in the
deposition rate used to make it.  The dilation of the structure gives rise to a phase
$\phi_m(r) = \phi(r/h)$ (of the ``manufactured'' lens in terms of the designed lens) which, from a Taylor expansion
based on (\ref{eq:phase-s}), yields a lowest-order term equal to $r^2/(2 f h^2)$. The
scaled MLL therefore has a focal length $f \,h^2$.  The scale error places the $n^\text{th}$ layer
at $h r_n$, or at the position $r_{n(m)}$ given by $r_{n(m)}/h = r_n$.  Each layer
provides an extra wavelength of path, with $\phi_m(h r_n) = \phi(r_n) = n\lambda$.  The
scaling changes the spacings of the layers and hence also the deflection angles of rays.
Since $\partial \phi(r/h)/\partial r = \phi'(r/h)/h$, it follows from
(\ref{eq:phase-gradient-flat}) that the periods of the scaled MLL are given by
$h d(r/h)$ (for the case of the flat lens).  That is, the period of the relocated
$n^\text{th}$ layer is scaled by $h$.  From (\ref{eq:d-plate}), this again shows that
the focal length becomes $h^2 f$.  The scaling also has an effect on higher-order
aberrations.  One way to conceptualise this case is to consider that scaling the lens and
the wavelength by $h$ similarly scales the focal length, as well as the OPD (in normalised
pupil coordinates), by the same factor.  That is, the spherical aberration of the scaled
lens will be zero for a wavelength $h\lambda$. Reverting to the original wavelength will
require scaling that by $1/h$ (giving a focal length $h$ times larger again), and will be
akin to the discussion of chromatic aberrations described in \sref{sec:chromatic},
where now $\Delta\lambda/\lambda = 1/h -1$.  That is, for an axi-symmetric lens originally
designed with focal length $f$ and radius $R$, the spherical aberration of the scaled lens
at the original design wavelength will be
\begin{equation}
  \label{eq:OPD-SA-scale}
  \text{OPD}_\text{SA}(\rho; h) = \frac{3 f}{8}\frac{1-h}{h}\left(1+\frac{1}{h}-\frac{2f}{R}\right)\rho^4.
\end{equation}
As with the chromatic aberration, the highest tolerance to scaling errors is achieved for
$R=f$, but the allowable scaling error is limited primarily by violating the Bragg
condition in the outermost layers, assuming that the layer tilts are also scaled (that is,
that the entire lens structure is dilated uniformly in all directions).  
For the lens parameters and wavelength considered in \sref{sec:2D-1nm}, a
scaling error of less than about \SI{1}{\percent} can be tolerated, as indicated by
figure~\ref{fig:transmission} (b).

\subsection{Deposition rate stability}
\label{sec:deposition}
A drift in the rate of deposition of materials in the fabrication of an MLL will lead to an
error in the $d$ spacing and the position of those layers.  Consider a deposition
recipe that assumes a constant rate $p$ of the accumulation of material which requires the
material pairs to be alternated at 
deposition times $T_n$ such that $r_n = p T_n$.  However, if that rate changes in time then those layers will occur at positions
\begin{equation}
  \label{eq:consumption-error}
  r_{n(m)} = \int_0^{T_n} p (1 + c(t)) \,\rmd t.
\end{equation}
A linear drift, $c(t) = c_1 t$, deposits layers according to
$r_{n(m)} = p T_n + (c_1/2) p T_n^2 = r_n + (\beta/2) r_n^2$, where $\beta = c_1/p$ is
the relative change in deposited thickness per thickness of material deposited (with units
of inverse length).  This coordinate error can be compared with the simpler scale error of
\sref{sec:scaling}, showing that
\begin{eqnarray}
  \label{eq:phase-consumption}
    \frac{\lambda}{2 \pi} \phi_m(r) &= \frac{\lambda}{2 \pi} \phi(r-\beta r^2/2) \nonumber\\
    &= s(r-\beta r^2/2) + \sqrt{(r-\beta r^2/2)^2 + (f-s(r-\beta r^2/2))^2} - f \\
    & \approx \frac{r^2}{2 f} - \frac{\beta r^3}{2 f} + \left(\frac{2}{R}-\frac{1}{f}
      \right) \frac{r^4}{8 f^2}. \nonumber
\end{eqnarray}
Here the inverse of the coordinate error was determined by solving
$r_n + \beta r_n^2 /2 = r_{n(m)}$ for $r_n$. 
When $\beta$ is
positive, the number of periods deposited over a height $r$ is reduced, and so
$\phi_m(r) < \phi(r)$ in agreement with (\ref{eq:phase-consumption}).

As with off-axis aberrations of ideal (non-distorted) lenses, the incident beam impinging
on the structure described by (\ref{eq:phase-consumption}) no longer matches a reference beam that
would give rise to a hologram described by $\phi_m$, and accordingly the Bragg diffraction
is modified due to the change in $d$ spacing. The OPD must therefore be calculated
according to (\ref{eq:OPD-R}), with a modified form of $\vec{q}$.  In particular,
since from (\ref{eq:phase-consumption}) $\partial \phi_m(r) / \partial r = (1-\beta r) \,\phi'(r-\beta r^2/2)$,
(\ref{eq:q}) must be replaced by $(1-\beta r)\,\vec{q}(r-\beta r^2/2)$.  Analogously to
the scaling error above, the period of the layer at $r$ is modified by a multiplicative factor
$1/(1-\beta r) \approx 1+\beta r$ and repositioned to $r_n+\beta r_n^2/2$. We find for
the on-axis field point of a curved axi-symmetric lens,
\begin{equation}
  \label{eq:OPD-consumption}
  \text{OPD}(\rho) \approx - \beta f^2 \rho^3, 
\end{equation}
for the normalised pupil coordinate $\rho=r/f$. For a 1D lens, a
similar scaling as for $\vec{q}$ must be applied to $\vec{q}_1$ in \tref{tab:crossed}. 
In this case we obtain the same result as (\ref{eq:OPD-consumption}) but
with $\rho$ replaced by the linear coordinate $\rho_x$ or $\rho_y$.   

Equation (\ref{eq:OPD-consumption}) shows that a linear drift of the deposition leads to a
third-order error. For an axi-symmetric lens this is not coma (which has a
direction and is not radial) and is a phase error that is approximately constant across
the field.  For a 1D lens, this does vary in a single direction as per the the coma terms
proportional to $\rho_x^3$ and $\rho_y^3$ in (\ref{eq:OPD4}).  Thus, in that case, a small degree of
deposition drift $\beta$ can be compensated by tilting the lens, such that $(f_1/R_1-1)
\alpha_y = \beta f_1$ (for $f_1 \neq R_1$). This compensation is only
practical for field angles $\alpha_y$ that are smaller than the Darwin-width acceptance
$w_\theta$ of the finest layers, which considerably limits
the range of $\beta$ that can be compensated.
The RMS OPD error for a 1D lens with coma given by (\ref{eq:OPD-consumption}) is equal
to $\beta f^2 \text{NA}^3 / (6 \sqrt{7})$, minimised here by a defocus.  The Marechal
condition is satisfied when the RMS OPD is less than $\lambda/14$, and so without any
compensation, we require
\begin{equation}
  \label{eq:alpha-tol}
  \beta < 1.13 \lambda / (f^2 \text{NA}^3) \approx 9 \delta / D^2
\end{equation}
where
$\delta = \lambda/(2 \text{NA})$ is the lens resolution and $D = 2\text{NA} f$ is the lens diameter.  For
$f=\SI{1}{\milli\meter}$, $\text{NA} = 0.037$, and $\lambda=\SI{0.075}{\nano\meter}$,
$\beta < \SI{1.6e-6}{\per\micro\meter}$ or a relative error of \num{1.5e-4} over the
course of the entire deposition.  This corresponds to an absolute error of
\SI{7.1}{\nano\meter} in the height of the lens.

Another example of a deposition error is a constant offset of the period,
$d_m(r) = d(r)+\Delta d$.  This can occur if the materials used to form the layers
interdiffuse and create a thin interface with a different density to either of the unmixed
materials~\cite{Petford-Long:1987}.  It can be seen that this is in fact equivalent to the
example just considered, since we have from (\ref{eq:d-plate}) that
$d(r) + \Delta d \approx \lambda f/r \,(1+ \Delta d \,r/(\lambda f))$, so that
$\beta = \Delta d/(\lambda f)$.  The magnitude of the coma generated by a given
contraction error $\Delta d$ increases as the wavelength is reduced, but from
(\ref{eq:alpha-tol}) a reduction in the focal length and the size of the lens relaxes
the tolerance of $\Delta d$.

\subsection{MLLs cut at an inclination}
\label{sec:inclination}
Multilayer Laue lenses are usually sliced to the desired thickness $\tau$ from an extended structure formed
by material deposition. The angle at
which a flat lens is cut might not be exactly normal to the optic axis, or the center of
curvature of a spherical surface might not exactly lie on that optic axis.  We consider
here the case of a flat lens cut at an angle that differs from the intended normal face by
an angle $\gamma$.  If the structure from which the lens is cut replicates the family of
paraboloids that are described by (\ref{eq:zp-thick}), then the resulting optical
element would be a tilted holographic lens, exhibiting zero OPD for the field point on the
optical axis.  Indeed, this will be the case for any arbitrary surface shape $z =
s(r)$. Thus, in this case, cutting a lens at an incline will only change the off-axis
aberrations.  However, if the deposited layers only match the paraboloid forms on a
particular surface (for example, $z = 0$) and differ in form away from that surface, then
a mis-cut will cause an aberrated image of the on-axis field point.  The aberration
depends on the form of the layers at the positions cut by the surface, and we consider
here the case of a flat 1D lens constructed from layers that are tilted planes.

\begin{figure}
  \centering
   \includegraphics[width=\linewidth]{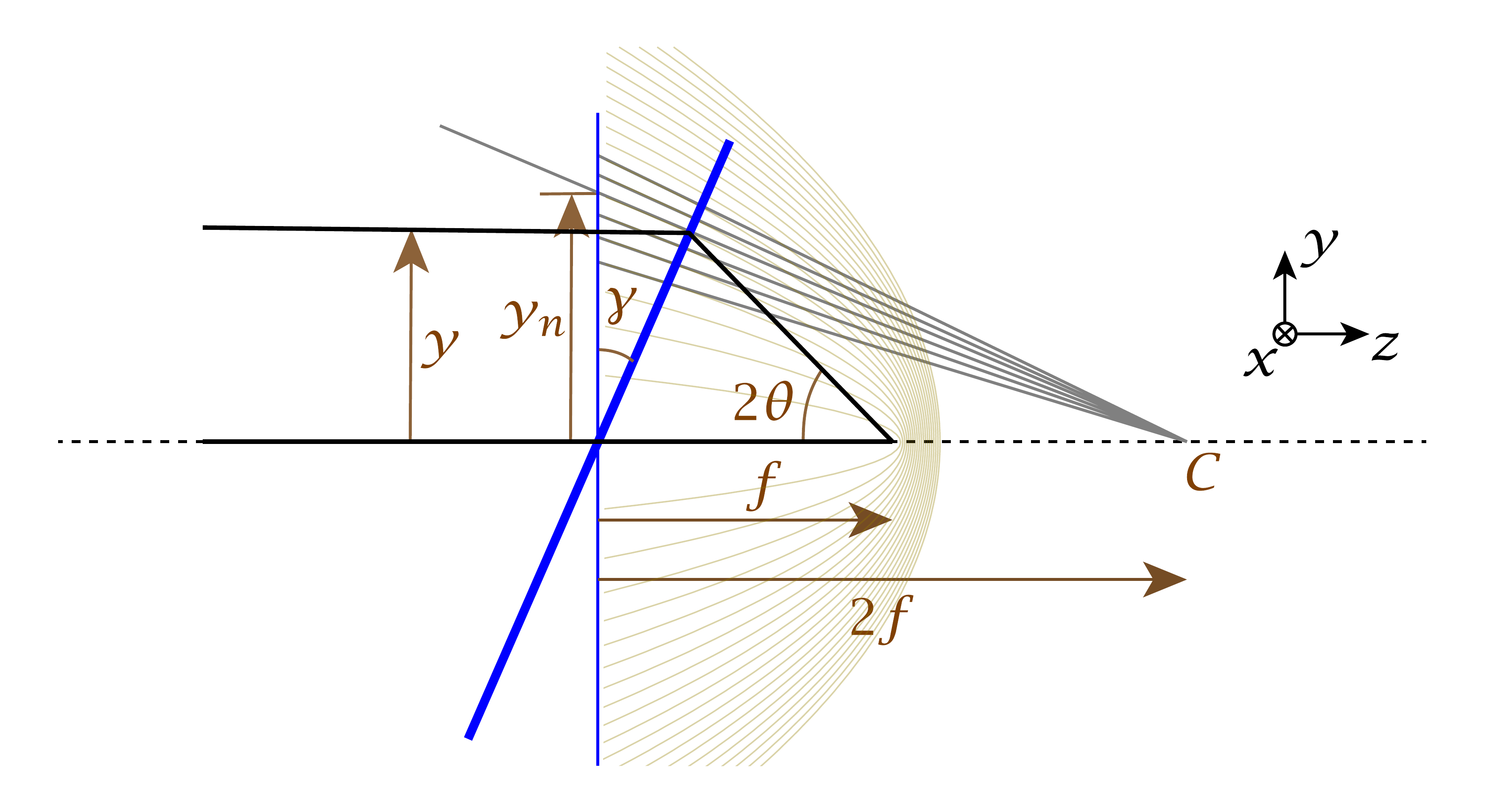}
   \caption{Geometry of a flat MLL of focal length $f$, cut at an incline with an angle
     $\gamma$. If the layers of the parent structure follow paraboloids (shown in brown)
     then rays (black lines) from an on-axis field angle will be focused without
     aberration.  Flat tilted layers (shown in grey) will cause aberrations for the
     on-axis field angle.  }
  \label{fig:incline}
\end{figure}

We consider a lens that focuses in the $y$ direction with a focal length $f$. The tilted
planes of the structure from which the lens is sliced all intersect the optic axis a
distance $2f$ from the lens vertex, as shown schematically in figure~\ref{fig:incline}. The
lens is cut on a plane that is tilted about the $x$ axis by an angle $\gamma$, with the
optic axis along $z$.  A ray from the on-axis field angle intersects the lens at a
height $y$ and a longitudinal distance $z=y \tan\gamma$ from the vertex.  The layer of the
MLL intersected by this ray, which we presume to be the $n^\text{th}$ period, passes
through the $z=0$ plane at a height $y_n = 2fy/(2f-y\tan\gamma)$. Since the lens was
intended to be cut normal to the optic axis, the position $y_n$ follows the prescription
of (\ref{eq:zp-thick}) for $z=0$.  We see from figure~\ref{fig:incline} that for positive
$\gamma$ the layers
the ray intersects are thinner and placed closer to the optic axis than intended, but have
almost the correct tilt to satisfy the Bragg condition.  With analogy to the analysis in
\sref{sec:scaling} the lens can be considered to be locally scaled by the factor
$y/y_n = 1-y/(2f) \tan \gamma$ so that
\begin{equation}
  \label{eq:phase-inclination}
  \frac{\lambda}{2 \pi} \phi_m(y) = \frac{\lambda}{2 \pi}
  \phi\left(\frac{y}{1-\nicefrac{y}{2f}\tan\gamma}\right) =
  \sqrt{f^2+\frac{y^2}{(1-\nicefrac{y}{2f}\tan\gamma)^2})} - f.
\end{equation}
Even though the surface of the MLL follows a tilted plane, we take $s(y) = 0$ in the
derivation of $\phi_m(y)$ from (\ref{eq:phase-s}) since the scaling accounts for this.  
Following the analyses of the previous sections,
\begin{equation}
  \label{eq:phase-gradient-inclination}
  \frac{\partial \phi_m(y)}{\partial y} = \frac{1}{\left(1-\nicefrac{y}{2f}\tan \gamma\right)^2}
  \phi'\left(\frac{y}{1-\nicefrac{y}{2f}\tan \gamma} \right),
\end{equation}
and the expression for $\vec{q}(y)$ must be scaled in a similar way.  The normal vector is
\begin{equation}
  \label{eq:normal-inclination}
  \hat{\vec{n}}(y) = \left(0,\,\sin \gamma,-\cos \gamma \right).
\end{equation}
For the on-axis field angle, we follow the ray-tracing procedure established in
\sref{sec:off-axis} with $l_1 = y\tan\gamma$ and $l_2\, \hat{\vec{r}}'\cdot
\hat{\vec{z}} = f-y\tan\gamma$.  Using a symbolic mathematics program the Taylor expansion
of the wavefront
aberration is evaluated to
\begin{equation}
  \label{eq:OPD-inclination}
  \text{OPD}^{(4)}(\rho_y) = \frac{f}{2}\tan \gamma \,\rho_y^3+\frac{f}{4}{\tan^2}\gamma \,\rho_y^4.
\end{equation}
The dominating term is coma, proportional to $\rho_y^3$.  This term depends linearly on
the tangent of the mis-cut angle $\gamma$.  To maintain the Marechal condition of an RMS wavefront error below
$\lambda/14$ requires a mis-cut angle
\begin{equation}
  \label{eq:gamma-tol}
  \gamma < 2.3 \frac{\lambda}{f\,\text{NA}^3},
\end{equation}
or less than \SI{3.3}{\milli\radian} for the 0.0375 NA lens considered in
\sref{sec:1nm}.

Perhaps more importantly, the linear dependence of coma on the
mis-cut angle gives a way to easily compensate for the main component of drift in the
deposition rate as explored in \sref{sec:deposition}.  The coma error of
(\ref{eq:OPD-consumption}) can be removed by cutting the lens at an angle given by
$\tan\gamma = 2\beta f$.  Setting a limit of $\tan \gamma < 1$ to ensure that the
additional spherical aberration term of (\ref{eq:OPD-inclination}) is kept manageable,
the largest drift in the deposition rate that can be tolerated is raised to $\beta =
1/(2f)$.  For a lens with $f=\SI{1}{\milli\meter}$, this corresponds to
\SI{5e-4}{\per\micro\meter} or a relative error of \SI{3.75}{\percent} over the height
of a \SI{75}{\micro\meter} diameter lens.  The range of angles at which the lens can be
cut may be limited by satisfying the Bragg condition.  The series expansion of the
expression of the deviation parameter determined
by solving $|\vec{r}'| = 1$ in the ray-trace procedure is given by
\begin{equation}
  \label{eq:epsilon-inclined}
  \epsilon = \frac{\tan \gamma}{2 \cos \gamma} \rho^3 + \frac{7 {\tan^2} \gamma}{8 \cos\gamma}\rho^4.
\end{equation}
Maintaining Bragg efficiency at the edge of the lens ($\rho = \text{NA}$) therefore
requires
\begin{equation}
  \label{eq:gamma-tol-vignette}
  \gamma < \frac{4 |\delta_1-\delta_2|}{\pi}\frac{1}{\text{NA}^3}.
\end{equation}
For the materials and parameters of the lens considered in \sref{sec:1D-MLL}, the largest tolerable cut angle
is thus about \SI{160}{\milli\radian}, which then limits the correctable deposition drift to no more than
\SI{0.8e-4}{\per\micro\meter}.

\section{Discussion}
\label{sec:conclusions}
The ray-tracing analyses of MLLs presented here, following procedures established for the
design and characterisation of holographic optical elements, can be used as a guide for the design and
manufacture of X-ray lenses by providing expressions for wavefront aberration maps of flat
and curved MLLs formed from axi-symmetric or linear multilayer structures.  These
expressions further allow an understanding of the tolerances required to fabricate such
lenses and the sizes of aplanatic areas of their image fields, as needed to design
full-field or scanning microscopes.  They also give insights into the potential
capabilities of MLLs.  As an example, we examined lenses of
\SI{1}{\milli\meter} focal length that give \SI{1}{\nano\meter} resolution at a wavelength
of \SI{0.075}{\nano\meter}.  The necessary tolerances to achieve diffraction-limited
imaging (defined here as reaching a Strehl ratio of 0.8) are quite stringent, as would be
expected in attempting to reach an RMS wavefront aberration of less than
\SI{0.005}{\nano\meter}.  Any drift of the deposition rate when fabricating the layered
structure must remain below a relative error of \num{1.5e-4}, for example, although errors about 50
times this amount can potentially be compensated by an appropriate slicing of the lens from the
structure at an angle (depending on details of the shape of the layers in the parent
structure).

The question of alignment of MLLs in their use in high-magnification full-field or scanning microscopes comes
down to the size of the aplanatic area of the image field of the lens.  For our example
cases for \SI{1}{\nano\meter} imaging, the maximum field angle for diffraction-limited
performance is about \SI{1}{\milli\radian} for a flat axi-symmetric lens, which can be
increased by a factor of six by using the aplanatic zone-plate design on a sphere of
radius equal to the focal length.  In this lens, the range of field angles is limited by
the $\pm \SI{0.3}{\milli\radian}$ acceptance of the Bragg reflection of the finest
layers, so cutting the lens to a spherical shape might not be that beneficial.  However,
for lenses of longer focal length, the analytical expressions of the aberrations indicate
that the aplanatic radius decreases with field angle, whereas the Bragg acceptance does
not.  Thus, for centimeter focal lengths there would be an advantage to a curved MLL.  We
note that curved lenses require that the layers are deposited to produce paraboloid
surfaces, at least in the volume of the structure intersected by the spherical shape of
the final lens.

Unlike thin zone plates which can be used over a large range of wavelengths limited by
spherical aberration (but with a narrow bandwidth for any given choice of wavelength),
thick MLLs of reasonably high NA can only be used over a narrow range of wavelengths that
are limited by the acceptance of Bragg diffraction.  This range depends inversely on the
square of the NA and for \SI{1}{\nano\meter} focusing at a wavelength of
\SI{0.075}{\nano\meter} with SiC/WC multilayers, this range is 
$\pm \SI{0.6}{\percent}$. The range does not depend on focal length, and so any MLL design
that achieves this resolution is limited to this wavelength range.  Other wavelengths
require require different tilts of the layers.

Most MLLs made to date are linear structures that focus only in one dimension, as a
cylindrical lens, and which are combined as a crossed pair to provide two-dimensional imaging.
While they are easier to fabricate than axi-symmetric lenses, a crossed pair does suffer from a lower
overall transmission than a single lens and cannot realise a completely null wavefront
aberration. Even when perfectly aligned, the rays from an on-axis field point that are
deflected by the first lens will not exactly match the Bragg condition in the second lens as
required to create a perfect spherical converging wave. Overcoming this requires a second
lens with layers that have a component of their tilt in the non-focusing direction. A pair
of perfect crossed 1D MLLs gives rise to an aberration on axis proportional to the product
of the squares of the pupil coordinate in each direction, which we refer to as an oblique
spherical aberration, with a minimum RMS wavefront aberration equal to $0.08 f\,\text{NA}^4$ for a
focal length $f$.  Diffraction-limited
imaging, in which this RMS aberration is smaller than $\lambda/14$ is therefore only
possible for crossed 1D MLLs of $\text{NA} < 0.016$ for focal lengths of
\SI{1}{\milli\meter} or more.  However, this aberration could be overcome in
higher NA lens pairs by conjugating the wavefront using a refractive phase plate, giving an aplanatic area similar
to achievable in an axi-symmetric MLL.
 We also
find that apodisation caused by this effect limits the highest NA of a lens system to a
value that only depends on the rocking-curve width of the multilayer system.  For the
SiC/WC system considered here, this limit is $\text{NA}=0.064$.  

Crossed lenses pose more difficulty for alignment than axi-symmetric lenses where the
aplanatic field simply tracks the region of the field that gives high diffraction
efficiency.  With crossed 1D MLLs, it is possible to tilt the lenses in such a way that
the maximum diffraction efficiency does not match the region of diffraction-limited
imaging. In practice, the alignment of such lenses may require wavefront sensing (such as
speckle tracking~\cite{Zdora:2018}) to optimise the aberration and transmission as the
relative tilts of the lenses are varied.  

The analysis presented here generally assumed on-axis lenses where the optic axis passes
through the center of the lens, but most 1D MLLs are constructed as off-axis portions of
such a lens with the optic axis outside of the lens pupil.  This does not change the
conclusions of the dependence of aberrations on parameters such as NA and focal length
since such lenses can be thought of as on-axis lenses with an additional phase tilt.
Where there is a difference is in the acceptance widths of Bragg layers.  For a given NA, the
layer periods are obviously thinner for an off-axis lens than for the on-axis version, and
these will have a narrower acceptance which will further restrict the field of view due to
vignetting.  This acceptance angle decreases linearly with pupil coordinate, so the field
size of a lens for which the aperture ranges from 0 to $D$ will be half of that which
ranges from $-D/2$ to $D/2$, for the same numerical aperture. 

The analysis presented here shows that producing MLLs at a resolution of \SI{1}{\nano\meter} is not
necessarily a challenge of making bigger lenses with more layers than currently
demonstrated or in achieving higher accuracy in the deposition of those layers, but 
which could be addressed by a judicious choice of lens parameters such as focal length and
assessing and compensating for errors that might be introduced in lens manufacture.  The
old adage of optical fabrication is that you cannot make what you cannot measure.  With
accurate metrology of lens aberrations, the analytical methods presented, combined with
an expanded design space that includes curving the lenses and ways to make corrections of
lenses, it is hoped that X-ray imaging at \SI{1}{\nano\meter} resolution will soon be realised.

\section*{Acknowledgments}
This work was supported by the Cluster of Excellence ``CUI: Advanced Imaging of Matter''
of the Deutsche Forschungsgemeinschaft (DFG) - EXC 2056 - project ID 390715994.

\bibliographystyle{iopart-num}
\section*{References}
\bibliography{meta-MLL-sub}

\end{document}